\documentclass[reqno,centertags,11pt]{amsart}

\usepackage{amsmath,amsthm,amscd,amssymb}
\usepackage{latexsym}
\usepackage {verbatim}

\usepackage{graphics}
\usepackage{amsmath}
\usepackage{amsfonts,amssymb,amsthm,eucal}
\usepackage{pstricks,multido,graphicx,psfrag}
\newcommand{\C}{\mathbb C}

\newcommand{\Z}{\mathbb Z}
\newcommand{\N}{\mathbb N}
\newcommand{\R}{\mathbb R}

\newcommand{\cal}{\mathcal}

\newcommand{\Hmm}[1]{\leavevmode{\marginpar{\tiny%
$\hbox to 0mm{\hspace*{-0.5mm}$\leftarrow$\hss}%
\vcenter{\vrule depth 0.1mm height 0.1mm width \the\marginparwidth}%
\hbox to
0mm{\hss$\rightarrow$\hspace*{-0.5mm}}$\\\relax\raggedright #1}}}

\newtheorem{theorem}{Theorem}
\newtheorem{proposition}[theorem]{Proposition}
\newtheorem{lemma}[theorem]{Lemma}
\newtheorem{corollary}[theorem]{Corollary}

\theoremstyle{definition}

\newtheorem{definition}[theorem]{Definition}

\newtheorem{remark}[theorem]{Remark}

\newcounter{theoremi}[theorem]

\numberwithin{theorem}{section} \numberwithin{equation}{section}

\def\defeq{{\mathchoice{%
\mathrel{\mskip\thickmuskip\raise.35pt\hbox{$\mathord{\displaystyle:}$}%
\hbox{$\mathord{\displaystyle=}$}\mskip\thickmuskip}}{%
\mathrel{\mskip\thickmuskip\raise.35pt\hbox{$\mathord{\displaystyle:}$}%
\hbox{$\mathord{\displaystyle=}$}\mskip\thickmuskip}}{%
\mathrel{\mskip.25\thinmuskip\raise.25pt\hbox{$\mathord{\scriptstyle:}$}%
\hbox{$\mathord{\scriptstyle=}$}\mskip.25\thinmuskip}}{%
\mathrel{\mskip.1\thinmuskip\raise.1pt\hbox{$\mathord{\scriptscriptstyle:}$}%
\hbox{$\mathord{\scriptscriptstyle=}$}\mskip.1\thinmuskip}%
}}}
\def\eqdef{{\mathchoice{%
\mathrel{\mskip\thickmuskip\hbox{$\mathord{\displaystyle=}$}%
\raise.35pt\hbox{$\mathord{\displaystyle:}$}\mskip\thickmuskip}}{%
\mathrel{\mskip\thickmuskip\hbox{$\mathord{\displaystyle=}$}%
\raise.35pt\hbox{$\mathord{\displaystyle:}$}\mskip\thickmuskip}}{%
\mathrel{\mskip.25\thinmuskip\hbox{$\mathord{\scriptstyle=}$}%
\raise.25pt\hbox{$\mathord{\scriptstyle:}$}\mskip.25\thinmuskip}}{%
\mathrel{\mskip.1\thinmuskip\hbox{$\mathord{\scriptscriptstyle=}$}%
\raise.1pt\hbox{$\mathord{\scriptscriptstyle:}$}\mskip.1\thinmuskip}%
}}}
\def\halmos{{%
\hspace*{\fill}\hbox to 18pt {\hfill\vrule width 9pt height 9pt
depth
0pt}%
}\par\ignorespaces}

\begin{document}

\title[Spectral Properties of a Polyharmonic Operator with Limit-Periodic Potential]
{Spectral Properties of  Polyharmonic Operators with Limit-Periodic Potential  in Dimension Two.}
\author[Yulia Karpeshina and Young-Ran Lee]{Yulia Karpeshina$^1$ and Young-Ran~Lee$^1$}
\address{Department of Mathematics, University of Alabama at Birmingham,
1300 University Boulevard, Birmingham, AL 35294.}
\email{karpeshi@math.uab.edu}
\address{Department of Mathematics, University of Illinois at Urbana-Champaign,
                1409 W.~Green Street, Urbana, IL 61801.}
\email{yrlee4@math.uiuc.edu}

\thanks{$^1$ Research partially supported by USNSF Grant DMS-0201383.}


%
\maketitle
\par
\noindent \hspace*{8cm}
\begin{minipage}[t]{7cm} {\em In memory of our colleague
and friend \centerline{Robert M. Kauffman.}}
\end{minipage}

\bigskip
\bigskip
\bigskip
\begin{abstract}
We consider a polyharmonic operator $H=(-\Delta)^l+V(x)$ in
dimension two with $l\geq 6$ and a limit-periodic potential
$V(x)$. We prove that the spectrum of $H$ contains a semiaxis and
there is a family of generalized eigenfunctions at every point of
this semiaxis with the following properties. First, the
eigenfunctions are close to plane waves $e^{i\langle \vec k,\vec
x\rangle }$ at the high energy region. Second, the isoenergetic
curves in the space of momenta $\vec k$ corresponding to these
eigenfunctions have a form of slightly distorted circles with
holes (Cantor type structure).
\end{abstract}

\section{Introduction} \setcounter{equation}{0}
 We study  an operator
    \begin{equation}
    H=(-\Delta)^l+V(x) \label{limper}
    \end{equation}
    in two dimensions, where $l\geq 6$, $V(x)$ is a limit-periodic potential:
    \begin{equation}V(x)=\sum _{r=1}^{\infty}V_r(x),\label{V}
    \end{equation}
    $\{V_r\}_{r=1}^{\infty}$ being a family of periodic potentials
with doubling periods and decreasing $L_{\infty}$-norms; namely,
$V_r$ has
 orthogonal periods $2^{r-1}\vec{b_1},\ 2^{r-1}\vec{b_2}$ and
$\|V_r\|_{\infty}<\hat Cexp(-2^{\eta r})$ for some $\eta>
2+64/(2l-11)$. Without loss of generality we assume that $\hat
C=1$ and $\int _{Q_r}V_r(x)dx=0$, $Q_r$ being the elementary cell
of periods corresponding to $V_r(x)$.

The one-dimensional analog of (\ref{limper}), (\ref{V}) with $l=1$
 is already thoroughly investigated. It
is proven in \cite{1}--\cite{9} that the spectrum of the operator
$H_1u=-u''+Vu$ is a Cantor type set. It has a positive Lebesgue
measure \cite{1,8}. The spectrum is absolutely continuous
\cite{1,3}, \cite{6}--\cite{7}. Generalized eigenfunctions can be
represented in the form of $e^{ikx}u(x)$, $u(x)$ being
limit-periodic \cite{6,8,9}. The case of a complex-valued
potential is studied in \cite{10}. Integrated density of states is
investigated in \cite{11}--\cite{14}. Properties of eigenfunctions
of discrete multidimensional limit-periodic Schr\"odinger
operators are studied in \cite{15}. As to the continuum
multidimensional case, it is proven \cite{14}, that the integrated
density of states for (\ref{limper}) is the limit of densities of
states for periodic operators.  Here we study properties of the
spectrum and eigenfunctions of (\ref{limper}), (\ref{V}) in the
high energy region. We prove the following results for the case
$d=2$, $l\geq 6$.
    \begin{enumerate}
    \item The spectrum of the operator (\ref{limper}), (\ref{V})
    contains a semiaxis. A proof of the analogous result by different means is to appear
    in the forthcoming paper \cite{20}. In \cite{20}, more general
    case, $8l>d+3$, $d\neq 1(\mbox{mod}4)$, is considered, however,
    under additional restriction on the potential:
       the lattices of periods  of all periodic potentials
      $V_r$ need to contain a nonzero vector $\vec \gamma$ in common,
      i.e., $V(x)$ is $\vec \gamma $-periodic.
    \item There are generalized eigenfunctions $\Psi_{\infty }(\vec k, \vec
    x)$,
    corresponding to the semiaxis, which are    close to plane waves:
    for every $\vec k $ in an extensive subset $\cal{G} _{\infty }$ of $\R^2$, there is
    a solution $\Psi_{\infty }(\vec k, \vec x)$ of the  equation
    $H\Psi _{\infty }=\lambda _{\infty }\Psi _{\infty }$ which can be described by
    the formula:
    \begin{equation}
    \Psi_{\infty }(\vec k, \vec x)
    =e^{i\langle \vec k, \vec x \rangle}\left(1+u_{\infty}(\vec k, \vec
    x)\right), \label{aplane}
    \end{equation}
    \begin{equation}
    \|u_{\infty}\|_{L_{\infty }(\R^2)}\underset{|\vec k| \rightarrow
     \infty}{=}O\left(|\vec k|^{-\gamma _1}\right),\ \ \ \gamma _1>0,
    \label{aplane1}
    \end{equation}
    where $u_{\infty}(\vec k, \vec x)$ is a limit-periodic
    function:
    \begin{equation}
    u_{\infty}(\vec k, \vec x)=\sum_{r=1}^{\infty}  u_r(\vec k, \vec
    x),\label{aplane2}
    \end{equation}
    $u_r(\vec k, \vec x)$ being periodic with periods $2^{r-1} \vec{b_1},\ 2^{r-1} \vec{b_2}$.
    The  eigenvalue $\lambda _{\infty }(\vec k)$ corresponding to
    $\Psi_{\infty }(\vec k, \vec x)$ is close to $|\vec k|^{2l}$:
    \begin{equation}
    \lambda _{\infty }(\vec k)\underset{|\vec k| \rightarrow
     \infty}{=}|\vec k|^{2l}+
    O\left(|\vec k|^{-\gamma _2}\right),\ \ \ \gamma _2>0. \label{16a}
    \end{equation}
     The ``non-resonant" set $\cal{G} _{\infty }$ of
       vectors $\vec k$, for which (\ref{aplane}) -- (\ref{16a}) hold, is
       an extensive
       Cantor type set: $\cal{G} _{\infty }=\cap _{n=1}^{\infty }\cal{G} _n$,
       where $\{\cal{G} _n\}_{n=1}^{\infty}$ is a decreasing sequence of sets in $\R^2$. Each $\cal{G} _n$ has a finite number of holes  in each bounded
       region. More and more holes appears when $n$ increases,
       however
       holes added at each step are of smaller and smaller size.
       The set $\cal{G} _{\infty }$ satisfies the estimate:
       \begin{equation}\left|\cal{G} _{\infty }\cap
        \bf B_R\right|\underset{R \rightarrow
     \infty}{=}|{\bf B_R}| \bigl(1+O(R^{-\gamma _3})\bigr),\ \ \ \gamma _3>0,\label{full}
       \end{equation}
       where $\bf B_R$ is the disk of radius $R$ centered at the
       origin, $|\cdot |$ is the Lebesgue measure in $\R^2$.

       \item The set $\cal{D}_{\infty}(\lambda)$,
defined as a level (isoenergetic) set for $\lambda _{\infty }(\vec
k)$,
$$ {\cal D} _{\infty}(\lambda)=\left\{ \vec k \in \cal{G} _{\infty }
 :\lambda _{\infty }(\vec k)=\lambda \right\},$$ is proven
to be a slightly distorted circle with infinite number of holes.
It can be described by  the formula: \begin{equation} {\cal
D}_{\infty}(\lambda)=\left\{\vec k:\vec k=\varkappa
_{\infty}(\lambda, \vec{\nu})\vec{\nu},
    \ \vec{\nu} \in {\cal B}_{\infty}(\lambda)\right\}, \label{D}
    \end{equation}
where ${\cal B}_{\infty }(\lambda )$ is a subset of the unit
circle $S_1$. The set ${\cal B}_{\infty }(\lambda )$ can be
interpreted as the set of possible  directions of propagation for
almost plane waves (\ref{aplane}). The set ${\cal B}_{\infty
}(\lambda )$ has a Cantor type structure and an asymptotically
full measure on $S_1$ as $\lambda \to \infty $:
\begin{equation}
L\bigl({\cal B}_{\infty }(\lambda )\bigr)\underset{\lambda
\rightarrow
     \infty}{=}2\pi +O\left(\lambda^{-\gamma _3/2l}\right),
\label{B}
\end{equation}
here and below $L(\cdot)$ is a length of a curve.
 The value $\varkappa _{\infty }(\lambda ,\vec \nu )$ in (\ref{D}) is the ``radius" of ${\cal D}_{\infty}(\lambda)$ in a direction $\vec \nu $.
 The function $\varkappa _{\infty }(\lambda ,\vec \nu
 )-\lambda^{1/2l}$ describes
the deviation of ${\cal D}_{\infty}(\lambda)$ from the perfect
circle of the radius $\lambda^{1/2l}$. It is proven that the
deviation is small:
\begin{equation}
\varkappa _{\infty }(\lambda ,\vec \nu )\underset{\lambda
\rightarrow
     \infty}{=}\lambda^{1/2l}+O\left(\lambda^{-\gamma _4 }\right),
 \ \ \ \gamma _4>0. \label{h}
 \end{equation}

\end{enumerate}

We will prove absolute continuity of the branch of the
    spectrum (the semiaxis) corresponding to $\Psi_{\infty }(\vec k, \vec x)$ in a
    forthcoming paper.

       To prove the results listed above we  develop a modification of the
Kolmogorov-Arnold-Moser (KAM) method. This paper is inspired by
\cite{21,22,24}, where the method is used for periodic problems.
In \cite{21} the KAM method is applied to classical Hamiltonian
systems. In \cite{22,24} the technique developed in \cite{21}  is
applied for semiclassical approximation for
         multidimensional periodic Schr\"{o}dinger operators at high
         energies.

\begin{figure}
\begin{minipage}[t]{8cm}
\centering
\includegraphics[totalheight=.25\textheight]{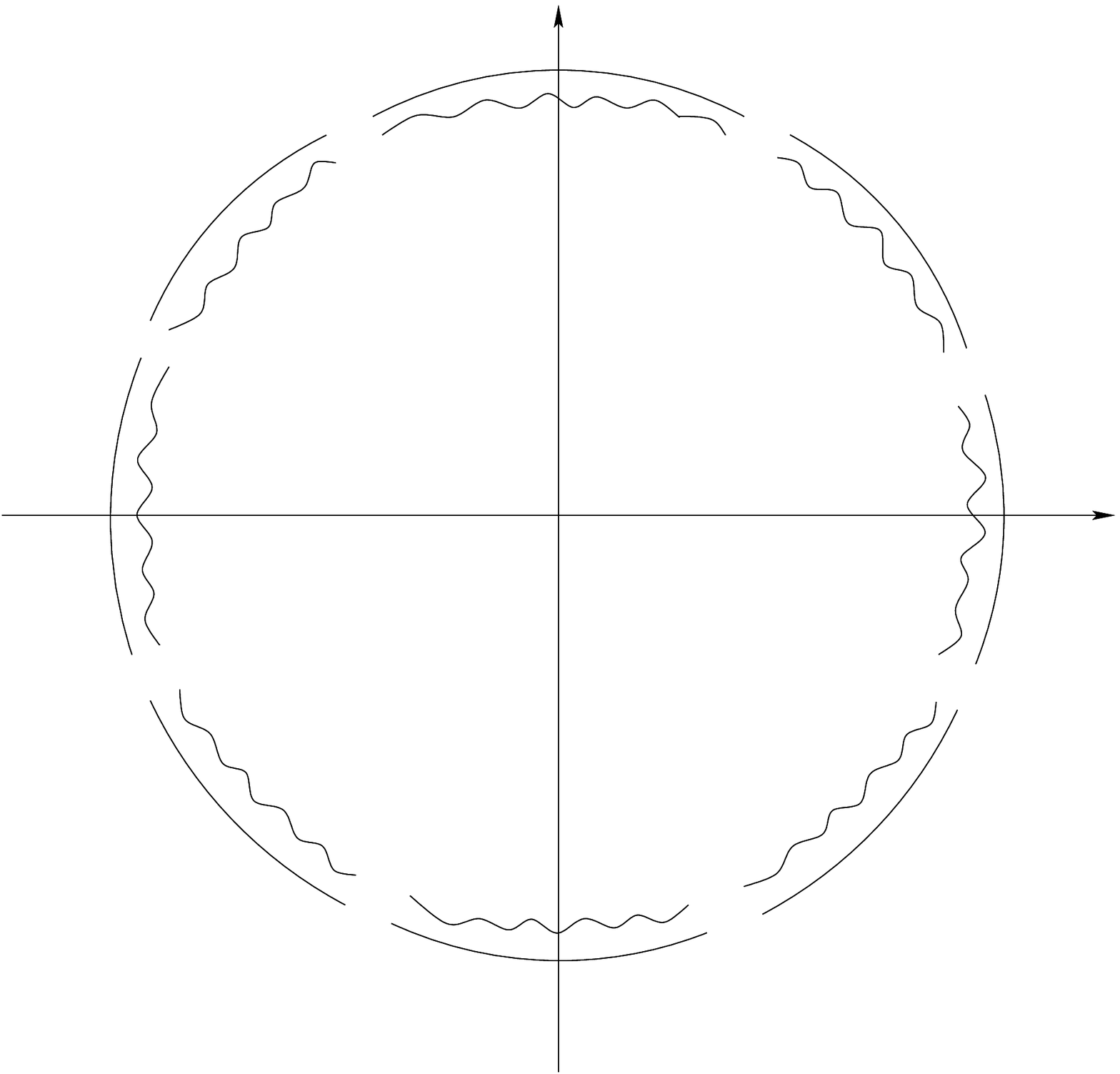}
\caption{Distorted circle with holes,
$\cal{D}_1(\lambda)$}\label{F:1}
\end{minipage}
\hfill
\begin{minipage}[t]{8cm}
\centering
\includegraphics[totalheight=.25\textheight]{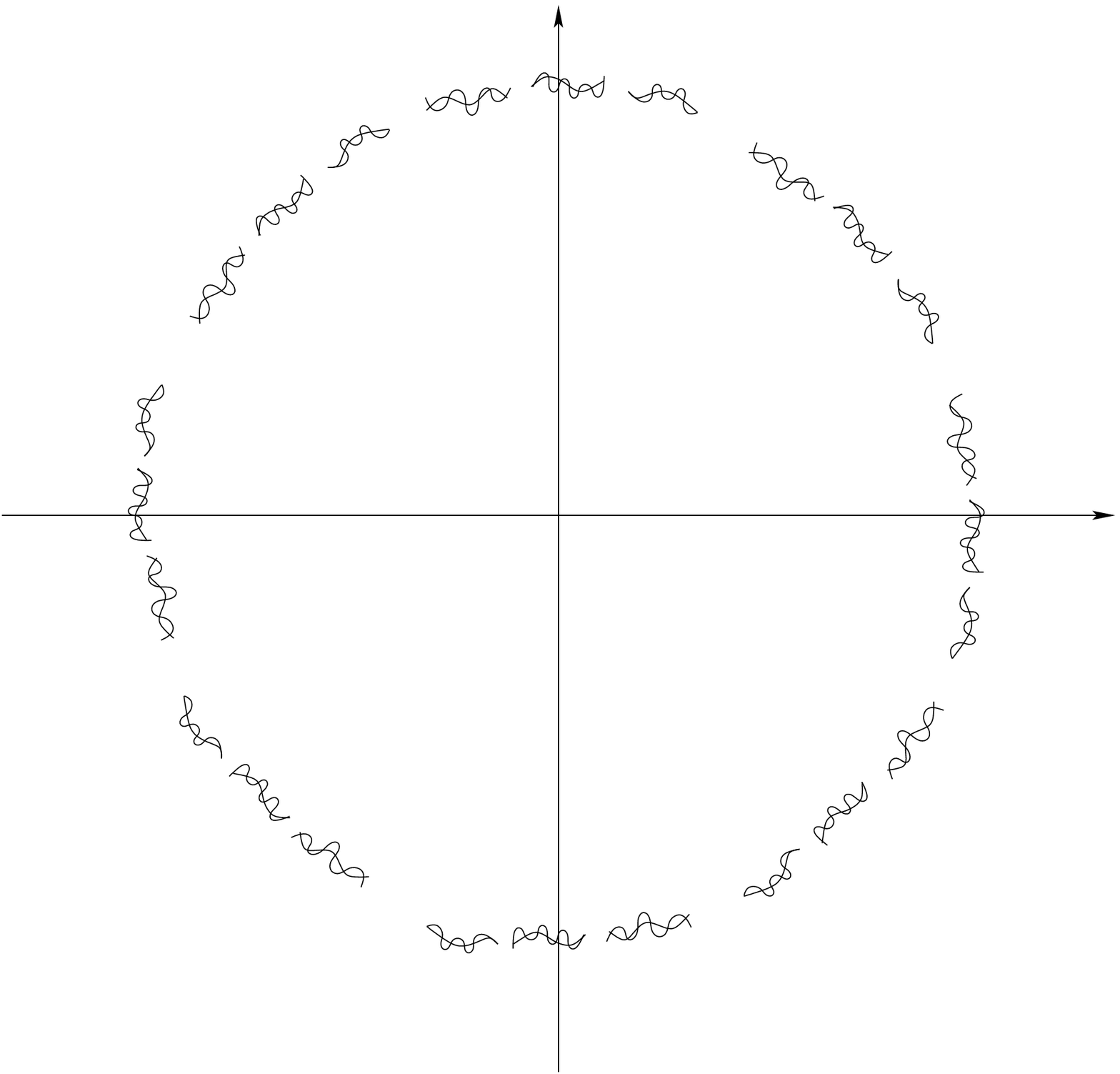}
 \caption{Distorted circle with holes,
$\cal{D}_2(\lambda)$}\label{F:2}
\end{minipage}
\hfill
\end{figure}

         We consider a sequence of operators
    $$ H_0=(-\Delta )^l, \ \ \ \ \ \
H^{(n)}=H_0+\sum_{r=1}^{M_n} V_r,\ \ \ n\geq 1, \ \ \ M_n \to
\infty \mbox{ as } n \to \infty .$$ Obviously, $\|H-H^{(n)}\|\to
0$ as $n\to \infty $ and $H^{(n)}=H^{(n-1)}+W_n$ where
$W_n=\sum_{r=M_{n-1}+1}^{M_n} V_r$. We treat each operator
$H^{(n)}$, $n\geq 1$, as a perturbation of the previous operator
 $H^{(n-1)}$. Every operator $H^{(n)}$ is periodic, however the
periods go to infinity as $n \to \infty$. We will show that there
is a $\lambda_*$, $\lambda_*=\lambda_*(V)$, such that the semiaxis
$[\lambda _*, \infty )$ is contained in the spectra of {\bf all}
operators $H^{(n)}$. For every operator $H^{(n)}$ there is a set
of eigenfunctions (corresponding to the semiaxis) close to plane
waves:
    for any $\vec k $ in an extensive subset $\cal{G} _n$ of $\R^2$, there is
    a solution $\Psi_{n}(\vec k, \vec x)$ of the differential equation
    $H^{(n)}\Psi _n=\lambda ^{(n)}\Psi _n$, which can be represented by
    the formula:
    \begin{equation}
    \Psi_n (\vec k, \vec x)
    =e^{i\langle \vec k, \vec x \rangle}\left(1+\tilde u_{n}(\vec k, \vec
    x)\right),\ \ \
\|\tilde u_{n}\|_{L_{\infty }(\R^2)}\underset{|\vec k| \rightarrow
     \infty}{=}O(|\vec k|^{-\gamma _1}),\ \
\ \gamma _1>0, \label{na}
\end{equation}
    where $\tilde u_{n}(\vec k, \vec x)$  has periods $2^{M_n-1}b_1,
    2^{M_n-1}b_2$.\footnote{Obviously, $\tilde u_{n}(\vec k, \vec x)$
    is simply related to functions $u_{r}(\vec k, \vec x)$
used in (\ref{aplane2}): $\tilde u_{n}(\vec k, \vec x)=\sum
_{r=M_{n-1}+1}^{M_n} u_{r}(\vec k, \vec x)$. } The corresponding
eigenvalue $\lambda ^{(n)}(\vec k)$ is close to $|\vec k|^{2l}$:
    $$ \lambda ^{(n)}(\vec k)\underset{|\vec k| \rightarrow
     \infty}{=}|\vec k|^{2l}+
    O\left(|\vec k|^{-\gamma _2}\right),\ \ \ \gamma _2>0.$$
     The non-resonant set $\cal{G} _{n}$
        is proven to be
       extensive in $\R^2$:
       \begin{equation}
       \left|\cal{G} _{n}\cap
       \bf B_R\right|\underset{R \rightarrow
     \infty}{=}|{\bf B_R}|\bigl(1+O(R^{-\gamma _3})\bigr). \label{16b}
       \end{equation}
The set ${\cal D}_{n}(\lambda)$ is defined as the level
(isoenergetic) set for non-resonant eigenvalue $\lambda
^{(n)}(\vec k)$:
$$ {\cal D} _{n}(\lambda)=\left\{ \vec k \in \cal{G} _n:\lambda ^{(n)}(\vec k)=\lambda \right\}.$$
 This set is proven to be a slightly distorted circle with a
finite number of holes (see Fig. \ref{F:1}, \ref{F:2}), the set
${\cal D} _{1}(\lambda)$ being strictly inside the circle of the
radius $\lambda^{1/2l}$ for sufficiently large $\lambda $. The set
${\cal D} _{n}(\lambda)$ can be described by the formula:
\begin{equation}
{\cal D}_{n}(\lambda)=\left\{\vec k:\vec k=
    \varkappa_{n}(\lambda, \vec{\nu})\vec{\nu},
    \ \vec{\nu} \in {\cal B}_{n}(\lambda)\right\}, \label{Dn}
    \end{equation}
where ${\cal B}_{n}(\lambda )$ is a subset  of the unit circle
$S_1$. The set ${\cal B}_{n}(\lambda )$ can be interpreted as the
set of possible directions of propagation for  almost plane waves
(\ref{na}). It has an asymptotically full measure on $S_1$ as
$\lambda \to \infty $:
\begin{equation}
L\bigl({\cal B}_{n}(\lambda )\bigr)\underset{\lambda \to \infty
}{=}2\pi +O\left(\lambda^{-\gamma _3/2l}\right). \label{Bn}
\end{equation}
 The set ${\cal
B}_{n}(\lambda)$ has only a finite number of holes, however their
number is growing with $n$. More and more holes of a smaller and
smaller size are added at each step. The value
$\varkappa_{n}(\lambda ,\vec \nu )-\lambda^{1/2l}$ gives the
deviation of ${\cal D}_{n}(\lambda)$ from the perfect circle of
the radius $\lambda^{1/2l}$  in the direction $\vec \nu $. It is
proven that the deviation is asymptotically small:
\begin{equation}
\varkappa_{n}(\lambda ,\vec \nu)
=\lambda^{1/2l}+O\left(\lambda^{-\gamma _4 }\right),\ \ \ \
\frac{\partial \varkappa_{n}(\lambda ,\vec \nu)}{\partial \varphi
}=O\left(\lambda^{-\gamma _5 }\right)
 \ \ \ \gamma _4, \gamma _5>0, \label{hn}
 \end{equation}
$\varphi $ being an angle variable, $\vec \nu =(\cos \varphi ,\sin
\varphi )$.

On each step more and more points are excluded from the
 non-resonant sets $\cal{G} _n$, thus $\{ \cal{G} _n \}_{n=1}^{\infty }$ is a
 decreasing sequence of sets. The set $\cal{G} _\infty $ is defined as the
 limit set: $\cal{G} _\infty=\cap _{n=1}^{\infty }\cal{G} _n $. It
 has an infinite number of holes, but nevertheless satisfies the
 relation (\ref{full}). For every $\vec
 k \in \cal{G} _\infty $ and every $n$, there is a generalized
 eigenfunction of $H^{(n)}$ of the type  (\ref{na}). It is
 proven that  the sequence of
 $\Psi _n(\vec k, \vec x)$ has a limit in $L_{\infty }(\R^2)$ when $\vec
 k \in \cal{G} _\infty $.
 The function $\Psi _{\infty }(\vec k, \vec x)
 =\lim _{n\to \infty }\Psi _n(\vec k, \vec x)$ is a generalized
 eigenfunction of $H$. It can be written in the form
 (\ref{aplane}) -- (\ref{aplane2}).
 Naturally, the corresponding eigenvalue $\lambda _{\infty }(\vec k) $ is
 the limit of $\lambda ^{(n)}(\vec k )$ as $n \to \infty $.

It is shown that $\{{\cal B}_n(\lambda)\}_{n=1}^{\infty }$  is a
decreasing sequence of sets,  on each step more and more
directions being excluded. We consider the limit ${\cal
B}_{\infty}(\lambda)$ of ${\cal B}_n(\lambda)$:
    \begin{equation}{\cal B}_{\infty}(\lambda)=\bigcap_{n=1}^{\infty} {\cal
    B}_n(\lambda).\label{Dec8a}
    \end{equation}
    This set has a Cantor type structure on the unit circle.
    It is proven that ${\cal B}_{\infty}(\lambda)$ has an asymptotically
    full measure on the unit circle (see (\ref{B})).
    We prove
    that the sequence $\varkappa _n(\lambda ,\vec \nu )$, $n=1,2,... ,$, describing the
     isoenergetic curves $\cal{D}_n(\lambda)$, quickly converges as $n\to \infty$. Hence,
 ${\cal D}_{\infty}(\lambda)$ can be described as the limit of  ${\cal
D}_n(\lambda)$ in the sense (\ref{D}), where $\varkappa
_{\infty}(\lambda, \vec{\nu})=\lim _{n \to \infty} \varkappa
_n(\lambda, \vec{\nu})$ for every $\vec{\nu} \in {\cal
B}_{\infty}(\lambda)$. It is shown that the derivatives of the
functions $\varkappa _n(\lambda, \vec{\nu})$ (with respect to the
angle variable on the unit circle) have a limit as $n\to \infty $
for every $\vec{\nu} \in {\cal B}_{\infty}(\lambda)$. We denote
this limit by $\frac{\partial \varkappa_{\infty}(\lambda ,\vec
\nu)}{\partial \varphi }$. It follows from (\ref{hn}) that
    \begin{equation}\frac{\partial \varkappa_{\infty}(\lambda ,\vec \nu)}{\partial
\varphi }=O\left(\lambda^{-\gamma _5 }\right).\label{Dec9a}
\end{equation} Thus, the limit curve ${\cal D}_{\infty}(\lambda)$ has a
tangent vector in spite of its Cantor type structure, the tangent
vector being the limit of corresponding tangent vectors for ${\cal
D}_n(\lambda)$ as $n\to \infty $.  The curve  ${\cal
D}_{\infty}(\lambda)$ looks as
  a slightly distorted circle with
infinite number of holes.

 The main technical difficulty overcome is construction of
   non-resonance sets
 $\cal{B} _n(\lambda)$ for every fixed sufficiently large $\lambda $, $\lambda >
\lambda_0 (V)$,
 where $\lambda_0(V)$ is the same for all $n$. The set
 $\cal{B} _n(\lambda)$ is obtained  by deleting a ``resonant" part
 from
 $\cal{B}_{n-1}(\lambda)$.
 Definition of  $\cal{B} _{n-1}(\lambda)\setminus \cal{B}_{n}(\lambda)$ includes Bloch
 eigenvalues of $H^{(n-1)}$. To describe  $\cal{B} _{n-1}(\lambda)\setminus \cal{B}_{n}(\lambda)$ one
 has to use
  not only non-resonant
 eigenvalues of the type (\ref{16a}), but also
 resonant
 eigenvalues, for which no suitable  formulae are known. Absence of
 formulae causes difficulties in estimating the size of $\cal{B} _{n-1}(\lambda)\setminus \cal{B}_n(\lambda)$.
       To deal with this problem we start with introducing
        an angle variable $\varphi \in [0,2\pi
       )$,  $\vec{\nu} = (\cos \varphi ,
       \sin \varphi )\in S_1$ and consider sets $\cal{B}_n(\lambda)$ in terms of this
       variable.
        Next, we show that
the resonant set  $\cal{B} _{n-1}(\lambda)\setminus
\cal{B}_{n}(\lambda)$ can be described as the set of zeros of
        determinants of the type $Det \bigl(I+S_n(\varphi )\bigr)$, $S_n(\varphi )$ being a
       trace type operator,
$$I+S_n(\varphi )=\Bigl(H^{(n-1)}\bigl(\vec \varkappa _{n-1}(\varphi )
       +\vec b\bigr)-\lambda-\epsilon \Bigr) \Bigl(H_{0}\bigl(\vec \varkappa _{n-1}(\varphi )
       +\vec b\bigr)+\lambda\Bigr)^{-1},$$
       where $\vec \varkappa _{n-1}(\varphi )$ is a vector-function describing $\cal{D}_{n-1}
       (\lambda)$: $\vec \varkappa _{n-1}(\varphi )=\varkappa _{n-1}(\lambda ,\vec \nu )\vec \nu $. To obtain $\cal{B} _{n-1}(\lambda)\setminus
       \cal{B}_{n}(\lambda)$ we take all values of $\epsilon $ in a small interval
       and vectors $\vec b$ in a finite set, $\vec b\neq 0$.
        Further, we extend our
       considerations to
        a complex neighborhood $\varPhi _0$ of $[0,2\pi
       )$. We show that the    determinants are analytic functions of
        $\varphi $ in $\varPhi _0$, and, by this,
         reduce the
        problem of estimating the size of the resonance set
         to
        a problem in complex analysis. We use theorems for analytic functions to count
          zeros of the determinants and to investigate how far  zeros move when
         $\varepsilon $ changes. It enables us to estimate the size
         of the zero set of the determinants, and, hence, the size of
         the non-resonance set $\varPhi _n\subset \varPhi _0$,  which is defined as a
         non-zero set for the determinants.
          Proving that the non-resonance set $\varPhi _n$
         is sufficiently large, we
          obtain estimates (\ref{16b}) for $\cal{G} _n$ and (\ref{Bn}) for
         $\cal{B}_n$, the set  $\cal{B}_n$ being the real part of
         $\varPhi _n$.
          To obtain $\varPhi _n$ we delete  from $\varPhi _0$ more and more holes of smaller and
         smaller radii at each step. Thus, the non-resonance set $\varPhi
         _n\subset \varPhi _0$ has a structure of Swiss
         Cheese (Fig. \ref{F:7}, \ref{F:8},  pages \pageref{F:7}, \pageref{F:8}). Deleting  resonance set from $\varPhi _0$ at each
         step of the recurrent procedure  we call a ``Swiss
         Cheese Method".  The essential
         difference of our method from those applied  in similar
         situations before (see e.g. \cite{21}--\cite{B3}) is that
 we construct a
         non-resonance set not only in the whole space of a parameter
         ($\vec k\in \R^2$ here), but also on   isoenergetic curves
         ${\cal D}_n(\lambda )$ in
         the space of the parameter, when $\lambda $ is sufficiently large. Estimates for the
         size of non-resonance sets on a curve require more subtle
         technical considerations than those sufficient for
         description of a non-resonant set in the whole space of
         the parameter.

         The restriction $l\geq 6
         $ is technical, it is needed only for
         the first two steps of the recurrent procedure. The
         requirement for super exponential decay of $\|V_r\|$ as
         $r \to \infty $ is more essential, since it is needed to
         ensure convergence of the recurrent procedure. It is not
         essential that the potential $V_r$ has doubling periods,
         the periods of the type $q^{r-1}\vec b_1,\ q^{r-1}\vec b_2, q \in
         \N,$ can be treated in the same way.
         First
         consideration of the problem was given in \cite{17}.

         The plan of the paper is the following. Sections 2 -- 6
         describe steps of the recurrent procedure. Discussion of
         convergence of the procedure and proofs of the results
         1 -- 3,
         listed above, are in Section 7.

\vspace{5mm} \noindent {\bf Acknowledgement} The authors are very
grateful to Prof. G. Stolz and Prof. G. Gallavotti for useful
discussions.



\section{The First Approximation}\label{chapt3}

 \setcounter{equation}{0}

\subsection{Operator $H^{(1)}$}

We introduce the first operator $H^{(1)}$, which corresponds to a
partial sum in the series (\ref{V}):
    \begin{equation}\label{2.1}
     H^{(1)}=(-\Delta )^l+W_1,
      \qquad
    W_1=\sum_{r=1}^{M_1}V_r,
     \end{equation}
     where $M_1$ is chosen in such a way that $
     2^{M_1}\approx k^{s_1}$ \footnote{We write $a(k)\approx b(k)$
when the inequalities $\frac{1}{2}b(k) \leq a(k) \leq 2b(k)
$ hold.} for a $k>1$, $s_1=(2l-11)/32$. Obviously, the periods of
$W_1$ are $(a_1,0)=2^{M_1-1} (b_1,0)$ and $(0,a_2)=2^{M_1-1}
(0,b_2)$, and
 $a_1\approx k^{s_1}b_1/2,$ $a_2\approx k^{s_1}b_2/2$.
     Note that
    $\|W_1\|_{\infty} \leq \sum_{r=1}^{ M_1}
    \|V_r\|_{\infty}
    =O(1)\ \text{as}\  k \to \infty. $

It is well-known (see e.g. \cite{RS})
 that
 spectral analysis of a periodic operator $H^{(1)}$ can be reduced
 to  analysis of a family of operators $H^{(1)}(t)$,
$t\in K_1$, where $K_1$ is the elementary cell of the dual
lattice, $K_1=[0,2\pi a_1^{-1})\times [0,2\pi a_2^{-1}).$ The
vector $t$ is called  $quasimomentum$. An operator $H^{(1)}(t)$,
$t\in K_1$, acts in $L_2(Q_1)$, $Q_1$ being the elementary cell of
the periods of the potential, $Q_1=[0,a_1]\times [0,a_2].$ The
operator $H^{(1)}(t)$ is described by formula (\ref{2.1}) and the
quasiperiodic conditions for a function itself and its
derivatives:
\begin{equation}
u(a_1,x_{2})= \exp (it_1a_1)u(0,x_{2}),\ \ u(x_1,a_{2})= \exp
(it_2a_2)u(x_{1},0), \label{quasi}
\end{equation}
$$u_{x_1}^{(j)}(a_1,x_{2})=\exp (it_1a_1)u_{x_1}^{(j)}(0,x_{2}), \ \
u_{x_2}^{(j)}(x_1,a_{2})=\exp (it_2a_2)u_{x_2}^{(j)}(x_{1},0),
$$
$0<j<2l$. Each operator $H^{(1)}(t)$, $t\in K_1$, has a discrete
bounded below spectrum $\Lambda ^{(1)}(t)$:
$$\Lambda ^{(1)}(t)=\cup _{n=1}^{\infty }\lambda _n^{(1)}(t),\
\lambda^{(1)} _n(t)\to _{n\to \infty }\infty .$$
The spectrum $\Lambda ^{(1)}$ of the operator $H^{(1)}$ is the
union of the spectra of
 the operators $H^{(1)}(t)$:
$\Lambda ^{(1)}=\cup _{t\in K_1}\Lambda (t)=\cup _{n\in N,t\in
K_1}\lambda _n^{(1)}(t).$ The functions $\lambda _n^{(1)}(t)$ are
continuous, so $\Lambda ^{(1)}$ has a band structure:
\begin{equation}\Lambda ^{(1)}=\cup _{n=1}^{\infty }[q_n^{(1)},Q_n^{(1)}],\ q_n^{(1)}
=\min _{t\in K_1}\lambda _n^{(1)}(t),\
 Q_n^{(1)}=\max _{t\in K_1}\lambda ^{(1)}_n(t). \label{bands}
 \end{equation}
Eigenfunctions of $H^{(1)}(t)$ and $H^{(1)}$ are simply related.
Extending all the eigenfunctions of the operators $H^{(1)}(t)$
quasiperiodically (see (\ref{quasi})) to $\R^2$, we obtain a
complete system of generalized eigenfunctions of
 $H^{(1)}$.

  Let $H_0^{(1)}$ be the operator (\ref{limper}) corresponding to  $V=0$.
  We consider that it has periods $a_1,a_2$
  and that operators $H_0^{(1)}(t)$, $t\in K_1$, are defined in $L_2(Q_1)$. Eigenfunctions of an
  operator $H_0^{(1)}(t)$, $t\in K_1$, are   plane waves
  satisfying (\ref{quasi}).
  They are naturally
 indexed by   points of $\Z^2$:
 $
 \Psi _j^0(t,x)=|Q_1|^{-1/2}\exp i \langle \vec{p}_j(t),x \rangle $, $j\in \Z^2$,
 the  eigenvalue corresponding to $\Psi _j^0(t,x)$ being equal to $p_j^{2l}(t)$,
  here and below $\vec{p}_j(t)=2\pi j/a+t$, $2\pi j/a=(2\pi j_1/a_1, 2\pi j_2/a_2),\ j \in
  \Z^2,$
 $|Q_1|=a_1a_2$ and $p_j^{2l}(t)=|\vec{p}_j(t)|^{2l}$.

\begin{figure}
\centering

\begin{pspicture}(0,0)(6.5,6.5)
\psset{unit=.9in}

\def\4row{
\pscircle(0.5,0.5){0.5} \qdisk(0.5,0.5){1.5pt}
\psline[linewidth=0.5pt](-0.2,0.5)(1.2,0.5)
\psline[linewidth=0.5pt](0.5,-0.2)(0.5,1.2)}

\multips(0,0)(0.4,0.0){4}{\4row}

\multido{\n=0+0.4}{4} {\multips(\n,0)(0.0,0.4){4}{\4row}}

\psline[linewidth=2pt](0.9,0.9)(1.3,0.9)
\psline[linewidth=2pt](0.9,0.9)(0.9,1.3)
\psline[linewidth=2pt](0.9,1.3)(1.3,1.3)
\psline[linewidth=2pt](1.3,0.9)(1.3,1.3)

\end{pspicture}

\hspace{5mm}

 \caption{The isoenergetic surface $S_0(\lambda)$ of
the free operator $H_0^{(1)}$}\label{F:3}
\end{figure}
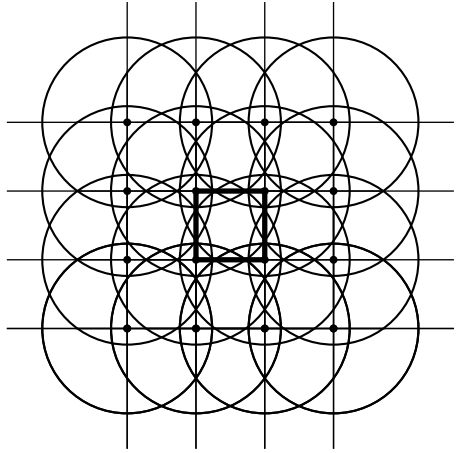

Next, we introduce an isoenergetic surface\footnote{``surface" is
a traditional term.
   In our case, it is a curve.} $S_0(\lambda)$ of the free
operator $H_0^{(1)}$. A point $t$ belongs to $S_0(\lambda)$ if and
only if $H_0^{(1)}(t)$ has an eigenvalue equal to $\lambda$, i.e.,
there exists a $j \in \Z^2$ such that $p_j^{2l}(t)=\lambda$. This
surface can be obtained as follows: the circle of radius
$k=\lambda ^{\frac{1}{2l}}$ centered at the origin is divided into
pieces by the dual lattice $\{\vec p_q(0)\}_{q \in \Z^2}$, and
then all pieces are translated in a parallel manner into the cell
$K_1$ of the dual lattice. We also can  get $S_0(\lambda)$ by
drawing sufficiently many circles of radii $k$ centered at the
dual lattice $\{\vec p_q(0)\}_{q \in \Z^2}$ and by looking at the
figure in the cell $K_1$. As a result of any of these two
procedures we obtain a circle of radius $k$ ``packed into the bag
$K_1$" as it is shown in the Fig.~\ref{F:3}. Note that each piece
of $S_0(\lambda )$ can be described by an equation
$p_j^{2l}(t)=\lambda$ for a fixed $j$. If $t\in S_0(\lambda )$,
then $j$ can be uniquely defined from the last equation, unless
$t$ is not the point of a self-intersection of the isoenergetic
surface. A point $t$ is a self-intersection of $S_0(\lambda )$ if
and only if
\begin{equation}
p_q^{2l}(t)=p_j^{2l}(t)=k^{2l}  \label{Apr2} \end{equation} for at
least on pair of indices $q,\ j$, $q\neq j$.

 Note that any vector $\vec \varkappa $ in $\R ^2$ can be uniquely represented
 in the form $\vec \varkappa =\vec{p}_j(t)$, where $j\in \Z ^2$ and
$t\in K_1$. Let $\cal K_1$ be the parallel shift into $K_1$:
${\cal K_1}:\R ^2\to K_1$, ${\cal K_1}\vec{p}_j(t)=t$.
Suppose $\Omega \subset \R ^2$. In order to obtain ${\cal
K}_1\Omega$ it is necessary to partition $\Omega $ by the lattice
with nodes at the points $\vec{p}_q(0)$,
 $q\in \Z ^2$ and to shift all parts in a parallel manner into a single cell.
It is obvious that $ \left|{\cal K_1}\Omega \right|\leq
\left|\Omega \right|$ for any $\Omega $. If $\Omega $ is a smooth
curve, then
\begin{equation}\label{2.51*}
L({\cal K_1}\Omega )\leq L(\Omega).
\end{equation}
 For any pair of
sets $\Omega _1$ and $\Omega _2$, ${\cal K_1}\bigl(\Omega _1\cup
\Omega _2\bigr)={\cal K_1}\Omega _1\cup {\cal K_1}\Omega _2$.
Obviously, ${\cal K_1}S_k=S_0(\lambda )$ and $L \bigl(S_0(\lambda
)\bigr)=L (S_k )=2\pi k$, $k=\lambda ^{\frac{1}{2l}}$, $S_k$ being
the circle of radius $k$ centered at the origin.

The operator $H^{(1)}$ has the following matrix representation in
the basis of plane waves: $H^{(1)}(t)_{mq}=p_m^{2l}(t)\delta
_{mq}+
    w_{m-q}$,\ \footnote{$\delta_{mq}$ is the Kronecker symbol.}
    here and below $w_{m-q}$ are Fourier coefficients of $W_1$, the
    coefficient $w_0$ being equal to zero.

\subsection{Perturbation Formulae} 
\begin{figure}
\centering
\includegraphics[totalheight=.2\textheight]{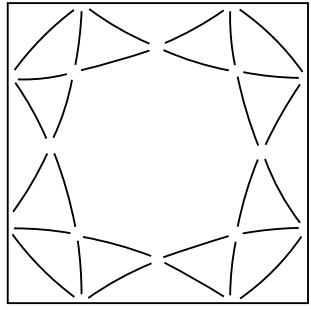}
\caption{The first non-resonance set $\chi _1(\lambda,\delta
)$}\label{F:4}
\end{figure}

In this section we consider
operator $H^{(1)}(t)$ as a perturbation of the free operator
$H^{(1)}_0(t)$. We show that for every sufficiently large $\lambda
$ there is a ``non-resonant" subset $\chi _1(\lambda )$ of
$S_0(\lambda )$ such that perturbation series for an eigenvalue
and a spectral projection of $H^{(1)}(t)$ converge when $t\in \chi
_1(\lambda )$. The set $\chi _1(\lambda )$ is obtained by deleting
small neighborhoods of selfintersections of $S_0(\lambda )$, see
Fig. \ref{F:4}. The selfintersections are described by
(\ref{Apr2}) and correspond to degenerated eigenvalues of
$H^{(1)}_0(t)$. The size of the neighborhood is $k^{-1-4s_1-\delta
}$, $k=\lambda ^{\frac{1}{2l}}$, $\delta $ being a small positive
number. The set $\chi _1(\lambda )$ is sufficiently large: its
relative measure with respect to $S_0(\lambda )$ tends to 1 as
$\lambda \to \infty $. The precise formulation of these results is
given in the next Geometric Lemma. The lemma is proven by
elementary geometric considerations in \cite{19}.

\begin{lemma}[Geometric Lemma]\label{L:2.1}
For an arbitrarily small positive $\delta $, $2\delta <2l-2-4s_1,$
and sufficiently large $\lambda $, $\lambda
>\lambda _0(\delta)$, there exists a nonresonance set
 $\chi _1(\lambda,\delta )\subset S_0(\lambda )$ such that:

\begin{enumerate}
\item For any point $t\in \chi_1$ the following conditions hold:
\begin{enumerate}
    \item There exists a unique $j\in \Z ^2$ such that $p_j(t)=k$, $k=\lambda ^{\frac{1}{2l}}$.
    \item The inequality holds:
    \begin{equation}\label{2.5}
    \min _{i\neq j}|p_j^{2}(t)-p_i^{2}(t)|
    >2k^{-4s_1-\delta }.
    \end{equation}
\end{enumerate}

\item For any $t$ in the $(2k^{-1-4s_1-2\delta })$-neighborhood of
the nonresonance set in $\C^2$, there exists a unique $j\in \Z ^2$
such that
    \begin{equation}\label{2.6}
    |p_j^2(t)-k^2| < 5k^{-4s_1-2\delta }
    \end{equation}
and  the inequality~(\ref{2.5}) holds.

\item The nonresonance set $\chi_1$ has an asymptotically full
measure on $S_0(\lambda)$ in the following sense:
\begin{equation}\label{2.7}
\frac{L\left(S_0(\lambda )\setminus \chi _1(\lambda,\delta
)\right)}{L\left(S_0(\lambda)\right)}=_{\lambda\to \infty }O(k^{-
\delta /2}).
\end{equation}
 \end{enumerate}
\end{lemma}

\begin{corollary}\label{C:2.2}
If $t$ belongs to the $(2k^{-1-4s_1-2\delta })$-neighborhood of
the nonresonance set $\chi _1(\lambda, \delta )$ in $\C^2$, then,
for any $z$ lying on the circle \begin{equation} C_1=\{ z:
|z-k^{2l}| =k^{2l-2-4s_1-\delta }\} \label{Apr3} \end{equation}
 and any $i$ in $ \Z^2$, the following inequality holds:
\begin{equation}\label{2.8}
2| p_i^{2l}(t)-z| >k^{2l-2-4s_1-\delta }.
\end{equation}
\end{corollary}

\begin{proof}[Corollary \ref{C:2.2}]
Let $t\in \chi _1(\lambda, \delta )$. Taking into account the
relation $p_j^{2l}(t)=k^{2l}$ and the definition of $C_1$, we see
that
\begin{equation}
 |p_j^{2l}(t)-z| =k^{2l-2-4s_1-\delta }. \label{2.9}
 \end{equation}
Thus, the estimate~(\ref{2.8}) is valid for $i=j$. From
(\ref{2.5}), it follows that $\min _{i\neq
j}|p_j^{2l}(t)-p_i^{2l}(t)|>     2k^{2l-2-4s_1-\delta }.$
%
%
Using the last inequality and (\ref{2.9}), we obtain
inequality~(\ref{2.8}) for $i\neq j$. It is easy to see that all
the estimates are stable under a perturbation of $t$ of order
$2k^{-1-4s_1-2\delta }$. Therefore, the estimate (\ref{2.8}) can
be extended to the $(2k^{-1-4s_1-2\delta })$-neighborhood of
$\chi_1$.
 \end{proof}

Let $E_j(t)$ be the spectral projection of the free operator,
corresponding to the eigenvalue $p_j^{2l}(t):$ $(E_j)_{rm}=\delta
_{jr}\delta _{jm} $. In the $(2k^{-1-4s_1-2\delta})$-neighborhood
of $\chi _1(\lambda ,\delta )$, we define functions
$g_r^{(1)}(k,t)$ and operator-valued functions $G_r^{(1)}(k,t)$,
$r=1, 2, \cdots$ as follows:
\begin{equation}\label{2.11}
 g_r^{(1)}(k,t)=\frac{(-1)^r}{2\pi ir}\mbox{Tr}\oint
 _{C_1}((H_0^{(1)}(t)-z)^{-1}W_1)^rdz,
 \end{equation}
\begin{equation}\label{2.12}
G_r^{(1)}(k,t)=\frac{(-1)^{r+1}}{2\pi i}\oint
_{C_1}((H_0^{(1)}(t)-z)^{-1}W_1)^r (H_0^{(1)}(t)-z)^{-1}dz.
 \end{equation}
To find $g_r^{(1)}(k,t)$ and $G_r^{(1)}(k,t)$ it is necessary to
compute the residues of a rational function of a simple structure,
whose numerator does not depend on $z$, while the denominator is a
product of factors of the type $(p_i^{2l}(t)-z)$. For all $t$ in
the nonresonance set within $C_1$ the integrand has a single pole
at the point $z=k^{2l}=p_j^{2l}(t).$  By computing the residue at
this point, we obtain explicit expressions for $g_r^{(1)}(k,t)$
and $G_r^{(1)}(k,t)$. For example, $g_1^{(1)}(k,t)=0$,
   \begin{align}
    g_2^{(1)}(k,t)&=\sum _{q\in \Z^2,q\neq 0}| w_q| ^2(p_j^{2l}(t)-
        p_{j+q}^{2l}(t))^{-1} \notag \\
    &=\sum _{q\in \Z^2,q\neq 0}\frac{| w_q| ^2(2p_j^{2l}(t)-
    p_{j+q}^{2l}(t)-p_{j-q}^{2l}(t))}{2(p_j^{2l}(t)-
    p_{j+q}^{2l}(t))(p_j^{2l}(t)-p_{j-q}^{2l}(t))},\label{2.14}
    \end{align}
\begin{equation}\label{2.15}
G_1^{(1)}(k,t)_{rm}=\frac{w_{j-m}}{p_j^{2l}(t)-p_m^{2l}(t)}\delta
_{rj}+ \frac{w_{r-j}}{p_j^{2l}(t)-p_r^{2l}(t)}\delta _{mj}, \quad
G_1^{(1)}(k,t)_{jj}=0.
\end{equation}
It is not difficult to show that $g_2^{(1)}(k,t)>0$ for
sufficiently large $\lambda $. For technical reasons it is
convenient to introduce parameter $\alpha $ in front of the
potential $W_1$. Namely, $H_{\alpha }^{(1)}=(-\Delta )^l +\alpha
W_1$, $0\leq \alpha \leq 1$, $H_1^{(1)}\equiv H^{(1)}.$
\begin{theorem}\label{T:2.3}
 Suppose $t$ belongs to the
$(2k^{-1-4s_1-2\delta })$-neighborhood in $K_1$ of the
nonresonance set $\chi _1(\lambda ,\delta )$, $0<2\delta
<2l-2-4s_1$. Then for sufficiently large $\lambda $, $\lambda
>\lambda _0(\|V\|,b_1,b_2,\delta )$, for all $\alpha $, $-1\leq \alpha \leq
1$, there exists a single eigenvalue of the operator $H_{\alpha
}^{(1)}(t)$
 in
the interval $\varepsilon_1 (k,\delta ):=
(k^{2l}-k^{2l-2-4s_1-\delta }, k^{2l}+k^{2l-2-4s_1-\delta })$. It
is given by the series:
\begin{equation}\label{2.16}
\lambda_j^{(1)} (\alpha ,t)=p_j^{2l}(t)+\sum _{r=2}^{\infty
}\alpha ^rg_r^{(1)}(k,t),
\end{equation}
converging absolutely in the disk $|\alpha|  \leq 1$, where the
index $j$ is  determined  according to articles 1(a) and 2 of
Geometric Lemma. The spectral projection, corresponding to
$\lambda_j^{(1)} (\alpha ,t)$, is given by the series:
\begin{equation}\label{2.17}
E_j^{(1)} (\alpha ,t)=E_j+\sum _{r=1}^{\infty }\alpha
^rG_r^{(1)}(k,t),
\end{equation}
which converges in the class $\mathbf{S_1}$ uniformly with respect
to $\alpha $ in the disk  $| \alpha | \leq 1$.

For coefficients $g_r^{(1)}(k,t)$,  $G_r^{(1)}(k,t)$ the following
estimates hold:
%
\begin{equation}\label{2.18}
| g_r^{(1)}(k,t) |<k^{2l-2-4s_1-\gamma _0r-\delta },\ \ \
%
 \| G_r^{(1)}(k,t)\| _1<k^{-\gamma _0r},
\end{equation}
where $\gamma _0=2l-2-4s_1-2\delta $.
\end{theorem}
\begin{corollary}\label{C:2.4}
For the perturbed eigenvalue and its spectral projection the
following estimates are valid:
\begin{equation}\label{2.20}
| \lambda_j^{(1)} (\alpha ,t)-p_j^{2l}(t)| \leq 2\alpha
^2k^{2l-2-4s_1-2\gamma _0-\delta },
\end{equation}
\begin{equation}\label{2.21}
\|E_j^{(1)} (\alpha ,t)-E_j\|_1\leq 2 | \alpha | k^{-\gamma _0}.
\end{equation}
\end{corollary}

\begin{remark} \label{R:May10} The theorem states that $\lambda_j^{(1)} (\alpha ,t)$ is a
single eigenvalue in the interval $\varepsilon_1 (k,\delta )$.
This means that $|\lambda_j^{(1)} (\alpha
,t)-k^{2l}|<k^{2l-2-4s_1-\delta }$. Formula (\ref{2.20}) provides
a stronger estimate on the location of $\lambda_j^{(1)} (\alpha
,t)$, the right-hand side of (\ref{2.20}) being smaller than the
size of $\varepsilon _1$.\end{remark}

%
\begin{proof}
The proof of the theorem is based on expanding the resolvent
$(H_{\alpha}^{(1)}(t)-z)^{-1}$ in a perturbation series for $z$,
belonging to the contour $C_1$ about the unperturbed eigenvalue
$p_j^{2l}(t)$. Then integrating the resolvent yields the formulae
for the perturbed eigenvalue and its spectral projection.

It is obvious that
\begin{equation}\label{2.22}
(H_{\alpha }^{(1)}(t)-z)^{-1}=(H_0^{(1)}(t)-z)^{-1}(I-\alpha
A_1)^{-1},\ \ \ A_1 := -W_1(H_0^{(1)}(t)-z)^{-1}.
\end{equation}
 Suppose
$z\in C_1$. Using~(\ref{2.8}), we obtain:
\begin{equation}\label{2.23}
\|(H_0^{(1)}(t)-z)^{-1}\| \leq 2 k^{-2l+2+4s_1+\delta},\ \ \
\|A_1\| \leq 2\|W_1\| k^{-\gamma _0-\delta}.
\end{equation}
%
%
Thus, $\|A_1\|\ll 1$ for sufficiently large $k$. Expanding
$(I-\alpha A_1)^{-1}$ in a series in powers of $\alpha A_1$ we
obtain:
\begin{equation}\label{2.25}
(H_{\alpha }^{(1)}(t)-z)^{-1}-(H_0^{(1)}(t)-z)^{-1}=\sum
_{r=1}^{\infty }\alpha ^r (H_0^{(1)}(t)-z)^{-1}A_1^r.
\end{equation}
Since $2l-2-2s_1>0$ it is not hard to show that
$(H_0^{(1)}(t)-z)^{-1}\in \mathbf{S_1}$. Taking into account
estimates~(\ref{2.23}), we see that the series~(\ref{2.25})
converges in the class $\mathbf{S_1}$ uniformly with respect to
$\alpha $ in the disk $|\alpha | \leq 1$. Substituting the series
into the formula for a spectral projection $E_j^{(1)} (\alpha
,t)=-\frac{1}{2\pi i}\oint _{C_1}(H_{\alpha }^{(1)}(t)-z)^{-1}dz$
and integrating termwise, we arrive at (\ref{2.17}). Formula
(\ref{2.16}) is obtained in the analogous way. Estimates
(\ref{2.18}) follow from (\ref{2.23}), for details see \cite{19}.

 \end{proof}

Next, we show that  the series~(\ref{2.16}),~(\ref{2.17}) can be
extended as holomorphic functions of $t$ in a complex neighborhood
of $\chi _1$; they can be differentiated  with respect to $t$ any
number of times with retaining their asymptotic character.

Let us introduce the notations:
$$T(m) :=\frac{\partial ^{|m|}}{\partial t_1^{m_1} \partial
t_2^{m_2}},\ \ m=(m_1,m_2),\ \ |m|:=m_1+m_2,\ \ m ! :=m_1 ! m_2
!,\
 T(0)f:=f.$$
%
\begin{lemma}\label{L:2.4.1/2a} The coefficients $g_r^{(1)}(k,t),$ and $G_r^{(1)}(k,t)$ can
be continued as holomorphic functions of two variables from the
real $(2k^{-1-4s_1-2\delta })$-neighborhood of the nonresonance
set $\chi_1$ to its complex $(2k^{-1-4s_1-2\delta })$-neighborhood
and the following estimates hold in the complex neighborhood:
%
\begin{equation}\label{2.43}
|T(m)g_r^{(1)}(k,t)| <m!k^{2l-2-4s_1-\delta -\gamma _0r+| m|
(1+4s_1+2\delta )},\end{equation}
\begin{equation}\label{2.44}
\| T(m)G_r^{(1)}(k,t)\|_1 <m!k^{-\gamma _0 r+| m| (1+4s_1+2\delta
)}.
\end{equation}
\end{lemma}

\begin{proof} Since~(\ref{2.8}) is valid in the complex $(2k^{-1-4s_1-2\delta
})$-neighborhood of the nonresonance set, it is not hard to see
that the coefficients $g_r^{(1)}(k,t),$ and $G_r^{(1)}(k,t)$ can
be continued from the real $(2k^{-1-4s_1-2\delta })$-neighborhood
of $t$ to the complex $(2k^{-1-4s_1-2\delta })$-neighborhood as
holomorphic functions of two variables, and
inequalities~(\ref{2.18}) are hereby preserved. Estimating by
means of the Cauchy integral formula, the value of the derivative
with respect to $t$ in terms of the value of the function itself
on the boundary of the $(2k^{-1-4s_1-2\delta })$-neighborhood of
$t$ (formulas~(\ref{2.18})), we obtain ~(\ref{2.43}) and
~(\ref{2.44}).
 \end{proof}

From this lemma the following theorem easily follows.

\begin{theorem}\label{T:2.5}
The series~(\ref{2.16}),~(\ref{2.17}) can be continued as
holomorphic functions of two variables from the real
$(2k^{-1-4s_1-2\delta })$-neighborhood of the nonresonance set
$\chi_1$ to its complex $(2k^{-1-4s_1-2\delta })$-neighborhood and
the following estimates hold in the complex neighborhood:
%
\begin{equation}\label{2.45}
| T(m)(\lambda_j^{(1)} (\alpha ,t)-p_j^{2l}(t))| <2m!\alpha
^2k^{2l-2-4s_1-\delta - 2\gamma _0+| m| (1+4s_1+2\delta )},
\end{equation}
\begin{equation}\label{2.46}
\| T(m)(E_j^{(1)} (\alpha ,t)-E_j)\|_1 <2m!\alpha k^{-\gamma
 _0+|m| (1+4s_1+2\delta )}.
\end{equation}
\end{theorem}
\begin{corollary}\label{C:2.6}
\begin{equation}\label{2.20*}
\nabla \lambda_j^{(1)} (\alpha ,t)=2lk^{2l-2}\vec k
+o\left(k^{2l-1}\right),\ \ \ \ \ \vec k=\vec p_j(t),
\end{equation}
\begin{equation}
 T(m)\lambda_j^{(1)} (\alpha ,t)<4l^2k^{2l-2},\ \ \ \mbox{if} \ \ \
|m|=2. \label{2.20a}
\end{equation}
\end{corollary}

The next lemma will be used in the second step of approximation
where the operator $H^{(1)}(t)$ will play a role of the initial
(unperturbed) operator.

\begin{lemma}\label{L:2.4.1/2}
For any $z$ on the circle $C_1$ given by (\ref{Apr3}) and $t$ in
the complex $(2k^{-1-4s_1-2\delta})$-neighborhood of $\chi_1$,
    \begin{equation}\label{2.40.2}
    \|(H^{(1)}(t)-z)^{-1}\| \leq 4k^{-2l+2+4s_1+\delta}.
    \end{equation}
\end{lemma}
\begin{proof}
Considering the following Hilbert relation
    $$(H^{(1)}(t)-z)^{-1}=(H_0^{(1)}(t)-z)^{-1}+(H_0^{(1)}(t)-z)^{-1}(-W_1)(H^{(1)}(t)-z)^{-1}$$
and (\ref{2.8}), i.e., $\|(H_0^{(1)}(t)-z)^{-1}\| \leq
2k^{-2l+2+4s_1+\delta}$, we obtain
    \begin{equation*}
    \|(H^{(1)}(t)-z)^{-1}\|  \leq
    \frac{\|(H_0^{(1)}(t)-z)^{-1}\|}{1-\|(H_0^{(1)}(t)-z)^{-1}W_1\|}
     \leq 2\|(H_0^{(1)}(t)-z)^{-1}\|
     \leq 4k^{-2l+2+4s_1+\delta}.
    \end{equation*}
 \end{proof}

\subsection{Nonresonant part of the Isoenergetic Set of $H^{(1)}$}

Let $S_1(\lambda )$ \footnote{$S_1(\lambda )$ definitely depends
on $\alpha W_1$, however we omit this to keep the notation
simple.} be the isoenergetic surface of the operator
$H_{\alpha}^{(1)}$
    \begin{equation}\label{2.65.1}
    S_1(\lambda )=\{t \in K_1 : \exists n \in \N\ \text{s.t.}\
    \lambda_n^{(1)}(\alpha,t)=\lambda  \},
    \end{equation}
    where $\{\lambda_n^{(1)}(\alpha,t)\}_{n=1}^{\infty }$ is the
    complete set of eigenvalues of $H_{\alpha}^{(1)}(t)$.
    Next, we construct a ``nonresonance" subset  $\chi _1^*(\lambda )$ of $S_1(\lambda
    )$, which
    corresponds to  nonresonance eigenvalues $\lambda
    _{j}^{(1)}(\alpha , t)$ given by the perturbation series.

Let us note that for every $t$ belonging to the non-resonant set
$\chi _1(\lambda, \delta )$ described by Lemma~\ref{L:2.1}, there
is a single $j\in \Z^2$ such that $p_j(t)=k$, $k=\lambda
^{\frac{1}{2l}}$. This means that the function $t\to \vec p_j(t)$
maps $\chi _1(\lambda,  \delta )$ into the circle $S_k$. The image
of $\chi _1(\lambda,  \delta )$ in $S_k$ we denote by
$\cal{D}_0(\lambda )_{nonres}$. Obviously,
\begin{equation}
\chi _1(\lambda,  \delta )={\cal
K_1}\cal{D}_0(\lambda)_{nonres},\label{p}
\end{equation}
 where
${\cal K_1}$ establishes one-to-one relation between two sets. Let
$\cal{B}_1(\lambda)$ is a set of unit vectors corresponding to
$\cal{D}_0(\lambda)_{nonres}$:
$$\cal{B}_1(\lambda)=\{\vec{\nu} \in S_1 :
k \vec{\nu} \in \cal{D}_0(\lambda)_{nonres}\}.$$ It is easy to see
that $\cal{B}_1(\lambda)$ is a unit circle with holes, centered at
the origin. We denote by $\Theta _1 (\lambda)$ the set of angles
$\varphi $ in polar coordinates, corresponding to
$\cal{B}_1(\lambda)$:
\begin{equation}
\Theta _1(\lambda)=\{\varphi \in [0,2\pi ):\ (\cos \varphi ,\sin
\varphi )\in \cal{B}_1(\lambda)\} \label{May17a}.
 \end{equation}

Let $\vec{\varkappa } \in \cal{D}_0(k)_{nonres}$. Then there is
$j\in \Z^2$, $t\in \chi _1(\lambda,\delta)$ such that
$\vec{\varkappa }=\vec p_j(t)$.\footnote{Usually the vector $\vec
p_j(t)$ is denoted by $\vec k$, the corresponding plane wave being
$e^{\langle \vec k,x \rangle }$. We use less common notation
$\vec{\varkappa }$, since we have already other $k$'s in the
text.} Obviously, $t={\cal K}_1\vec{\varkappa }$ and, by
(\ref{p}), $t\in \chi _1(\lambda, \delta)$. According to Theorem
\ref{T:2.3}, for sufficiently large $k$, there exists an
eigenvalue of the operator $H^{(1)}_{\alpha}(t)$, $t={\cal K_1}
\vec \varkappa $, $0 \leq \alpha \leq 1$, given by (\ref{2.16}).
It is convenient here to denote $\lambda ^{(1)}_j(\alpha , t)$ by
$\lambda^{(1)}(\alpha,\vec{\varkappa})$, we can do this since
there is one-to-one correspondence between $\vec{\varkappa}$ and
the pair $(t,j)$. We rewrite (\ref{2.16}) in the form:
    \begin{equation}\label{2.66}
    \lambda^{(1)}(\alpha,\vec{\varkappa})=\varkappa ^{2l}+f_1(\alpha,\vec{\varkappa}),
    \ \ \varkappa =|\vec \varkappa |, \ \ \ f_1(\alpha,\vec{\varkappa})=\sum _{r=2}^{\infty}
\alpha^{r}g_r^{(1)}(\vec{\varkappa}),
    \end{equation}
    $g_r^{(1)}(\vec{\varkappa})$ being defined by (\ref{2.11}) with
$j$ and $t$ such that $\vec{p}_j(t)=\vec{\varkappa}$, and
$f_1(\alpha,\vec{\varkappa})$ satisfying the estimates:
\begin{equation}\label{2.67}
    |f_1(\alpha,\vec \varkappa)|\leq 2
    \alpha^2k^{2l-2-4s_1-2\gamma_0-\delta},\end{equation}
   \begin{equation}\label{2.67a}
    |T(m)f_1(\alpha,\vec \varkappa)|\leq 2
    \alpha^2k^{2l-2-4s_1-2\gamma_0-\delta+|m|(1+4s_1+2\delta)}.
    \end{equation}
By Theorem \ref{T:2.3}, the formulae (\ref{2.66}) -- (\ref{2.67})
hold in  the $(2k^{-1-4s_1-2\delta})$-neighborhood of
$\cal{D}_0(\lambda )_{nonres}$, i.e., they hold for any $\varkappa
\vec{\nu}$ such that $\vec{\nu} \in \cal{B}_1(\lambda),\
|\varkappa-k|<k^{-1-4s_1-2\delta}$. We define $\cal{D}_1(\lambda
)$ as the level set of the function
$\lambda^{(1)}(\alpha,\vec{\varkappa})$ in this neighborhood:
    \begin{equation}\label{2.68}
    \cal{D}_1(\lambda ):=\{\vec{\varkappa }=\varkappa \vec{\nu}:
    \vec{\nu}\in
    \cal{B}_1(\lambda), \  |\varkappa-k|<k^{-1-4s_1-2\delta},\
    \lambda^{(1)}(\alpha,\vec{\varkappa})=\lambda \}
    .
    \end{equation}
 We prove in Lemma \ref{L:2.13} that $\cal{D}_1(\lambda )$ is a
distorted circle with holes, which is close to and inside of the
circle of radius $k$, Fig. \ref{F:1}. First, we prove that the
equation $\lambda^{(1)}(\alpha,\vec{\varkappa})=\lambda $ is
solvable with respect to $\varkappa =|\vec \varkappa |$ for any
$\vec \nu=\frac{\vec \varkappa}{\varkappa} \in
    \cal{B}_1(\lambda)$.
%
%
\begin{lemma}\label{L:2.12}
For every $\vec{\nu} \in \cal{B}_1(\lambda)$ and every $\alpha$,
$0 \leq \alpha \leq 1$ and sufficiently large $\lambda $, there is
a unique $\varkappa _1 =\varkappa _1(\lambda ,\vec{\nu})$ in the
interval $I_1:=[k-k^{-1-4s_1-2\delta},k+k^{-1-4s_1-2\delta}]$,
$k^{2l}=\lambda $, such that
    \begin{equation}\label{2.70}
    \lambda^{(1)}(\alpha,\varkappa _1\vec{\nu})=\lambda .
    \end{equation}
Furthermore, $|\varkappa _1 - k| \leq
ck^{-1-4s_1-2\gamma_0-\delta}$.
\end{lemma}
%
%
\begin{proof}
1) Let us show that there exists such an $\varkappa _1 $. Given
$\vec{\nu} \in \cal{B}_1(\lambda)$, by Theorem \ref{T:2.3},
 there exists an eigenvalue $\lambda ^{(1)} (\alpha, \varkappa
\vec{\nu})$ given by (\ref{2.66}) for all $\varkappa$ in the
interval $I_1$ with a sufficiently large $k$. Obviously,
   $ \lambda^{(1)}(0,k\vec{\nu})=p_j^{2l}(t_0)=k^{2l}=\lambda,\
    \lambda^{(1)}(0,\varkappa\vec{\nu})=p_j^{2l}(t)=\varkappa^{2l},\
    \vec{p}_j(t_0)=k\vec{\nu},\ \vec{p}_j(t)=\varkappa \vec{\nu}$,
where $j,\ t$, and $t_0$ are uniquely determined by $\varkappa
\vec{\nu} $ and $k \vec{\nu}$. Let
    $\cal L^{(1)}(\alpha,\vec{\nu}):=
    \{\lambda=\lambda^{(1)}(\alpha,\varkappa \vec{\nu}) : \varkappa  \in I_1\}.$
Using the definition of $I_1$ and considering that
$\lambda^{(1)}(0,\varkappa \vec{\nu})=\varkappa ^{2l}$, we easily
obtain
    $
    \cal L^{(1)}(0,\vec{\nu}) \supset
    [k^{2l}-\delta _1,k^{2l}+\delta _1]$,
    $\delta _1=c_1k^{2l-2-4s_1-2\delta}$, $
    0<c_1 \neq c_1(k).
    $
     Since $\lambda^{(1)}(\alpha,\varkappa \vec{\nu})$ is
continuous in $\varkappa $ and ~(\ref{2.67}) is valid, we have
$\cal L^{(1)}(\alpha,\vec{\nu})\supset [k^{2l}-\delta _1+\delta
_2,k^{2l}+\delta _1-\delta _2]$, $\delta
_2=c_2k^{2l-2-4s_1-2\gamma_0-\delta}$.
Since $2\gamma_0-\delta>0$, we have $\cal
L^{(1)}(\alpha,\vec{\nu}) \supset [k^{2l}-\delta
_1/2,k^{2l}+\delta _1/2]$.
Thus for every $\vec{\nu} \in \cal{B}_1(\lambda)$, there exists a
$\varkappa _1$ such that $\lambda^{(1)}(\alpha,\varkappa
_1\vec{\nu})=k^{2l}$, $\varkappa _1 \in I_1$.

2) Now we show that there is only one $\varkappa _1$ satisfying
(\ref{2.70}) in the interval $I_1$. Differentiating ~(\ref{2.66})
with respect to $\varkappa $ and using (\ref{2.67a}), we get:
    \begin{equation}\label{2.72.1}
     \frac{\partial \lambda^{(1)}(\alpha, \varkappa \vec{\nu})}{\partial
    \varkappa }=2l\varkappa ^{2l-1}+O(k^{2l-1-2\gamma_0+\delta }).
    \end{equation}
Therefore,
    \begin{equation}\label{2.73}
    \dfrac{\partial \lambda^{(1)}(\alpha,
    \varkappa \vec{\nu})}{\partial \varkappa } \geq 2lk^{2l-1}\bigl( 1+o(1) \bigr),
    \end{equation}
and this implies that $\lambda^{(1)}(\alpha,\varkappa \vec{\nu})$
is monotone with respect to $\varkappa $ in $I_1$. Thus, there is
a single  $\varkappa _1\in I_1$ satisfying ~(\ref{2.70}).


3) Now let us estimate $|\varkappa _1-k|$. By~(\ref{2.66}) and
(\ref{2.67}) we have
    \begin{equation}\label{2.74}
    \lambda^{(1)}(\alpha,k\vec{\nu})=k^{2l}+f_1(\alpha, k\vec{\nu})=k^{2l}+
    O(k^{2l-2-4s_1-2\gamma_0-\delta}).
    \end{equation}
Using (\ref{2.74}), (\ref{2.70}), (\ref{2.73}), and the Mean Value
Theorem, we obtain
    $$
    ck^{2l-2-4s_1-2\gamma_0-\delta} \geq
    |\lambda^{(1)}(\alpha, \varkappa _1\vec{\nu})-\lambda^{(1)}(\alpha, k\vec{\nu})|
    \geq $$ $$
    \underset{\varkappa \in I_1}
    {\text{min}} \left| \frac{\partial \lambda^{(1)}(\alpha, \varkappa \vec{\nu})}{\partial
    \varkappa }\right| |\varkappa _1-k|
    \geq ck^{2l-1}|\varkappa _1-k|,
    $$
and hence we arrive at $|\varkappa _1 - k| \leq
ck^{-1-4s_1-2\gamma_0-\delta}$.
 \end{proof}
Here and below, we use the notations $\varkappa _1
(\varphi):=\varkappa _1 (\lambda, \vec{\nu}),\ \vec{\nu}= (\cos
\varphi , \sin \varphi )$, $\vec \varkappa _1 (\varphi) =\varkappa
_1(\varphi) \vec{\nu},\ h_1(\varphi )=\varkappa _1(\varphi) -k$.
%
\begin{lemma}\label{L:2.13}
\begin{enumerate}
\item The set $\cal{D}_1(\lambda )$ is a distorted circle with
holes which is strictly inside the circle of the radius $k$: it
can be described by the formula:
\begin{equation}
\cal{D}_1(\lambda )=\bigl\{\vec \varkappa \in \R^2: \vec \varkappa
=\varkappa_1 (\varphi)\vec \nu,\ \ \vec{\nu} \in
\cal{B}_1(\lambda) \},\label{May20} \end{equation} where
    $ \varkappa_1 (\varphi)=k+h_1(\varphi) $
and $h_1(\varphi)$ obeys the inequalities:
    \begin{equation}\label{2.75}
    \ \ \ |h_1|<ck^{-1-4s_1-2\gamma_0-\delta},\ \ \
    \left|\frac{\partial h_1}{\partial \varphi} \right| <
    ck^{-2\gamma_0+1},
\end{equation}
$h_1(\varphi )<0$ when $W_1\neq 0$.
    \item The total length of $\cal{B}_1(\lambda)$ satisfies the estimate:
\begin{equation}\label{theta1}
    L\left(\cal{B}_1\right)=2\pi(1+O(k^{-\delta/2})).
    \end{equation}
\item Function $h_1(\varphi)$ can be extended as a holomorphic
function of $\varphi $ to the complex $(2k^{-2-4s_1-2\delta
})$-neighborhood of each connected component of $\Theta _1
(\lambda)$ and estimates (\ref{2.75}) hold. \item The curve
$\cal{D}_1(\lambda)$ has a length which is asymptotically close to
that of the whole circle in the following sense:
    \begin{equation}\label{2.77}
     L\bigl(\cal{D}_1(\lambda)\bigr)\underset{\lambda \rightarrow
     \infty}{=}2\pi k\bigl(1+O(k^{-\delta/2})\bigr),\ \lambda =k^{2l}.
     \end{equation}
     \end{enumerate}
\end{lemma}
%
%

\begin{proof}\begin{enumerate}
\item Let $\vec \varkappa \in \cal{D}_1(\lambda )$, then $\vec
\varkappa =\varkappa_1 \vec{\nu}$.
Using Lemma \ref{L:2.12}, we get
     $|h_1(\varphi)|= |\varkappa _1 -k| \leq
     ck^{-1-4s_1-2\gamma_0-\delta}$,
and
     $\dfrac{\partial h_1}{\partial \varphi}=\dfrac{\partial \varkappa _1}{\partial \varphi}$.
  From
(\ref{2.66}) and the definition of $\cal{D}_1(\lambda )$, we have
    \begin{equation}\label{2.78}
     \lambda = \varkappa _1^{2l}+f_1(\varkappa_1 \cos \varphi, \varkappa _1 \sin \varphi),
     \end{equation}
and, therefore,  by implicit differentiation,
    \begin{equation}\label{2.79}
     \left| \frac{\partial h_1}{\partial \varphi} \right|=
    \left| \frac{\partial \varkappa _1}{\partial \varphi} \right|=
    \left| \frac{\frac{\partial f_1}{\partial
    \varphi}}{ 2l \varkappa _1 ^{2l-1}+ \langle \nabla f,\vec \nu \rangle } \right|.
    \end{equation}
Using (\ref{2.67a}), we easily obtain $ |\nabla f_1| \leq
ck^{2l-2\gamma_0-1}$,
     and, hence,
 (\ref{2.75}) for the derivative. It is easy to show that
 $g_2^{(1)}(k,t)$ given by (\ref{2.14}) is positive when
 $W_1(x)$ is non-zero, and, moreover, $f_1(\alpha,\vec \varkappa)>0$.
 It easily follows that $h_1<0$.
 \item By definition, $\cal{B}_1(\lambda)$ is the set of directions corresponding to
 $\cal{D}_0(\lambda)_{nonres}$, the latter set being a subset of the sphere of radius k. Formula (\ref{p}) establishes
 one-to-one correspondence between $\chi _1(\lambda,  \delta )$ and
 $\cal{D}_0(\lambda)_{nonres}$, their length being equal.
 Considering (\ref{2.7}), we obtain
 $L\left(\cal{D}_0(\lambda)_{nonres}\right)=2\pi
 k\bigl(1+O(k^{-\delta/2})\bigr)$. Hence, (\ref{theta1}) holds.
\item This part of the lemma easily follows from the Implicit
 Function Theorem applied to (\ref{2.78}), the above estimate for $|\nabla
 f_1|$
  and the fact that $f_1(\alpha,\vec \varkappa )$ is a
 holomorphic function of two variables (Theorem
 \ref{T:2.5}).
\item Considering
 that
    $
     L(\cal{D}_1(\lambda ))=\int_{\varTheta_1 (\lambda)} \sqrt{\varkappa _1 ^2(\varphi)+\varkappa _1^{\prime}(\varphi)^2}\,d\varphi
     $
and taking into account (\ref{theta1}) and (\ref{2.79}), we obtain
    \begin{align*}
    \lefteqn{
    L(\cal{D}_0(\lambda))-L(\cal{D}_1(\lambda))
    =\int_{[0,2\pi] \setminus \varTheta_1(\lambda)}k \,d\varphi
    +\int_{\varTheta_1} \Bigl(k
    -\sqrt{\varkappa _1^2(\varphi)+\varkappa _1^{\prime}(\varphi)^2}\Bigr)\,d\varphi}&\\
    &\hspace{1cm} \leq \int_{[0,2\pi] \setminus \varTheta_1(\lambda)}k \,d\varphi
    +\int_{\varTheta_1(\lambda)}\bigl( h_1(\varphi)+h^{\prime}_1(\varphi)\bigr)\\
    &\hspace{1cm}
    <ck \cdot
    k^{-\delta/2}+O(k^{-1-4s_1-2\gamma_0-\delta})+O(k^{-2\gamma_0+1})
    =k\cdot O(k^{-\delta/2}),
    \end{align*}
here and below, $\cal{D}_0(\lambda)$, corresponding to the free
operator, is the perfect circle of radius $\lambda ^{1/2l}$.
Hence, (\ref{2.77}) is valid.
\end{enumerate}
 \end{proof}


Next, we define a nonresonance subset $\chi_1^*(\lambda )$ of
isoenergetic set $S_1(\lambda )$ as the parallel shift of
$\cal{D}_1(\lambda )$ into $K_1$ (Fig. \ref{F:5}):
\begin{figure}
    \centering
\includegraphics[totalheight=.2\textheight]{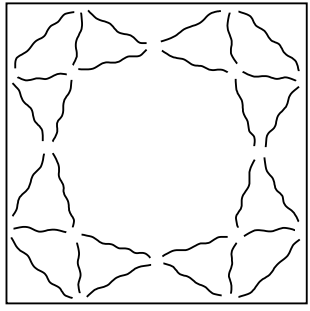}
\caption{The  set $\chi ^*_1(\lambda,\delta )$}\label{F:5}
\end{figure}
    \begin{equation}\label{2.81}
    \chi_1^*(\lambda ):=\cal{K}_1\cal{D}_1(\lambda ).
    \end{equation}
    \begin{lemma} \label{Apr4}
    The set $\chi_1^*(\lambda )$ belongs to the
    $(ck^{-1-4s_1-2\gamma _0-\delta })$-neighborhood of $\chi_1(\lambda
    )$ in $K_1$. If $t\in \chi_1^*(\lambda )$, then the operator
    $H^{(1)}_{\alpha }(t)$ has a simple eigenvalue $\lambda
   _n^{(1)}(\alpha ,t)$, $n\in \N$,
   equal to $\lambda $, no other eigenvalues being in the interval
   $\varepsilon_1 (k,\delta )$, $\varepsilon_1 (k,\delta ):=
(k^{2l}-k^{2l-2-4s_1-\delta }, k^{2l}+k^{2l-2-4s_1-\delta })$.
This
    eigenvalue is given by the perturbation series (\ref{2.16}),
    $j $ being uniquely defined by $t$ from the relation
    $p_j^{2l}(t)\in \varepsilon_1 (k,\delta )$.
\end{lemma}

\begin{proof} By Lemma \ref{L:2.13}, $\cal{D}_1(\lambda )$ is in the
$\left(ck^{-1-4s_1-2\gamma_0-\delta}\right)$-neighborhood of
$\cal{D}_0(\lambda )$. Considering that $\chi_1(\lambda )=\cal
K_1\cal{D}_0(\lambda )$ and $\chi_1^*(\lambda )=\cal
K_1\cal{D}_1(\lambda )$, we immediately obtain that
$\chi_1^*(\lambda )$ is in the
    $(ck^{-1-4s_1-2\gamma _0-\delta })$-neighborhood of $\chi_1(\lambda
    )$. The size of this neighborhood is
less than $k^{-1-4s_1-2\delta }$, hence Theorem~\ref{T:2.3} holds
in it, i.e. for any $t\in \chi_1^*(\lambda )$ there is a single
eigenvalue of $H^{(1)}_{\alpha }(t)$ in the interval
$\varepsilon_1 (k,\delta )$. Since $\chi_1^*(\lambda )\subset
S_1(\lambda )$, this eigenvalue is equal to $\lambda $. By the
theorem, the eigenvalue is given by the series (\ref{2.16}),
    $j $ being uniquely defined by $t$ from the relation
    $p_j^{2l}(t)\in \varepsilon_1 (k,\delta )$.
      \end{proof}

    \begin{lemma}\label{L:May10a} Formula (\ref{2.81}) establishes
    one-to-one correspondence between $\chi_1^*(\lambda )$ and $\cal{D}_1(\lambda
    )$.
    \end{lemma}
    \begin{remark} From geometric point of view this means that $\chi_1^*(\lambda
    )$ does not have self-intersections.
    \end{remark}
    \begin{proof} Suppose there is a pair $\vec \varkappa _{1,1}, \vec \varkappa
    _{1,2} \in \cal{D}_1(\lambda)$, such that $\cal K_1\vec \varkappa _{1,1}=\cal K_1\vec \varkappa
    _{1,2}=t$, $t\in \chi_1^*(\lambda )$. By the definition (\ref{2.68}) of $\cal{D}_1(\lambda
    )$, we have $\lambda^{(1)}(\alpha,\vec \varkappa _{1,1})=\lambda^{(1)}(\alpha,\vec \varkappa _{1,2})=
    \lambda $, i.e., the eigenvalue $\lambda $ of
    $H^{(1)}_{\alpha }(t)$ is not simple. This contradicts to the
    previous lemma.
      \end{proof}


\section{The Second Approximation} \label{chapt4}

\setcounter{equation}{0}

\subsection{The Operator $H_{\alpha}^{(2)}$}

Choosing $s_2=2s_1$, we define the second operator
$H_{\alpha}^{(2)}$ by the formula:
    \begin{equation}\label{3.1}
     H_{\alpha }^{(2)}=H^{(1)}+\alpha W_2,\quad     ( 0\leq \alpha \leq
     1), \qquad
     W_2=\sum_{r=M_1+1}^{M_2}V_r,
     \end{equation}
where $H^{(1)} $ is defined by (\ref{2.1}), $M_2$ is chosen in
such a way that $2^{M_2} \approx k^{s_2}$. Obviously, the periods
of $W_2$ are $2^{M_2-1} (b_1,0)$ and $2^{M_2-1} (0,b_2)$. We will
write them in the form: $N_1(a_1,0)$ and $N_1(0,a_2)$, where
$a_1,a_2$ are the periods of $W_1$ and $N_1=2^{M_2-M_1},\
\frac{1}{4}k^{s_2-s_1} < N_1 < 4 k^{s_2-s_1}$. Note that
    \begin{equation}\|W_2\|_{\infty} \leq \sum_{r=M_1+1}^{M_2}
    \|V_r\|_{\infty} \leq \sum_{r=M_1+1}^{M_2}\exp(-2^{\eta r})
    <\exp(-k^{\eta s_1}). \label{W2}
    \end{equation}
\subsubsection{Multiple Periods of $W_1(x)$}\label{S:2.2}

 The operator $H^{(1)}=H_0 +
 W_1(x)$
%
%
has the periods $a_1, a_2$. The corresponding family of operators,
$\{H^{(1)}(t)\}_{t \in K_1}$, acts in $L_2(Q_1)$, where
$Q_1=[0,a_1] \times [0,a_2]$ and $K_1=[0, 2\pi/a_1)\times [0,
2\pi/a_2)$. Eigenvalues of $H^{(1)}(t)$ are denoted by
$\lambda_n^{(1)}(t)$, $n \in \N$, and its spectrum by $\Lambda
^{(1)}(t)$. Now let us consider the same $W_1(x)$ as a periodic
function with the periods $N_1a_1, N_1a_2$. Obviously, the
definition of the operator $H^{(1)}$ does not depend on the way
how we define the periods of $W_1$. However, the family of
operators $\{H^{(1)}(t)\}_{t \in K_1} $ does change, when we
replace the periods $a_1, a_2$ by $N_1a_1, N_1a_2$. The family of
operators $\{H^{(1)}(t)\}_{t \in K_1}$ has to be  replaced by a
family of operators $\{ \tilde{H}^{(1)}(\tau) \}_{\tau \in K_2}$
acting in $L_2(Q_2)$, where $Q_2=[0,N_1a_1] \times [0,N_1a_2]$ and
$K_2=[0, 2\pi/N_1a_1)\times [0, 2\pi/N_1a_2)$. We denote
eigenvalues of $\tilde{H}^{(1)}(\tau)$ by
$\tilde{\lambda}_n^{(1)}(\tau)$, $n \in \N$ and its spectrum by
$\tilde{\Lambda}^{(1)}(\tau)$. The next lemma establishes a
connection between spectra of operators $H^{(1)}(t)$ and
$\tilde{H}^{(1)}(\tau)$. It easily follows from Bloch theory (see
e.g. \cite{RS}).
\begin{lemma}\label{L:3.1}
For any $\tau \in K_2$,
    \begin{equation}\label{3.3}
    \tilde{\Lambda}^{(1)}(\tau)=\bigcup_{p \in P} \Lambda^{(1)}(t_p),
    \end{equation}
 where \begin{equation}
  P=\{p=(p_1,p_2) \in \Z^2 : 0 \leq p_1 \leq N_1-1,\ 0 \leq p_2
\leq N_1-1\} \label{May10a}
\end{equation}
 and $t_p=(t_{p,1}, t_{p,2})=(\tau _1 +2\pi
p_1/N_1a_1, \tau_2+2\pi p_2/N_1a_2) \in K_1$, see Fig. 6.
\end{lemma}
\begin{figure}\label{F:6}
\centering
    \psfrag{2p/a1}{\hspace{2mm}\small{$\frac{2\pi}{a_1}$}} \psfrag{2p/a2}{\small{$\frac{2\pi}{a_2}$}}
    \psfrag{2p/Na1}{\small{$\frac{2\pi}{N_1a_1}$}}
    \psfrag{2p/Na2}{\hspace{2mm}\small{$\frac{2\pi}{N_1a_2}$}}
\includegraphics[totalheight=.2\textheight]{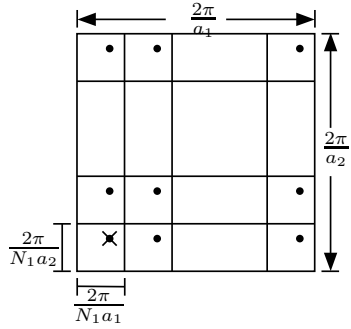}
 \caption{Relation between $\tau \ (\times )$ and $t_p \ (\cdot )$}
\end{figure}

We defined  isoenergetic set $S_1(\lambda ) \subset K_1$ of
$H^{(1)}$ by formula (\ref{2.65.1}). Obviously, this definition is
directly associated with the family of operators $H^{(1)}(t)$ and,
therefore, with periods $a_1, a_2$, which we assigned to $W_1(x)$.
Now, assuming that the periods are equal to $N_1a_1, N_1a_2$, we
give an analogous definition of the isoenergetic set
$\tilde{S}_1(\lambda )$ in $K_2$:
   $\tilde{S}_1(\lambda ):=\{\tau \in K_2: \exists n \in \N:\ \
    \tilde{\lambda}_n^{(1)}(\tau)=\lambda \}.$
By Lemma \ref{L:3.1}, $\tilde{S}_1(\lambda )$ can be expressed as
follows:
   $$ \tilde{S}_1(\lambda )=\Bigl\{\tau \in K_2: \exists n \in \N,\ p \in P:\
    \lambda_n^{(1)}\bigl(\tau+2\pi p/N_1a\bigr)=\lambda \Bigr\},$$
 $2\pi
    p/N_1a=\left(\frac{2\pi p_1}{N_1a_1}, \frac{2\pi
    p_2}{N_1a_2}\right).$
The relation between $S_1 (\lambda)$ and $\tilde{S}_1 (\lambda)$
can be easily understood from the geometric point of view as
    $\tilde{S}_1 (\lambda)=\cal{K}_2 S_1 (\lambda), $
where $\cal{K}_2$ is the parallel shift into $K_2$, i.e.,
    $
    \cal{K}_2:\R^2 \rightarrow K_2,\   \cal{K}_2(\tau+2\pi
    m/N_1a)=\tau,\ m \in \Z^2,\ \tau \in K_2.
    $
    Thus, $\tilde{S}_1(\lambda )$ is obtained from $S_1 (\lambda)$ by
    cutting $S_1 (\lambda)$ into pieces of the size $K_2$ and shifting them
    together in $K_2$.

     \begin{definition} \label{D:May6}
    We say that $\tau $ is a point of
    self-intersection of $\tilde{S}_1(\lambda )$,  if there is a
    pair $n,\hat n\in N$, $n\neq \hat n$ such that $\tilde{\lambda}_n^{(1)}(\tau)=
    \tilde{\lambda}_{\hat n}^{(1)}(\tau)=\lambda $.\end{definition}
    \begin{remark} \label{R:May6} By Lemma \ref{L:3.1}, $\tau $ is a point of
    self-intersection of $\tilde{S}_1(\lambda )$,  if there is a
    pair $p,\hat p\in P$ and a pair $n,\hat n \in N$ such that $|p-\hat
    p|+|n-\hat n|\neq 0$ and $\lambda_n^{(1)}(\tau+2\pi
    p/N_1a)=
    \lambda_{\hat n}^{(1)}(\tau+2\pi
    \hat p/N_1a)=\lambda $.
    \end{remark}
Now let us recall that the isoenergetic set $S_1(\lambda )$
consists of two parts:
    $S_1(\lambda )=\chi_1^*(\lambda )\cup \bigl(S_1(\lambda )
    \setminus \chi_1^*(\lambda ) \bigr),$
where $\chi_1^*(\lambda )$ is the first non-resonance set given by
(\ref{2.81}). Obviously $\cal{K}_2\chi_1^*(\lambda )\subset
\cal{K}_2S_1(\lambda)=\tilde S_1(\lambda )$ and can be described
by the formula:
$$\cal{K}_2\chi_1^*(\lambda )=\left\{\tau \in
K_2:\ \exists p\in P: \tau +2\pi p/N_1a \in \chi_1^*(\lambda
)\right\}.$$
Let us consider only those self-intersections of $\tilde{S}_1$
which belong to $\cal{K}_2\chi_1^*(\lambda )$, i.e., we consider
the points of intersection of $\cal{K}_2\chi_1^*(\lambda )$ both
with itself and with $\tilde S_1(\lambda )\setminus
\cal{K}_2\chi_1^*(\lambda )$.
\begin{lemma}\label{L:May6} A self-intersection $\tau $ of $\tilde S_1(\lambda )$
belongs to $\cal{K}_2\chi_1^*(\lambda )$ if and only if there are
a pair $p,\hat p\in P$, $p\neq \hat p$ and a pair $n,\hat n\in N$
such that $\tau +2\pi p/N_1a \in \chi_1^*(\lambda )$ and $\lambda
_n^{(1)}(\tau +2\pi p/N_1a)=\lambda _{\hat n}^{(1)}(\tau +2\pi
\hat p/N_1a)=\lambda$, the eigenvalue $\lambda _n^{(1)}(\tau +2\pi
p/N_1a)$ being given by the series (\ref{2.16}) with $t=\tau +2\pi
p/N_1a $ and
    $j $  uniquely defined by $t$ from the relation
    $p_j^{2l}(t)\in \varepsilon _1$, $\varepsilon _1=
    (k^{2l}-k^{2l-2-4s_1-\delta }, k^{2l}+k^{2l-2-4s_1-\delta })$.
    \end{lemma}
    \begin{proof} Suppose $\tau $ is a point of
     self-intersection  of $\tilde S_1(\lambda )$
belonging to $\cal{K}_2\chi_1^*(\lambda )$. Since, $\tau \in
\cal{K}_2\chi_1^*(\lambda )$, there is a $p\in P$ such that $\tau
+2\pi p/N_1a \in \chi_1^*(\lambda )$. By Lemma \ref{Apr4}, there
is an eigenvalue $\lambda _n^{(1)}(\tau +2\pi p/N_1a)=\lambda$,
which is given by the series (\ref{2.16}) with $t=\tau +2\pi
p/N_1a $ and
    $j $  uniquely defined by $t$ from the relation
    $p_j^{2l}(t)\in \varepsilon _1$. By the same lemma $\lambda _n^{(1)}(\tau +2\pi
    p/N_1a)$
    is a unique eigenvalue of $H^{(1)}\left(\tau +2\pi p/N_1a \right)$
    in
    $\varepsilon _1$. This means
    \begin{equation}
    \left|\lambda _n^{(1)}(\tau +2\pi
    p/N_1a)-\lambda _{\hat n}^{(1)}(\tau +2\pi
 p/N_1a)\right|\geq k^{2l-2-4s_1-\delta }, \ \ \mbox{when}\ \ \hat
 n\neq n. \label{july13a}
 \end{equation}
 Since $\tau $ is a point of self-intersection of $\tilde S_1(\lambda )$,
 $\lambda
_n^{(1)}(\tau +2\pi p/N_1a)=\lambda _{\hat n}^{(1)}(\tau +2\pi
\hat p/N_1a)=\lambda$ for some $\hat n, \hat p$: $|p-\hat
p|+|n-\hat n |\neq 0$. Considering the above inequality, we obtain
that the equality holds for some $\hat p\neq p$. The converse part
of the lemma is trivial.  \end{proof}


To obtain a new non-resonance set $\chi_2(\lambda )$ we remove
from $\cal{K}_2\chi_1^*(\lambda )$, shortly speaking,  a
neighborhood of its intersections with the whole
$\tilde{S}_1(\lambda )$. More precisely, we remove from
$\cal{K}_2\chi_1^*(\lambda )$ the following set:
    \begin{multline} \label{3.2.1}
     \Omega _1(\lambda )=\{\tau \in \cal{K}_2\chi_1^*(\lambda ): \exists n,\hat{n} \in
    \N,\ p,\hat{p} \in P,\  p\neq \hat{p}:\ \lambda_{n}^{(1)}(\tau+2\pi
    p/N_1a)=\lambda ,\\ \tau+2\pi p/N_1a \in \chi_1^*(\lambda ),\
     |\lambda_{n}^{(1)}(\tau+2\pi p/N_1a)-\lambda_{\hat{n}}^{(1)}(\tau+2\pi
    \hat{p}/N_1a)| \leq \epsilon_1  \},
    \end{multline}
    where $\ \epsilon_1 =e^{-\frac{1}{4}k^{\eta
    s_1}}$.
We define $ \chi_2(\lambda )$  by the formula:
    \begin{equation}\label{3.5.1}
     \chi_2(\lambda )=\cal{K}_2\bigl( \chi _1^*(\lambda )
    \bigr)\setminus \Omega _1(\lambda ).
    \end{equation}

\subsection{Perturbation Formulae}\label{S:3.3}

Before proving the main result, we formulate the Geometric Lemma:
\begin{lemma}[Geometric Lemma]\label{L:3.2}
For an arbitrarily small positive $\delta $, $7\delta
<2l-11-16s_1$ and sufficiently large $\lambda $, $\lambda
>\lambda _0(V,\delta)$, there exists a non-resonance set $\chi
_2(\lambda ,\delta ) \subset \cal{K}_2\chi_1^*$ such that:
\begin{enumerate}
\item For any $\tau \in \chi _2$, the following conditions hold:
\begin{enumerate}
    \item There exists a unique $p \in P$
    such that $\tau + 2\pi p/N_1a \in \chi_1^*$.
    \footnote{From geometric point of view this means that $\chi _2 (\lambda)$
    does not have self-intersections.}
    \item The following relation holds:
    $$ \lambda _j^{(1)}(\tau + 2\pi p/N_1a)=k^{2l},$$
    where $\lambda _j^{(1)}(\tau + 2\pi p/N_1a)$ is  given  by the perturbation series
    (\ref{2.16}) with $\alpha
    =1$, $j$ being uniquely defined by $t=\tau + 2\pi p/N_1a $ as it is described
    in Part 2 of
    Geometric Lemma for the previous step.
    \item The eigenvalue $\lambda _j^{(1)}(\tau + 2\pi p/N_1a)$ is
    a simple eigenvalue of $\tilde H^{(1)}(\tau )$ and
its distance from all other eigenvalues $\lambda _{\hat
n}^{(1)}(\tau + 2\pi \hat{p}/N_1a),\
    {\hat n} \in
    \N$ of $\tilde{H}_1(\tau )$ is greater than $\epsilon_1 = e^{-\frac{1}{4}k^{\eta s_1}}$:
    \begin{equation}\label{3.6}
    |\lambda _j^{(1)}(\tau + 2\pi p/N_1a)-\lambda _{\hat n}^{(1)}(\tau + 2\pi
    \hat{p}/N_1a)|>\epsilon_1.
    \end{equation}
\end{enumerate}

\item For any $\tau$ in the $(\epsilon_1 k^{-2l+1-\delta
})$-neighborhood  in $\C^2$ of $\chi_2$, there exists a unique $p
\in P$ such that $\tau +2\pi p/N_1a$ is in the $(\epsilon_1
k^{-2l+1-\delta })$-neighborhood in $\C^2$ of $\chi_1^*$ and
    \begin{equation}\label{3.7}
    | \lambda_j^{(1)}(\tau +2\pi p/N_1a)-k^{2l}| < \epsilon_1 k^{-\delta
    },
    \end{equation}
    $j$ being uniquely defined by $\tau + 2\pi p/N_1a $ as it is described
    in Part 2 of
    Geometric Lemma for the previous step.
    \item The second non-resonance set $\chi_2$ has an asymptotically
full measure in $\chi_1^*$ in the following sense:
\begin{equation}\label{3.9}
\frac{L(\cal{K}_2\chi_1^* \setminus \chi
_2))}{L(\chi_1^*)}<k^{-2-2s_1}.
\end{equation}
\end{enumerate}
\end{lemma}

 The proof of the lemma will be presented in section~\ref{S:3.2}.

\begin{corollary}\label{C:3.3}
If $\tau$ belongs to the complex $(\epsilon_1
k^{-2l+1-\delta})-$neighborhood of the second non-resonance set
$\chi_2(\lambda,\delta)$, then for any  $z$ lying on the circle
$C_2=\{ z: |z-k^{2l}| =\epsilon_1/2 \} $, the following
inequalities hold:
\begin{align}
&\|(\tilde{H}^{(1)}(\tau)-z)^{-1}\| \leq \frac{4}{\epsilon_1},\label{3.10} \\
&\|(\tilde{H}^{(1)}(\tau)-z)^{-1}\|_1 \leq
\frac{c_0k^{2+2s_2}}{\epsilon_1}, \ \ \ c_0=32b_1b_2. \label{3.11}
\end{align}
\end{corollary}
Corollary is proven in Appendix \ref{A:1}.
\begin{remark} \label{R:11} Note that every point $2\pi m/N_1a$ ($m\in \Z^2$) of a
dual lattice corresponding to the larger periods $N_1a_1, N_1a_2$
can be uniquely represented in the form $2\pi m/N_1a=2\pi j/a+2\pi
p/N_1a$, where $m=N_1j+p$ and $2\pi j/a$ is a point of a dual
lattice for periods $a_1$, $a_2$, while $p\in P$ is responsible
for refining the lattice.

Let us consider a normalized eigenfunction $\psi_n(t,x)$ of
$H^{(1)}(t)$ in $L_2(Q_1)$. We extend it quasiperiodically to
$Q_2$, renormalize in $L_2(Q_2)$ and denote the new function by
$\tilde{\psi}_n(\tau,x)$, $\tau ={\cal K}_2t$. The Fourier
representations of $\psi_n(t,x)$ in $L_2(Q_1)$ and
$\tilde{\psi}_n(\tau,x)$ in $L_2(Q_2)$ are simply related. If we
denote Fourier coefficients of $\psi_n(t,x)$ with respect to the
basis of exponential functions
    $|Q_1|^{-1/2}e^{i \langle t+2 \pi j/a,x \rangle}$, $j \in \Z^2,$
in $L_2(Q_1)$  by $C_{nj}$, then,  the Fourier coefficients
$\tilde{C}_{nm}$ of $\tilde{\psi}_n(\tau,x)$ with respect to the
basis of exponential functions
    $|Q_2|^{-1/2}e^{i \langle \tau + 2\pi m/N_1a ,x \rangle}$, $m \in \Z^2,$
in $L_2(Q_2)$ are given by the formula:
    \begin{equation*}
    \tilde{C}_{nm}=
        \begin{cases}
        C_{nj},  &\text{if $m=jN_1+p$;}\\
        0,       &\text{otherwise},
        \end{cases}
    \end{equation*}
    $p$ being defined from the relation
$t=\tau+2\pi p/N_1a,\ p \in P$. Hence, matrices of the projections
on $\psi_n(t,x)$ and $\tilde{\psi}_n(\tau,x)$ with respect to the
above bases are simply related:
    \begin{equation*}
    (\tilde{E}_n)_{j\hat{j}}=
        \begin{cases}
        (E_n)_{m\hat{m}},  &\text{if $m=jN_1+p,\ \hat{m}=\hat{j}N_1+p$;}\\
        0,       &\text{otherwise},
        \end{cases}
    \end{equation*}
$\tilde{E}_n$ and $E_n$ being projections in $L_2(Q_2)$ and
$L_2(Q_1)$, respectively.

Let us denote by $\tilde{E}_j^{(1)}\bigl(\tau+2\pi p/N_1a \bigr)$
the spectral projection ${E}_j^{(1)}(\alpha, t)$ (see
(\ref{2.17})) with $\alpha =1$ and $t=\tau+2\pi p/N_1a $,
``extended" from $L_2(Q_1)$ to $L_2(Q_2)$.
\end{remark}


 By analogy with (\ref{2.11}), (\ref{2.12}), we define functions $g_r^{(2)}(k,\tau)$ and
operator-valued functions $G_r^{(2)}(k,\tau)$, $r=1, 2, \cdots$,
as follows:
\begin{equation}\label{3.13}
 g_r^{(2)}(k,\tau)=\frac{(-1)^r}{2\pi ir}\mbox{Tr}\oint _{C_2}
 \Bigl(\bigl(\tilde{H}^{(1)}(\tau)-z\bigr)^{-1}W_2\Bigr)^rdz,
 \end{equation}
\begin{equation}\label{3.14}
G_r^{(2)}(k,\tau)=\frac{(-1)^{r+1}}{2\pi i}\oint
_{C_2}\Bigl(\bigl(\tilde{H}^{(1)}(\tau)-z\bigr)^{-1}W_2\Bigr)^r
\bigl(\tilde{H}^{(1)}(\tau)-z\bigr)^{-1}dz.
 \end{equation}
We consider the operators $H_{\alpha}^{(2)}=H^{(1)}+\alpha W_2$
and the family $H_{\alpha}^{(2)}(\tau)$, $\tau \in K_2$, acting in
$L_2(Q_2)$.
\begin{theorem}\label{T:3.4}
 Suppose $\tau$ belongs to the
$(\epsilon_1 k^{-2l+1-\delta })$-neighborhood in $K_2$ of the
 second non-resonance set $\chi _2(\lambda ,\delta)$, $0<7\delta
<2l-11-16s_1$, $\epsilon_1=e^{-\frac{1}{4}k^{\eta s_1}}$. Then,
for sufficiently large $\lambda $, $\lambda
>\lambda _0(V,\delta)$ and for all $\alpha $, $0 \leq \alpha \leq 1$,
there exists a unique eigenvalue of the operator
$H_{\alpha}^{(2)}(\tau)$
 in
the interval $\varepsilon_2 (k):= (k^{2l}-\epsilon_1 /2,
k^{2l}+\epsilon_1 /2)$. It is given by the series:
\begin{equation}\label{3.15}
\lambda_{\tilde{j}}^{(2)} (\alpha ,\tau)=\lambda_j^{(1)}\bigl(
\tau+2\pi p/N_1a \bigr)+\sum _{r=1}^{\infty }\alpha ^r
g_r^{(2)}(k,\tau),\ \ \ \ \tilde j=j+p/N_1,
\end{equation}
converging absolutely in the disk $|\alpha|  \leq 1$, where $p\in
P$ and $j\in \Z^2$ are described as in Geometric Lemma
\ref{L:3.2}. The spectral projection corresponding to
$\lambda_{\tilde{j}}^{(2)} (\alpha ,\tau)$ is given by the series:
\begin{equation}\label{3.16}
E_{\tilde{j}}^{(2)} (\alpha
,\tau)=\tilde{E}_j^{(1)}\bigl(\tau+2\pi p/N_1a \bigr)+\sum
_{r=1}^{\infty }\alpha ^rG_r^{(2)}(k,\tau),
\end{equation}
which converges in the trace class $\mathbf{S_1}$ uniformly with
respect to $\alpha $ in the disk  $| \alpha | \leq 1$.

 The
following estimates hold for coefficients $g_r^{(2)}(k,\tau)$,
$G_r^{(2)}(k,\tau)$, $r\geq 1$:
%
\begin{equation}\label{3.17}
\bigl| g_r^{(2)}(k,\tau)
\bigr|<\frac{3\epsilon_1}{2}(4\epsilon_1^3)^r,\ \ \ \
%
\bigl\| G_r^{(2)}(k,\tau)\bigr\| _1< 6r (4 \epsilon_1^3)^r.
\end{equation}
%

\end{theorem}
\begin{corollary}\label{C:3.5}
The following estimates hold for the perturbed eigenvalue and its
spectral projection:
\begin{equation}\label{3.19}
\Bigl| \lambda_{\tilde{j}}^{(2)} (\alpha ,\tau)-\lambda_j^{(1)}
\bigl(\tau+2\pi p/N_1a\bigr)\Bigr| \leq  12 \alpha \epsilon_1 ^4,
\end{equation}
\begin{equation}\label{3.20}
\Bigl\|E_{\tilde{j}}^{(2)} (\alpha
,\tau)-\tilde{E}_j^{(1)}\bigl(\tau+2\pi p/N_1a\bigr) \Bigr\|_1\leq
48 \alpha \epsilon_1^3 .
\end{equation}
\end{corollary}
\begin{remark} \label{R:May10a} The theorem states that
$\lambda_{\tilde{j}}^{(2)} (\alpha ,\tau )$ is a single eigenvalue
in the interval $\varepsilon _2 (k,\delta )$. This means that
$\bigl|\lambda_{\tilde{j}}^{(2)} (\alpha ,\tau
)-k^{2l}\bigr|<\epsilon _1/2$. Formula (\ref{3.19}) provides a
stronger estimate on the location of $\lambda_{\tilde{j}}^{(2)}
(\alpha ,\tau )$.\end{remark}
%
\begin{proof}
The proof of the theorem is analogous to that of Theorem
\ref{T:2.3} and it is based on expanding the resolvent
$(H_{\alpha}^{(2)}(\tau)-z)^{-1}$ in a perturbation series for $z
\in C_2 $. Integrating the resolvent yields the formulae for an
eigenvalue of $H_{\alpha}^{(2)}$ and its spectral projection. In
fact, it is obvious that
\begin{equation}\label{3.21}
(H_{\alpha
}^{(2)}(\tau)-z)^{-1}=(\tilde{H}^{(1)}(\tau)-z)^{-1}(I-\alpha
A_2)^{-1}, \ \  A_2 := -W_2(\tilde{H}^{(1)}(\tau)-z)^{-1}.
\end{equation}
Suppose $z\in C_2$. Using Corollary \ref{C:3.3} and the estimate
$\|W_2\|<e^{-k^{\eta s_1}}=\epsilon_1^4$, we obtain:
\begin{equation}\label{3.22}
\|(\tilde{H}^{(1)}(\tau)-z)^{-1}\| \leq \dfrac{4}{\epsilon_1},\ \
\ \|A_2\| \leq \dfrac{4 \|W_2\|}{\epsilon_1}<4\epsilon_1^3<1.
\end{equation}
%
%
The last inequality makes it possible to expand $(I-\alpha
A_2)^{-1}$ in the series in powers of $\alpha A_2$.  Integrating
the series for the resolvent, we obtain formulae (\ref{3.15}),
(\ref{3.16}). Estimates (\ref{3.17}) follow from the estimates
(\ref{3.22}).  \end{proof}

Next, we show that  the series~(\ref{3.15}),~(\ref{3.16}) can be
extended as holomorphic functions of $\tau $ in a complex
neighborhood of $\chi _2$; they can be differentiated   any number
of times with respect to $\tau$ and retain their asymptotic
character.

\begin{lemma}\label{T:3.6} The following estimates
hold for the coefficients $g_r^{(2)}(k,\tau)$ and
$G_r^{(2)}(k,\tau)$ in the complex $(\frac{1}{2}\epsilon_1
k^{-2l+1-\delta })$-neighborhood  of the non-resonance set
$\chi_2$:
    \begin{align}
    |T(m)g_r^{(2)}(k,\tau)| &< m!\cdot 3 \cdot 2^{2r-1+|m|}
    \epsilon_1^{3r+1-|m|}k^{|m|(2l-1+\delta)},\label{3.39}\\
    \| T(m)G_r^{(2)}(k,\tau)\|_1 &<
    m!\cdot 3r \cdot 2^{2r+1+|m|} \epsilon_1^{3r-|m|} k^{|m|(2l-1+\delta)}.\label{3.40}
    \end{align}
\end{lemma}
\begin{proof} Since~(\ref{3.10}) is valid in the complex
$(\epsilon_1 k^{-2l+1-\delta })$- neighborhood of the second
non-resonance set, it is not hard to see that the coefficients
$g_r^{(2)}(k,\tau)$ and $G_r^{(2)}(k,\tau)$ can be continued from
the real $(\epsilon_1 k^{-2l+1-\delta })$-neighborhood of $\tau$
to the complex $(\epsilon_1 k^{-2l+1-\delta })$-neighborhood as
holomorphic functions of two variables and
inequalities~(\ref{3.17}) are hereby preserved. Estimating, by
means of the Cauchy integral formula, the value of the derivative
with respect to $\tau$ in terms of the value of the function
itself on the boundary of the $(\frac{1}{2}\epsilon_1
k^{-2l+1-\delta })$-neighborhood of $\tau$
(formulas~(\ref{3.17})), we obtain~(\ref{3.39}) and~(\ref{3.40}).
 \end{proof}

From this lemma the following theorem easily follows.

\begin{theorem}\label{T:3.5a}
Under the conditions of Theorem \ref{T:3.4} the
series~(\ref{3.15}), ~(\ref{3.16})  can be continued as
holomorphic functions of two variables from the real
$(\epsilon_1k^{-2l+1-\delta})$-neighborhood of the non-resonance
set $\chi_2$ to its complex
$(\epsilon_1k^{-2l+1-\delta})$-neighborhood and the following
estimates hold in the complex neighborhood:
%
\begin{align}
    \left| T(m)\left(\lambda_{\tilde{j}}^{(2)} (\alpha ,\tau)
    -\lambda_j^{(1)}(\tau+2\pi p/N_1a)\right)\right| &<\alpha
    C_m\epsilon_1^{4-|m|}k^{|m|(2l-1+\delta)},
    \label{3.41}\\
   \left \| T(m)\left(E_{\tilde{j}}^{(2)} (\alpha ,\tau)-\tilde{E}_j^{(1)}(\tau+2\pi p/N_1a)
   \right)\right\|_1
    &< \alpha C_m\epsilon_1^{3-|m|}k^{|m|(2l-1+\delta)},
    \label{3.42}
    \end{align}
    here and below $C_m=48m!2^{|m|}$.
\end{theorem}
\begin{corollary}\label{C:3.6}
\begin{equation}\label{2.20**}
\nabla \lambda_j^{(2)} (\alpha ,\tau )=2lk^{2l-2}\vec
k+o(k^{2l-1}),\ \ \ \vec k=\vec p_j(\tau +2\pi p/N_1a).
\end{equation}
\begin{equation}
T(m)\lambda_j^{(2)}(\alpha ,\tau )=O\bigl(k^{2l-2}\bigr)\ \ \ \
\mbox{if}\ \ \ |m|=2. \label{2.20**a}
\end{equation}
\end{corollary}

The next lemma will be used in the third step of approximation
where the operator $H^{(2)}(\tau)=H^{(2)}_1(\tau)$ will play a
role of the initial (unperturbed) operator.

%

\begin{lemma}\label{L:3.5.1/2}
For any $z$ on the circle $C_2$ and $\tau$
 in the complex $(\epsilon_1k^{-2l+1-\delta})-$ neighborhood of $\chi_2$,
    \begin{equation}\label{3.40.2}
    \|(H^{(2)}(\tau )-z)^{-1}\| \leq \frac{8}{\epsilon_1}.
    \end{equation}
\end{lemma}
\begin{proof}
Considering the  Hilbert relation
    $$(H^{(2)}(\tau)-z)^{-1}=(\tilde{H}^{(1)}(\tau)-z)^{-1}
    +(\tilde{H}^{(1)}(\tau)-z)^{-1}(-W_2)(H^{(2)}(\tau)-z)^{-1},$$
and  the estimate (\ref{3.10}),
together with the estimate $\|W_2\| < \epsilon_1^4$,  we obtain:
    \begin{equation}
    \|(H^{(2)}(\tau)-z)^{-1}\|  \leq
    \frac{\|(\tilde{H}^{(1)}(\tau)-z)^{-1}\|}{1-\|(\tilde{H}^{(1)}(\tau)-z)^{-1}W_2\|}
     \leq 2\|(\tilde{H}^{(1)}(\tau)-z)^{-1}\|
     \leq \frac{8}{\epsilon_1}.\label{3.40.4}
    \end{equation}
 \end{proof}

\subsection{Proof of the Geometric Lemma}\label{S:3.2}

\subsubsection{Proof of Statement 1}
At the end of Section \ref{S:2.2} (page \pageref{3.2.1}) we
defined the second non-resonance set, $ \chi_2(\lambda )\subset
K_2$, by the formula $\chi_2(\lambda )=\cal{K}_2(\chi_1^*)
\setminus
    \Omega _1$, the set $\Omega _1$ being given by (\ref{3.2.1}).
Suppose  $\tau \in \chi_2(\lambda )$. Since $\chi_2(\lambda
)\subset \cal{K}_2(\chi_1^*)$, there is a $p\in P$ such that $\tau
+2\pi p/N_1a\in \chi_1^*$.  By Lemma \ref{Apr4}, there is an
eigenvalue $\lambda _n^{(1)}(\tau +2\pi p/N_1a)=\lambda$, which is
given by perturbation series~(\ref{2.16}).
 By the same lemma $\lambda _n^{(1)}(\tau +2\pi
    p/N_1a)$
    is a unique eigenvalue of $H^{(1)}\left(\tau +2\pi p/N_1a \right)$
    in
    $\varepsilon _1=(k^{2l}-k^{2l-2-4s_1-\delta }, k^{2l}+k^{2l-2-4s_1-\delta })$. This means
    \begin{equation}
    \left|\lambda _n^{(1)}(\tau +2\pi
    p/N_1a)-\lambda _{\hat n}^{(1)}(\tau +2\pi
 p/N_1a)\right|\geq k^{2l-2-4s_1-\delta }, \ \ \mbox{when}\ \ \hat
 n\neq n. \label{july13a*}
 \end{equation}
 Using the definition of $\Omega _1$, we obtain:
 $$|\lambda_{n}^{(1)}(\tau+2\pi p/N_1a)-\lambda_{\hat{n}}^{(1)}(\tau+2\pi
    \hat{p}/N_1a)|>\epsilon_1\ \ \ \mbox{when}\ \ \ p\neq \hat
    p.$$
    Combining the last two inequalities, we obtain Statement 1(c).
    It remains to prove that $p$ is defined uniquely by the
    relation $\tau
+2\pi p/N_1a\in \chi_1^*$. Suppose it is not so, and there is
$\hat p\neq p$: $\tau +2\pi \hat p/N_1a\in \chi_1^*$.  Applying
Lemma \ref{Apr4} to both points in $\chi_1^*$, we obtain $\lambda
_n^{(1)}(\tau +2\pi
    p/N_1a)=\lambda _{\hat n}^{(1)}(\tau +2\pi
    \hat p/N_1a),$ the eigenvalues being given by perturbation
    series. The last equality contradicts to (\ref{july13a*}).
    \subsubsection{Proof of the Statement 2}
    If $\tau$ is in the complex $(\epsilon_1 k^{-2l+1-\delta
})$-neighborhood of $\chi_2$, then there exists a $\tau _0$ in
$\chi_2$ such that $|\tau-\tau_0|<\epsilon_1 k^{-2l+1-\delta}$ and
$\tau_0$ satisfies Statement 1 of the lemma. Obviously $\tau+2\pi
p/N_1a$ is in the $(k^{-1-4s_1-2\delta})$-neighborhood of
$\chi_1$,
where Theorem \ref{T:2.3} holds. By Corollary \ref{C:3.6},\\
    $\left|\lambda _j^{(1)}(\tau + 2\pi p/N_1a)-\lambda _j^{(1)}(\tau_0 + 2\pi p/N_1a)\right|
    \leq \sup _{\tilde{\tau}} \left|\nabla_{\tau} \lambda_j^{(1)}(\tilde{\tau}+2\pi p/N_1a)
    \right|
    |\tau-\tau_0|$
    $=O(\epsilon_1 k^{-\delta}).$
 Considering that $\lambda _j^{(1)}(\tau_0 + 2\pi p/N_1a)=k^{2l}$, we get (\ref{3.7}).
\subsubsection{Determinants. Intersections and Quasi-intersections.
Description of the set $\Omega _1$ in terms of determinants}
\label{sep20} We considered self-intersections of $\tilde S_1
(\lambda)$ belonging to ${\cal K}_2\chi _1^*$
    (see
     Definition \ref{D:May6} and Lemma \ref{L:May6}, page \pageref{D:May6}).
    Here we describe self-intersections as  zeros of
    determinants of  operators of the type $I+A$, $A \in
\mathbf{S_1}$ (see e.g. \cite{RS}). In fact,
%
%
%
let us represent  the operator $(H^{(1)}(t)-\lambda
)(H_0(t)+\lambda )^{-1}$ in the form $I+A_1$:
    \begin{equation}
    (H^{(1)}(t)-\lambda )(H_0(t)+\lambda )^{-1}=I+A_1(t),\ \ \ A_1(t)=(W_1-2\lambda )
    (H_0(t)+\lambda )^{-1}. \label{july9a}
    \end{equation}
%
%
%
%
Obviously, $A_1(t) \in \mathbf{S_1}$. \begin{remark}
\label{R:May22} From properties of determinants and  the
definition of $S_1(\lambda )$ it easily follows that the
isoenergetic set $S_1(\lambda )$  is the zero set  of
$\det\Bigl(I+A_1(t)\Bigr)$ in $K_1$.
\end{remark}
\begin{remark} \label{R:May20} Matrices $H_0(t)$, $H^{(1)}(t)$ and $A_1(t)$ can
be extended as holomorphic functions of two variables to a
vicinity of $\R^2$. If $\vec y=\vec p_j(0)+t$, then matrices
$H_0(\vec y)$, $H^{(1)}(\vec y)$ and $A_1(\vec y)$ differ from
$H_0(t)$, $H^{(1)}(t)$ and $A_1(t)$ only by a shift of indices:
$H_0(\vec y)_{mm}=H_0(t)_{m+j,m+j}$, etc. Hence,
$\left\|(H^{(1)}(\vec
y)-z)^{-1}\right\|=\left\|(H^{(1)}(t)-z)^{-1}\right\|$ and $\det
\bigl(I+A_1(\vec y)\bigr)=\det \bigl(I+A_1(t)\bigr).$ \end{remark}

 Next, we recall that the set ${\cal D}_1(\lambda )$ can be
 described in terms of vectors $\vec \varkappa (\varphi )=\varkappa (\lambda ,\vec \nu )\vec
 \nu $, $\vec \nu =(\cos \varphi ,\sin \varphi )$, $\varphi \in
 \Theta _1$, see (\ref{May17a}), (\ref{May20}). By definition, $\chi
_1^*(\lambda )={\cal K}_1{\cal D}_1(\lambda )$. By Lemma
\ref{Apr4}, page \pageref{Apr4}, $\chi _1^*(\lambda )$ does not
have self-intersections, i.e., for every $t\in \chi _1^*(\lambda
)$, there is a single $\vec \varkappa (\varphi )\in {\cal
D}_1(\lambda )$, such that $t={\cal K}_1\vec \varkappa (\varphi
)$. Next, if $\tau \in {\cal K}_2\chi _1^*(\lambda )$, then there
is $p\in P$
  such that $\tau +2\pi p/N_1a \in \chi _1^*(\lambda
 )$.  Note that $p$  is not uniquely defined by $\tau $, since ${\cal K}_2\chi _1^*(\lambda
 )$
 may have self-intersections. Hence, every $\tau \in {\cal K}_2\chi _1^*(\lambda
 )$ can be represented  as $\tau ={\cal K}_2\vec \varkappa (\varphi
 )$, here $\vec \varkappa (\varphi
 )$ is not necessary uniquely defined.
The next lemma describes self-intersection of $\tilde S_1$
belonging to ${\cal K}_2\chi _1^*(\lambda )$  as
 zeros of a group of determinants.

 \begin{lemma} \label{L:May24} If $\tau $ is a point of
self-intersection of $\tilde S_1 (\lambda)$ belonging to ${\cal
K}_2\chi _1^*(\lambda )$, then $\tau $ can be represented in the
form $\tau = {\cal K}_2\vec \varkappa (\varphi
 )$, where $\varphi \in
\Theta _1$ and  satisfies the equation
\begin{equation}
\det \Bigl(I+A_1\bigl(\vec y(\varphi)\bigr)\Bigr)=0, \ \ \ \vec
y(\varphi)=\vec \varkappa (\varphi )+\vec b,\  \ \vec b=2\pi
p/N_1a, \label{May18c}
\end{equation}
 for some $p\in P\setminus \{0\}$. Conversely, if (\ref{May18c})
is satisfied for  some $p\in P\setminus \{0\}$, then $\tau = {\cal
K}_2\vec \varkappa (\varphi )$ is a point of self-intersection.
\end{lemma}
\begin{proof} Suppose, $\tau $ is a point of self-intersection of $\tilde S_1
(\lambda)$ belonging to ${\cal K}_2\chi _1^*(\lambda )$. By Lemma
\ref{L:May6}, page \pageref{L:May6}, there is $\tilde p\in P$
(here we  use $\tilde p$ instead of $p$ in Lemma \ref{L:May6}),
such that $\tau +2\pi \tilde p/N_1a\in \chi _1^*$, and $\hat p \in
P$, $\hat p\neq \tilde p$, such that $\tau +2\pi \hat p/N_1a\in
S_1(\lambda )$. This means that
\begin{equation}
\det \Bigl(I+A_1\bigl(\tau +2\pi \tilde p/N_1a\bigr)\Bigr)=0\ \ \
\mbox{and\ \ \ } \det \Bigl(I+A_1\bigl(\tau +2\pi \hat
p/N_1a\bigr)\Bigr)=0.\label{vecher19}
\end{equation}
By Lemma \ref{L:May10a}, page \pageref{L:May10a}, there is a
single $\vec \varkappa (\varphi) \in \cal D_1(\lambda )$: ${\cal
K}_1\vec \varkappa =\tau +2\pi \tilde p/N_1a $ (this means ${\cal
K}_2\vec \varkappa =\tau$). According to Remark \ref{R:May20} and
formulae (\ref{vecher19}), $\det \Bigl(I+A_1(\vec \varkappa
)\Bigr)=0$, $\det\Bigl(I+A_1(\vec \varkappa +\vec b)\Bigr)=0$,
where $\vec b=\bigl[2\pi (\hat p-\tilde p)/N_1a\bigr]_{\text{mod
}K_1}$. Thus, we have obtained (\ref{May18c}).

Suppose now that (\ref{May18c}) holds for some $p\in P$. Let us
show that $\tau ={\cal K}_2\vec \varkappa (\varphi )$ is a point
of self-intersection.
 Let $\tilde p\in P$: $\tau +2\pi
\tilde p/N_1a={\cal K}_1\vec \varkappa $; such $\tilde p$,
obviously, exists, since ${\cal K}_1\vec \varkappa  \in K_1$,
$\tau \in K_2$. Since $\vec \varkappa (\varphi )\in {\cal
D}_1(\lambda )$, we have $\tau +2\pi \tilde p/N_1a \in \chi
_1^*\subset S_1(\lambda )$. Therefore, there is $n\in \N$:
\begin{equation}\lambda _n^{(1)}(\tau +2\pi
\tilde p/N_1a)=\lambda . \label{**1}
\end{equation}
Next, it follows from (\ref{May18c}) and Remarks \ref{R:May22},
\ref{R:May20}
 that ${\cal K}_1\vec y(\varphi )\in S_1(\lambda)$.
Hence, there is $\hat \tau \in K_2$, $\hat p \in P$ such that
$\hat \tau + 2\pi \hat p/N_1a={\cal K}_1\vec y(\varphi )\in
S_1(\lambda )$. This means that there is an $\hat n\in \N$ such
that
\begin{equation}
\lambda _{\hat n}^{(1)}(\hat \tau + 2\pi \hat p/N_1a)=\lambda .
\label{**2}
\end{equation}
Note that $\tau -\hat \tau = {\cal K}_2\vec y(\varphi )-{\cal
K}_2\vec \varkappa (\varphi )={\cal K}_2\vec b=0$, i.e., $\tau
=\hat \tau $. Considering this equality together with (\ref{**1}),
(\ref{**2}), we see that $\tau $ is a point of self-intersection,
provided that $p\neq \hat p$. It remains to show that $p\neq \hat
p$. In fact, $\tau +2\pi p/N_1a={\cal K}_1\vec \varkappa $, $\tau
+ 2\pi \hat p/N_1a={\cal K}_1\vec y$, hence $2\pi (\hat
p-p)/N_1a={\cal K}_1\bigl(\vec y-\vec \varkappa )={\cal K}_1\vec
b=\vec b \neq 0 $, the last equality holding, because $\vec b\in
K_1$. By Lemma \ref{L:May6}, $\tau $ is a point of
self-intersection of $\tilde S_1$ belonging to ${\cal K}_2\chi
_1^*(\lambda )$.
 \end{proof}
\begin{definition} \label{D:July3a} Let $\varPhi _1$ be the complex $\bigl(k^{-2-4s_1-2\delta
}\bigr)$-neighborhood of $\Theta _1$. \end{definition} By Lemma
\ref{L:2.13}, $\vec \varkappa (\varphi )$ is an analytic function
in $\varPhi _1$, and, hence, $\det \Bigl(I+A\bigl(\vec y (\varphi
\bigr)\Bigr)$, $\vec y= \vec \varkappa (\varphi )+\vec b$, $\vec
b\in K_1$, is analytic too.
\begin{definition}We say that  $\varphi
\in \varPhi _1$ is a quasi-intersection of ${\cal K}_2\chi _1^*$
with $\tilde S_1(\lambda )$ if the relation (\ref{May18c})
 holds for some $p\in P\setminus \{0\}.$
\end{definition}

Thus, real intersections correspond to real zeros of the
determinant, while quasi-intersections may have a small imaginary
part (quasi-intersections include intersections).

Next we describe the non-resonance set in terms of determinants.
    \begin{lemma} \label{L:Sept25}
    If  $\tau \in \Omega _1$, then $\tau = {\cal K}_2\vec \varkappa (\varphi
 )$ where $\varphi \in
\Theta _1$ and satisfies the equation
\begin{equation}
\det \left(\dfrac{H^{(1)}\bigl(\vec
y(\varphi)\bigr)-k^{2l}-\epsilon }{H_0\bigl(\vec
y(\varphi)\bigr)+k^{2l}}\right)=0,\ \ \  \vec y(\varphi)=\vec
\varkappa (\varphi )+\vec b,\ \ \vec b=2\pi p/N_1a,
\label{May18c+}
\end{equation}
 for some $p\in P\setminus \{0\}$ and $|\epsilon |<\epsilon _1$.
 Conversely, if (\ref{May18c+})
is satisfied for  some $p\in P\setminus \{0\}$, and $|\epsilon
|<\epsilon _1$, then $\tau = {\cal K}_2\vec \varkappa (\varphi )$
belongs to $\Omega _{1}$.
\end{lemma}
\begin{proof} The proof is analogous to the proof of the previous
lemma after we notice that  the set of points $\tau $ satisfying
the inequality\\ $\left|\lambda_{n}^{(1)}(\tau+2\pi
p/N_1a)-\lambda_{\hat{n}}^{(1)}(\tau+2\pi
\hat{p}/N_1a)\right|<\epsilon_1 $ can be described as the set of
points satisfying $\lambda_{n}^{(1)}(\tau+2\pi
p/N_1a)-\lambda_{\hat{n}}^{(1)}(\tau+2\pi
    \hat{p}/N_1a)=\epsilon $ for some $|\epsilon |<\epsilon_1
    $. \end{proof}


Let us discuss separately the case $W_1=0$. If $\varphi $ is a
point of intersection, then (\ref{May18c}) means that
\begin{equation}
\left|\vec \varkappa (\varphi )+2\pi j/a+2\pi p/N_1a\right|^2=k^2
\label{july25} \end{equation} for some $j\in \Z^2$ and $p\in P$,
and vise versa. Thus, (\ref{May18c}) means that $\vec \varkappa
(\varphi )$ is a point of intersection of the circle of the radius
$k$ centered at the origin ($\left|\vec \varkappa (\varphi
)\right|^2=k^2$ when $W_1=0$) and the circle of the same radius
centered at a point $2\pi j/a+2\pi p/N_1a$ of a dual lattice
corresponding to the periods $N_1a_1$, $N_1a_2$. If there is a
quasi-intersection, then (\ref{july25}) holds for some $\varphi
\in \varPhi _1$. We will refer to such $\varphi $ as a point of
quasi-intersection of two circles. Geometrically, such terminology
is justified, since  validity of (\ref{july25}) for some $\varphi
$ with a small imaginary part means, that two circles come close
together, even if they do not intersect.

We denote  by $\omega _1$ the set of $\varphi \in \Theta _1$
corresponding to $\Omega _1$, i.e., $\omega _1=\{\varphi \in
\Theta _1: {\cal K}_2\vec \varkappa_1 (\varphi
 )\in \Omega _1\} \subset [0, 2\pi )$.

\subsubsection{Complex resonant set}\label{CRS}
Further we consider a complex resonance set $\omega _1^*(\lambda
)$, which is the set of zeros of the determinants (\ref{May18c+})
in $\varPhi _1$ ($p\in P\setminus \{0\}$, $|\epsilon |<\epsilon
_1$). By Lemma \ref{L:Sept25}, $\omega _1=\omega _1^*\cap \Theta
_1$. We prefer to consider quasi-intersections, instead of
intersections, and the complex resonance set, instead of just the
real one, for the following reason: the determinants
(\ref{May18c}) and (\ref{May18c+}), involved in the definitions of
quasi-intersections and the complex resonance set $\omega _1^*$
are holomorphic functions of $\varphi $ in $\varPhi _1$. We can
apply complex analysis theorem to these determinants. Rouch\'{e}'s
theorem is particularly important here, since it states stability
of  zeros of a holomorphic function with respect to  small
perturbations of the function. We take the determinant
(\ref{May18c}) as a holomorphic function, its zeros being
quasi-intersections: the initial determinant corresponds to the
case $W_1=0$, the perturbation  is obtained by ``switching on" a
potential $W_1$. Since there is no analog of Rouch\'{e}'s theorem
(e.g. see \cite{Rouche}) for real functions on the real axis,
introducing the region $\varPhi _1$ and analytic extension of the
determinants into this region is in the core of our
considerations. We will use also a well-known inequality for the
determinants (e.g. see \cite{RS}):
    \begin{equation}\label{3.2.27}
    \Bigl|\det(I+A)-\det(I+B)\Bigr| \leq \|A-B\|_1
    \text{exp}\Bigl(\|A\|_1+\|B\|_1+1\Bigr),\ A,B \in \mathbf{S}_1.
    \end{equation}
    Note that $\omega _1^*=\cup _{p\in P\setminus \{0\}}\omega
    _{1,p}^*$, where $\omega
    _{1,p}^*$ corresponds to a fixed $p$ in (\ref{May18c+}); and
    similarly, $\omega _1=\cup _{p\in P\setminus \{0\}}\omega
    _{1,p}$.
    We fix $p\in P$ and study each $\omega
    _{1,p}^*$ separately.
We start with the case $W_1=0$. The
    corresponding determinant (\ref{May18c}) is
    \begin{equation}
    \det\Bigl(I+A_0\bigl(\vec
    y_0(\varphi)\bigr)\Bigr), \ I+ A_0\bigl(\vec
    y_0(\varphi)\bigr)=\Bigl(H_0\bigl(\vec y_0(\varphi)\bigr)-
    \lambda \Bigr)\Bigl(H_0\bigl(\vec
    y_0(\varphi)\bigr)+\lambda \Bigr)^{-1},\label{eee} \end{equation}
$ \vec
    y_0(\varphi)=k(\cos \varphi ,\sin
\varphi )+\vec b$. This determinant  can be investigated by
elementary means. We easily check, that the number of zeros of
this determinant in $\varPhi _1$ does not exceed $c_0k^{2+2s_1}$
(Proposition \ref{L:3.10}, Corollary \ref{C:May22c}). The
resolvent $ \bigl(H_0\bigl(\vec
y_0(\varphi)\bigr)-k^{2l}\Bigr)^{-1}$ has poles at zeros of the
determinant. The resolvent  norm at $\varphi \in \varPhi _1$ can
be estimated by the distance, which $\varphi $ has to the nearest
zero of the determinant (Lemma \ref{L:3.12}). Next, we introduce a
set ${\cal O}(\vec b)$, which is the union of all disks of the
radius $r=k^{-4-6s_1-3\delta }$ surrounding zeros of the
determinant (\ref{eee}) in $\varPhi _1$. Obviously, any
    $\varphi \in \varPhi _1\setminus {\cal O}(\vec b)$ is
    separated from zeros of the determinant (\ref{eee}) by the
    distance no less than $r$. This estimate on the distance
    yields an estimate for the norm of the
    resolvent $\Bigl(H_0\bigl(\vec y_0
    (\varphi)\bigr)-k^{2l}\Bigr)^{-1},$ when
     $\varphi \in \varPhi _1\setminus {\cal O}(\vec b)$
    (Remark \ref{R:1}). Further, we introduce a potential $W_1$. Our
    goal in this section is to prove that the number of zeros of each determinant
(\ref{May18c+}) is preserved in each connected component $\Gamma
(\vec b)$ of ${\cal O}(\vec b)$, when we switch from the case
$W_1=0, A_1=A_0$ to the case of nonzero $W_1$ and from $\epsilon
=0$ to $|\epsilon |<\epsilon _1$. We also  show that estimates for
the resolvent are stable under such change when  $\varphi \in
\varPhi _1\setminus {\cal O}(\vec b)$. We ``switch on" potential
$W_1$ in two steps. First, we replace $\vec y_0(\varphi )$ by
$\vec y(\varphi )$, i.e., we consider $\det\Bigl(I+A_0\bigl(\vec
    y(\varphi)\bigr)\Bigr)$ and $\Bigl(H_0\bigl(\vec y(\varphi)\bigr)-k^{2l}\Bigr)^{-1}$
    in $\varPhi _1$. We take into account that $\vec y(\varphi)-\vec
    y_0(\varphi)$ is small and holomorphic in $\varPhi _1$ (Lemma
    \ref{L:2.13}),  use (\ref{3.2.27}) on the boundary of $\Gamma
    $ and apply Rouch\'{e}'s theorem. This leads us  to the conclusion, that the number of
    zeros of the determinant in $\Gamma (\vec b)$ is preserved, when we
    replace $\vec y_0(\varphi)$ by $\vec
    y(\varphi)$ (Lemma \ref{L:3.13}). Applying Hilbert relation for resolvents, we
    show that  estimates for the resolvent in $\varPhi _1\setminus {\cal O}(\vec b)$
     are also stable under such change
    (Lemma \ref{L:3.13}, Corollary \ref{C:May22a}).
    In the second step we replace $H_0\bigl(\vec
    y(\varphi)\bigr)$ by $H^{(1)}\bigl(\vec
    y(\varphi)\bigr)$ and prove similar results (Lemmas
    \ref{L:3.14}, \ref{L:*1}, Corollaries \ref{May22d}, \ref{C:May23}). Further
    we consider the determinant (\ref{May18c+}). Obviously it is
    equal to (\ref{May18c}) when $\epsilon =0$. Since, $\epsilon
    $ is very small, its properties are similar to those of
    (\ref{May18c}).
    In particular, it has the same number of zeros (for a fixed
    $\epsilon $) as (\ref{May18c}) at each connected component
    $\Gamma (\vec b)$ of ${\cal O}(\vec b)$ (Lemma \ref{L:3.17}).
    From this lemma we see that  $\omega^*_{1,p}\subset {\cal O}(\vec b)$, $\vec b=2\pi
    p/N_1a$. Taking the real parts of the sets, we conclude:
    $\omega_{1,p}\subset {\cal O}(\vec b)\cap \Theta _1$. We
    show
    that ${\cal O}(\vec b)$ is formed by no more than $c_0k^{2+2s_1}$  disks of the radius
    $r=k^{-4-6s_1-3\delta }$. From this we easily obtain, that the total length of
    $\omega_{1,p}$ does not exceed $k^{-2-4s_1-3\delta }$. Since
    the set $P$ contains no more than $4k^{2s_2-2s_1}$ elements,
    $s_2=2s_1$, the total length of $\Omega_{1}$ does not exceed $k^{-2-2s_1-3\delta
    }$. This estimate is proven in Section \ref{St3}.

\begin{figure}
    \centering
\psfrag{Phi_2}{$\Phi_2$}
\includegraphics[totalheight=.2\textheight]{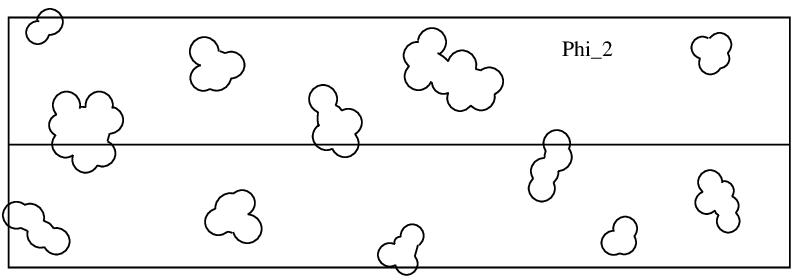}
\caption{The set $\varPhi
    _2$.}\label{F:7}
\end{figure}

    Let us introduce new notations:
    \begin{equation}
    {\cal O}^{}_*=\cup _{p^{}\in P^{}\setminus \{0\}}{\cal
    O}^{}\bigl(2\pi p^{}/N_1a\bigr), \ \ \ \ \varPhi
    _2=\varPhi _1\setminus {\cal O}^{}_*,
     \label{ut1}
    \end{equation}
    $\varPhi _1$ being given by Definitions \ref{D:July3a}.
     Note that the set ${\cal O}_*$  is formed by small disks
 centered at  points $\varphi _{m,p}^{\pm }\in \varPhi _1$, which are
 quasi-intersections of the circle of the
 radius $k$
around the origin   with the circles of the same radius $k$ around
 points $\vec p_m(0)+2\pi p/N_1a $, $p\neq 0$, that is around points of the dual lattice corresponding to the
periods $N_1a_1, N_1a_2$. We will show (Corollary \ref{C:ut}) that
$\omega _1^*\subset {\cal
    O}^{}_*$ and $\omega _1 \subset {\cal
    O}^{}_*\cap \Theta _1.$ From now on we call ${\cal
    O}^{}_*$ the first complex resonance set in $\varPhi
    _1$ and ${\cal
    O}^{}_*\cap \Theta _1$ the first resonance set in
    $\Theta _1$. Obviously,
    to obtain $\varPhi _2$ we produce round holes in each connected
    component of $\varPhi _1$. The set $\varPhi _2$ has a
    structure of Swiss cheese (Fig. \ref{F:7}); we will add more holes of a smaller
    size at each  step of approximation.

In the definition of the non-resonance set we take $\vec b =2\pi
p/N_1a$, $p\neq 0$. Let us consider now arbitrary $\vec b\in K_1$
and investigate ${\cal O}(\vec b)$.\footnote{We need arbitrary
$\vec b $ for next steps of approximation.} We denote by  $b_0$ be
the distance from $\vec b\in K_1$ to the nearest vertex of $K_1$ :
    \begin{equation}\label{3.2.3.1}
   b_0=\min_{m=(0,0),(0,1),(1,0),(1,1)}\left|\vec b-2\pi m/a\right|
    \end{equation}
    Further we assume $b_0\neq 0$, i.e., $\vec b$ is not a vertex
    of $K_1$.
We denote by $\varPhi _0$ a narrow strip in the complex plane
around the interval $[0,2\pi )$: \begin{align*} \varPhi
_0=\left\{\varphi : \Re \varphi \in [0,2\pi ), \left|\Im \varphi
\right| <k^{-2-4s_1}\right\}.\end{align*} Obviously, $\varPhi
_1\subset \varPhi _0$.
 Let $| \vec y_0 (\varphi)-\vec x_0
|_*^2$ be an analytic function of $\varphi$ defined by the
formula:
    $ | \vec y_0 (\varphi)-\vec x_0 |_*^2 :=(k\cos \varphi+b_1-x_{01})^2+
    (k \sin \varphi+b_2-x_{02})^2,$
$\vec x_0$ being a vector in $\R^2$.
 Note that
     $| \vec y_0 (\varphi)-\vec b |_*^2 =k^2 \cos^2 \varphi+
    k^2 \sin^2 \varphi=k^2$.
    Let
    $ | \vec y_0 (\varphi)-\vec x_0 |_*^{2l} :=
    \Bigl(| \vec y_0 (\varphi)-\vec x_0 |_*^2\Bigr)^l.$
    Obviously, $\det\Bigl(I+ A_0\bigl(\vec y_0(\varphi)\bigr)\Bigr)=0$ if and only if
    $| \vec y_0 (\varphi)+\vec p_m(0) |_*^{2l}=k^{2l}$ for some $m \in
\Z^2 $.


\begin{proposition}\label{L:3.10} Suppose $\vec b$ is not a vertex of
$K_1$. If $p_m(0)<4k$, then the number of solutions of the
equation
    \begin{equation}\label{3.2.4}
    | \vec y_0 (\varphi)+\vec p_m(0) |_*^{2l}=k^{2l}
    \end{equation}
in $\varPhi _0$ does not exceed two. The solutions of
(\ref{3.2.4}) are, in fact, the solutions $\varphi_m^{\pm}$ of the
equation
    \begin{equation}\label{3.2.5}
    | \vec y_0 (\varphi)+\vec p_m(0) |_*^2=k^2.
    \end{equation}
If the distance from $\varphi \in \varPhi _0$ to the nearest
solution $\varphi_m^{\pm}$ of (\ref{3.2.5}) is greater than
$\epsilon _0$, $\epsilon _0>0$, then
    \begin{equation}\label{3.2.5.1}
     \Bigl| | \vec y_0 (\varphi)+\vec p_m(0) |_*^{2l}-k^{2l} \Bigr| > c_lb_0 \epsilon_0 ^2
    k^{2l-1},
    \end{equation}
    $c_l$ being a positive number depending only on $l$.
If $p_m(0) \geq 4k$, then the equation (\ref{3.2.4}) does not have
solutions in $\varPhi _0$. Moreover,
    \begin{equation}\label{3.2.5.2}
     \Bigl| | \vec y_0 (\varphi)+\vec p_m(0) |_*^{2l}-k^{2l} \Bigr| >
    \frac{1}{2^l}p_m^{2l}(0).
    \end{equation}
\end{proposition}
The proof is elementary and based on the fact that (\ref{3.2.5})
is a quadratic equation with respect to $e^{i\varphi }$.

\begin{corollary} \label{C:May22c}
For each $\vec b$, the number of $\varphi \in \varPhi _0$
satisfying (\ref{3.2.4}) for at least one $m\in \Z^2$ does not
exceed $c_0 k^{2+2s_1}$, $c_0=32b_1b_2$.
\end{corollary}
\begin{proof}[Corollary \ref{C:May22c}]
The equation (\ref{3.2.4}) can have solutions in $\varPhi _0$ only
if $p_m(0)<4k$. The number of such $m$'s does not exceed $16b_1b_2
k^{2+2s_1}$. For a fixed $m$ there exist no more than $2$
solutions, $\varphi_m^{\pm}$. This gives the statement of the
corollary.  \end{proof} According to the definition at the end of
Section \ref{sep20}, points $\varphi_m^{\pm}$ are points of
quasi-intersections of the circle of the radius $k$ centered at
the origin and the circle of the same radius centered at $-\vec
p_m(0)-\vec b$.
\begin{lemma}\label{L:3.12}
If $\varphi \in \varPhi _0$ satisfies the inequality
    $\ \ \min_{m\in \Z^2}|\varphi-\varphi_m^{\pm}|\geq \epsilon_0 \ \ $ for some $0<\epsilon_0<1,$
then
the following estimates hold:
    \begin{align}\label{3.2.6}
         \Bigl\| \Bigl(H_0\bigl(\vec y_0(\varphi)\bigr)-k^{2l}\Bigr)^{-1}\Bigr\| &\leq
     c_lk^{-2l+1}b_0^{-1}\epsilon_0^{-2},
    \\
     \Bigl\| \Bigl(H_0\bigl(\vec y_0(\varphi)\bigr)-k^{2l}\Bigr)^{-1}\Bigr\|_1 &\leq
     c_lk^{-2l+3+2s_1}b_0^{-1}\epsilon_0^{-2}.\label{3.2.7}
     \end{align}
\end{lemma}
\begin{proof}
Estimate (\ref{3.2.6}) immediately follows from (\ref{3.2.5.1})
and (\ref{3.2.5.2}). Considering that $H_0\bigl(\vec
y_0(\varphi)\bigr)$ is a diagonal operator and summarizing
(\ref{3.2.5.1}), (\ref{3.2.5.2}) over $m$, we arrive at
(\ref{3.2.7}).  \end{proof}


 Now let us construct  contours in $\varPhi _1\subset \varPhi _0$ whose distance
from all the solutions of the equations
   $ | \vec y_0 (\varphi)+\vec p_m(0)|_*^{2l}=k^{2l}$, $m \in \Z^2
   $,
is at least $k^{-4-6s_1-3\delta}$. We recall that  there are no
more than two solutions $\varphi_m^{\pm}$  in $\varPhi _0$ if
$p_m(0)<4k$, and there are no solutions if $p_m(0) \geq 4k$
(Proposition \ref{L:3.10}).

\begin{definition} \label{D:gamma} For each $\varphi_m^{+}\ (\text{or}\ \varphi_m^{-})$ in
$\varPhi _0$ we make an open disk of the radius
$k^{-4-6s_1-3\delta}$ centered at $\varphi_m^{+}\ (\text{or}\
\varphi_m^{-})$ and denote it by ${\cal O}_m^{+}\ (\text{or}\
{\cal O}_m^{-})$. Let ${\cal O}'$ the union of all such disks:
${\cal O}'=\cup _{m,\pm}{\cal O}_m^{\pm}$. We denote by ${\cal O}$
the part of ${\cal O}'$ related to $\varPhi _1$. Namely, ${\cal
O}$ is the union of connected components of ${\cal O}'$ not being
completely outside $\varPhi _1$.
 To stress that ${\cal O}$  depends on $\vec
b$, we often will write ${\cal O}(\vec b)$.\end{definition}

\begin{remark} \label{R:Nov11} Obviously, ${\cal O}(\vec b)$ consists of no more than
$c_0k^{2+2s_1}$ disks (see Corollary \ref{C:May22c}). Thus set
${\cal O}_*$ given by (\ref{ut1}) contains no more than
$4c_0k^{2+2s_2}$ disks, since $P$ contains less than
$4k^{s_2-s_1}$ elements.\end{remark}

\begin{definition} \label{D:May17a}
Let $\Gamma(\vec b)$ be a connected component of
 ${\cal O}(\vec b)$ and  $\gamma=\partial \Gamma$. \end{definition}
\begin{remark}\label{R:May18} The set ${\cal O}(\vec b)$ consists of  several components
    $\Gamma(\vec b)$. Since the number of points $\varphi_m^{\pm}$ in
$\varPhi _0$ does not exceed $c_0k^{2+2s_1}$, the size of any
$\Gamma$ is less than
    $ c_0k^{2+2s_1} \cdot  2k^{-4-6s_1-3\delta}=2c_0k^{-2-4s_1-{\bf{3}\delta}}.$
    Clearly it is much smaller than the size of $\varPhi
    _1$ (see Definition \ref{D:July3a}).
 Some  $\Gamma(\vec b)$ are strictly inside $\varPhi
    _1$, while others can have their parts outside. However, any $\Gamma(\vec b)$ is in the
    $(2c_0k^{-2-4s_1-{\bf{3}\delta}})-$
    neighborhood of $\varPhi
    _1$. This neighborhood has  properties completely analogous to those of $\varPhi
    _1$, since $ck^{-2-4s_1-{\bf{3}\delta}}<< k^{-2-4s_1-{\bf
    2}\delta}$, the latter value being the size of $\varPhi
    _1$.
     Hence, even if a component $\Gamma(\vec b)$ is not
    strictly inside $\varPhi
    _1$ it has the same properties as those inside $\varPhi
    _1$. Further we consider $\Gamma(\vec b)\subset \varPhi
    _1$, ${\cal O}(\vec b)\subset \varPhi
    _1$.
    \end{remark}
\begin{remark}\label{R:1}
For any $\varphi \in \varPhi _1\setminus {\cal O}(\vec b)$  and
any $\varphi_m^{\pm}$ the estimate $|\varphi-\varphi_m^{\pm}| \geq
k^{-4-6s_1-3\delta}$ holds.
     Therefore, the estimates (\ref{3.2.5.1}), (\ref{3.2.6})
      and (\ref{3.2.7}) hold  with
      $\epsilon_0=k^{-4-6s_1-3\delta}$ for such $\varphi $. \end{remark}

      Obviously,  zeros of $\det\Bigl(I+A_0\bigl(\vec y(\varphi )
      \bigr)\Bigr)$
      are described by the equations $|\vec \varkappa (\varphi )+
      \vec b +\vec
      p_m(0)|_*^{2l}=k^{2l}$, $m\in \Z^2$, $\varphi \in \varPhi _1$, which can be rewritten as
 $|\vec y_0 (\varphi)+h_1(\varphi )\vec\nu +\vec
    p_m(0)|_*^{2l}=k^{2l}$, the functions
    $\vec \varkappa (\varphi )\equiv \varkappa (\lambda,\vec \nu )$ and
    $h_1(\varphi )\equiv h_1(\lambda,\vec \nu )$ being defined in Lemma \ref{L:2.13}.
      It is easy to see that the vectors $\vec \varkappa (\varphi  )$
      satisfying this equation
with a real
      $\varphi $  are just intersections of $\cal D_1(\lambda )$ with
      the circle of radius $k$ centered at $-\vec b -\vec
      p_m(0)$.
      Since $\vec y(\varphi )$ differ from $\vec
      y_0(\varphi)$ by a small value $h_1(\varphi )$, the estimates
      (\ref{3.2.5.1}),
      (\ref{3.2.5.2})
and (\ref{3.2.6})
       (\ref{3.2.7}) with
      $\epsilon_0=k^{-4-6s_1-3\delta}$
      are preserved for $\varphi \in \varPhi _1\setminus {\cal O}$, when we replace
     $\vec y_0(\varphi )$ by $\vec
      y(\varphi)$. We prove this result in the following lemma.

\begin{lemma}\label{L:3.13}
If  $b_0 > k^{7+8s_1-2\gamma_0+6\delta}$, $\gamma
_0=2l-2-4s_1-2\delta $, then:
\begin{enumerate} \item
The number of zeros of $\det \Bigl(I+A_0\bigl(\vec
y(\varphi)\bigr)\Bigr)$ in  any $\Gamma (\vec b)$ is the same as
that of $\det \Bigl(I+A_0\bigl(\vec y_0(\varphi)\bigr)\Bigr)$.
\item The following estimates hold for $\varphi \in \varPhi _1
\setminus {\cal O}(\vec b)$:
    \begin{align}
    \Bigl\| \Bigl(H_0\bigl(\vec y (\varphi)\bigr)-k^{2l}\Bigr)^{-1}\Bigr\| &
    <cb_0^{-1}k^{-2l+9+12s_1+6\delta},
    \label{3.2.28}\\
    \Bigl\| \Bigl(H_0\bigl(\vec y (\varphi)\bigr)-k^{2l}\Bigr)^{-1}\Bigr\|_1 &
    <cb_0^{-1}k^{-2l+11+14s_1+6\delta}.
    \label{3.2.29}
    \end{align}
    \end{enumerate}
\end{lemma}
\begin{corollary} \label{C:May22a}
There are no zeros of $\det \Bigl(I+A_0\bigl(\vec
y(\varphi)\bigr)\Bigr)$ in $\varPhi _1\setminus {\cal O}(\vec b)$.
\end{corollary}
\begin{proof} [Lemma \ref{L:3.13}] Note that $\det \Bigl(I+A_0\bigl(\vec
y_0(\varphi)\bigr)\Bigr)=0$ if and only if $|\vec
y_0(\varphi)+\vec p_m(0)|_*^{2l}=k^{2l}$ for at least one $m\in
\Z^2$. The analogous statement holds for $\det
\Bigl(I+A_0\bigl(\vec y(\varphi)\bigr)\Bigr)$. It is easy to show
(Appendix \ref{A:2}) that
\begin{align}\label{3.2.30}
        \Bigl| |\vec y_0(\varphi)+\vec p_m(0)|_*^{2l}
        -|\vec y(\varphi)+\vec p_m(0)|_*^{2l}\Bigr|<
        \frac{1}{2}\Bigl| |\vec y_0(\varphi)+\vec p_m(0)|_*^{2l}-k^{2l}\Bigr|
        \end{align}
        for
    any $\varphi$ in $\varPhi _1\setminus {\cal
O}(\vec b)$ and $m\in \Z^2$. By  Rouch\'{e}'s theorem
\cite{Rouche}, the number of zeros of $|\vec y_0(\varphi)+\vec
p_m(0)|_*^{2l}-k^{2l}$ in every $\Gamma (\vec b)$ is the same as
that of $|\vec y(\varphi)+\vec p_m(0)|_*^{2l}-k^{2l}$. Therefore,
the first statement of the lemma is proved.

To prove the second statement, we use the Hilbert relation:
\begin{align*}
     \lefteqn{
     \Bigl(H_0\bigl(\vec y(\varphi)\bigr)-k^{2l}\Bigr)^{-1}
     =\Bigl(H_0\bigl(\vec y_0
    (\varphi)\bigr)-k^{2l}\Bigr)^{-1}}& \notag\\
    &\hspace{1cm}
    +\Bigl(H_0\bigl(\vec y(\varphi)\bigr)-k^{2l}\Bigr)^{-1}
    \Bigl(H_0\bigl(\vec y_0(\varphi)\bigr)-H_0\bigl(\vec y(\varphi)\bigr)\Bigr)
    \Bigl(H_0\bigl(\vec y_0(\varphi)\bigr)-k^{2l}\Bigr)^{-1}.
    \end{align*}
    By Lemma \ref{L:3.12} with
$\epsilon_0=k^{-4-6s_1-3\delta}$,
    \begin{align*}
    \Bigl\|\Bigl(H_0\bigl(\vec
    y_0(\varphi)\bigr)-k^{2l}\Bigr)^{-1}\Bigr\|&<cb_0^{-1}k^{-2l+9+12s_1+6\delta},\\
    \Bigl\|\Bigl(H_0\bigl(\vec
    y_0(\varphi)\bigr)-k^{2l}\Bigr)^{-1}\Bigr\|_1&<cb_0^{-1}k^{-2l+11+14s_1+6\delta}.
    \end{align*}
    We easily obtain from (\ref{3.2.30}) that
    \begin{equation}
    \Bigl\|\Bigl( H_0\bigl(\vec y_0(\varphi)\bigr)-H_0\bigl(\vec y(\varphi)\bigr) \Bigr)
    \Bigl(H_0\bigl(\vec
    y_0(\varphi)\bigr)-k^{2l}\Bigr)^{-1}\Bigr\|<1/2.\label{May18b}
    \end{equation}
 Now,   using a standard argument as in Lemma \ref{L:2.4.1/2}, we
 arrive at (\ref{3.2.28}) and (\ref{3.2.29}).
  \end{proof}
\begin{lemma}\label{L:3.14}
If $b_0 > k^{-2l+9+12s_1+7\delta}$, then the following estimates
hold for any $\varphi \in \varPhi _1\setminus {\cal O}(\vec b)$:
    \begin{align}
    \left\|(H^{(1)}(\vec y(\varphi))-k^{2l})^{-1}\right\|
    &<cb_0^{-1}k^{-2l+9+12s_1+6\delta}, \label{3.2.34.1}\\
       \left \|(H^{(1)}(\vec y(\varphi))-k^{2l})^{-1}\right\|_1
    &<cb_0^{-1}k^{-2l+11+14s_1+6\delta}. \label{3.2.34.2}
    \end{align}
\end{lemma}
\begin{corollary}\label{May22d} The determinant of $I+A_1\bigl(\vec
y(\varphi )\bigr)$ has no zeros in $\varPhi _1\setminus {\cal
O}(\vec b)$.
\end{corollary}
\begin{proof}
We use Hilbert relation $\Bigl(H^{(1)}\bigl(\vec
y(\varphi)\bigr)-k^{2l}\Bigr)^{-1}
     =$
    $$
     \Bigl(H_0\bigl(\vec y
    (\varphi)\bigr)-k^{2l}\Bigr)^{-1}+\Bigl(H^{(1)}\bigl(\vec y(\varphi)\bigr)-k^{2l}\Bigr)^{-1}
    (-W_1)\Bigl(H_0\bigl(\vec y(\varphi)\bigr)-k^{2l}\Bigr)^{-1}.
    $$
Since $b_0 > k^{-2l+9+12s_1+7\delta }>k^{7+8s_1-2\gamma _0+6\delta
}$ \ $(\gamma _0=2l-2-4s_1-2\delta )$, the previous lemma can be
applied. It yields: $$\Bigl\|\Bigl(H_0\bigl(\vec
y(\varphi)\bigr)-k^{2l}\Bigr)^{-1}\Bigr\|=O\left(b_0^{-1}k^{-2l+9+12s_1+6\delta}\right)=o(1).$$
Using a standard argument as in Lemma \ref{L:2.4.1/2}, we
 arrive at (\ref{3.2.34.1}) and (\ref{3.2.34.2}). \end{proof}
\begin{remark}
Obviously (\ref{3.2.34.1}) and (\ref{3.2.34.2}) hold for any
$H_{\alpha}^{(1)}=H_0+\alpha W_1$, $0 \leq \alpha \leq 1$
uniformly in $\alpha$.
\end{remark}
%
%
 Further,  we will use the
 notation $\frac{T_1}{T_2}=T_1T_2^{-1}$ for a pair of operators $T_1$, $T_2$.
\begin{lemma}\label{L:*1}
If $b_0 > k^{-2l+9+12s_1+7\delta}$, then the number of zeros of
$\det \Bigl(I+A_1\bigl(\vec y(\varphi )\bigr)\Bigr)$ in any
$\Gamma (\vec b)$ is the same as that of $\det
\Bigl(I+A_0\bigl(\vec y(\varphi )\bigr)\Bigr)$.
\end{lemma}
\begin{proof}
Let us consider $\det\left(\dfrac{H_0\bigl(\vec y(\varphi
)\bigr)+\alpha_1 W_1-k^{2l}}{H_0\bigl(\vec y(\varphi
)\bigr)+\alpha_2 W_1-k^{2l}}\right)$, where $0 \leq \alpha_1 \leq
\alpha_2 \leq 1$. Obviously,
    $$\det\left(\frac{H_0\bigl(\vec y(\varphi
)\bigr)+\alpha_1 W_1-k^{2l}}{H_0\bigl(\vec y(\varphi
)\bigr)+\alpha_2 W_1-k^{2l}}\right)
    =\det\left(I+\frac{(\alpha_1-\alpha_2) W_1}{H_0\bigl(\vec y(\varphi
)\bigr)+\alpha_2
    W_1-k^{2l}}\right).$$
Using (\ref{3.2.34.2}), which is uniform with respect to $\alpha$,
we obtain, by Rouch\'{e}'s theorem, that this determinant has the
same number of poles and zeros if $|\alpha_2-\alpha_1|$ is
sufficiently small:
$|\alpha_2-\alpha_1|<b_0k^{2l-11-14s_1-7\delta}$. We  make a
finite number of steps between $\alpha=0$ and $\alpha=1$. Indeed,
$\det\left(\dfrac{H_0\bigl(\vec y(\varphi )\bigr)+\alpha_1
W_1-k^{2l}}{H_0\bigl(\vec y(\varphi )\bigr)+\alpha_2
W_1-k^{2l}}\right)$ has the same number of zeros and poles and so
does \\ $\det\left(\dfrac{H_0\bigl(\vec y(\varphi )\bigr)+\alpha_2
W_1-k^{2l}}{H_0\bigl(\vec y(\varphi )\bigr)+\alpha_3
W_1-k^{2l}}\right)$. Both determinants can be represented as
fractions of holomorphic functions:
$\dfrac{f_1(\varphi)}{f_2(\varphi)}$ and
$\dfrac{g_1(\varphi)}{g_2(\varphi)}$, respectively. The number of
zeros being the same for the pairs $f_1$, $f_2$ and  $g_1$, $g_2$.
Note that
    $\det\left(\dfrac{H_0(\vec y(\varphi )\bigr)+\alpha_1 W_1
    -k^{2l}}{H_0(\vec y(\varphi )\bigr)+\alpha_3
    W_1-k^{2l}}\right)=\dfrac{f_1g_1}{f_2g_2}.$
Obviously, it also has the same number of zeros and poles. Hence,
$\det\left(\dfrac{H_0\bigl(\vec y(\varphi
)\bigr)+W_1-k^{2l}}{H_0(\vec y(\varphi )\bigr)-k^{2l}}\right)$ has
the same number of zeros and poles inside $\Gamma (\vec b)$.
Considering that the last determinant is the quotient of $\det
\Bigl(I+A_1\bigl(\vec y(\varphi )\bigr)\Bigr)$ and  $\det
\Bigl(I+A_0\bigl(\vec y(\varphi )\bigr)\Bigr)$ and
these two determinants are holomorphic functions, we obtain, that
they have the same number of zeros inside $\Gamma (\vec b)$.
\end{proof}
\begin{lemma}\label{L:*1*}
\begin{enumerate}
\item
If $b_0 > k^{-2l+9+12s_1+7\delta}$, then the number of zeros of\\
$\det \Bigl(I+A_1\bigl(\vec y(\varphi )\bigr)\Bigr)$ in  any
$\Gamma (\vec b)$ is the same as that of $\det
\Bigl(I+A_0\bigl(\vec y_0(\varphi )\bigr)\Bigr)$. \item The
distance from a zero of  $\det \Bigl(I+A_1\bigl(\vec y(\varphi
)\bigr)\Bigr)$ to the nearest zero of $\det \Bigl(I+A_0\bigl(\vec
y_0(\varphi )\bigr)\Bigr)$ does not exceed $r/2$.
\end{enumerate}
\end{lemma}
\begin{corollary} \label{C:May23}
The  total number of zeros of $\det \Bigl(I+A_1\bigl(\vec
y(\varphi )\bigr)\Bigr)$ in  ${\cal O}(\vec b)$ does not exceed
$c_0k^{2+2s_1}$.
\end{corollary}
The Corollary immediately follows from the first statement of the
lemma, the definition of ${\cal O}(\vec b)$ as a collection of
disks surrounding zeros of $\det \Bigl(I+A_0\bigl(\vec y_0(\varphi
)\bigr)\Bigr)$ and   Remark \ref{R:Nov11}, page \pageref{R:Nov11}.
\begin{corollary} \label{C:May24} The number of self-intersections of
${\cal K}_2\chi _1^*(\lambda )$ does not exceed $c_0k^{2+2s_1}$
and $L\left({\cal K}_2\chi _1^*(\lambda )\right)=L\left(\chi
_1^*(\lambda )\right)$.  \end{corollary}
 The
latter equality holds, since ${\cal K}_2\chi _1^*(\lambda )$ has
only a finite number of self-intersections.
\begin{proof} [Lemma \ref{L:*1*}] The first statement immediately
follows  Lemmas \ref{L:3.13} (Part 1) and \ref{L:*1}. Next, note
that Lemmas \ref{L:3.13} -- \ref{L:*1} hold not only for ${\cal
O}(\vec b)$, but also for a set $\tilde{\cal O}(\vec b)$,
constructed from disks of twice smaller radius $r/2$ with the same
centers, since all estimates in the lemmas preserved under such
change up to some insignificant constants. This proves that all
zeros of $\det \Bigl(I+A_1\bigl(\vec y(\varphi )\bigr)\Bigr)$ in
${\cal O}(\vec b)$ are, in fact, in the smaller set $\tilde {\cal
O}(\vec b)$, i.e., the second statement of the lemma holds.
\end{proof}
Let us recall that  $\omega _{1,p}^*$  is the set of zeros in
$\varPhi _1$ of the determinant  (\ref{May18c+}) when $|\epsilon
|<\epsilon _1$, here $p\in P\setminus \{0\}$ and fixed. The set
$\omega _{1,p}$ is the real part of $\omega _{1,p}^*$: $\omega
_{1,p}=\omega _{1,p}^*\cap \Theta _1.$
\begin{lemma}\label{L:3.17}
If $b_0 > k^{-2l+9+12s_1+7\delta}$, $|\epsilon
|<b_0k^{2l-11-14s_1-7\delta}$, then
\begin{enumerate} \item The determinant of $\ \dfrac{H^{(1)}(\vec
y(\varphi))-k^{2l}-\epsilon }{H_0(\vec y(\varphi))+k^{2l}}$
 has no zeros in $\varPhi _1\setminus {\cal O}(\vec b)$. \item
The number of zeros of  $\det\left(\dfrac{H^{(1)}(\vec
y(\varphi))-k^{2l}-\epsilon }{H_0(\vec y(\varphi))+k^{2l}}\right)$
in $\Gamma (\vec b)$ is the same as that of \\
$\det\left(\dfrac{H^{(1)}(\vec y(\varphi))-k^{2l}}{H_0(\vec
y(\varphi))+k^{2l}}\right)$.
\end{enumerate}
\end{lemma}
\begin{corollary}\label{C:ut} If $\vec b=2\pi p/N_1a$,
$\omega _{1,p}^*\subset {\cal O}(\vec b), \ \ \ \omega
_{1,p}\subset
 {\cal O}(\vec b)\cap \Theta _1.$ \end{corollary}
\begin{proof}   To prove Lemmas
\ref{L:3.14} and \ref{L:*1}, we used the estimate $\|W_1\|=O(1)$.
Obviously, $\|W_1-\epsilon \|=O(1)$. Hence, the analogous lemmas
hold when we replace $W_1$ by $W_1-\epsilon $, i.e., when we
consider $H^{(1)}(\vec
    y(\varphi))-k^{2l}-\epsilon $ instead of just $H^{(1)}(\vec
    y(\varphi))-k^{2l}$. Modified Lemmas \ref{L:3.14} and \ref{L:*1} immediately yield the
    statement of the present lemma.
     \end{proof}

\subsubsection{Proof of statement 3 of Geometric Lemma}\label{St3}
By Corollary \ref{C:ut},
    $\omega_{1,p}\subset {\cal O}(\vec b)\cap \Theta _1$, $\vec b=2\pi p/N_1a$.
    It is easy to see that $b_0 \geq c_bk^{-s_2}$,
$c_b=\frac{1}{4}\min \{b_1,b_2\}$. Such $\vec b$ satisfies
conditions of Lemmas \ref{L:3.13} -- \ref{L:3.17}
    for any $p\in P \setminus \{ 0 \}$.
   Taking
into
    account that
    that ${\cal O}(\vec b)$ is formed by no more than $c_0k^{2+2s_1}$  disks of the radius
    $r=k^{-4-6s_1-3\delta }$, we obtain that the total length of
    $\omega_{1,p}$ does not exceed $2c_0k^{-2-4s_1-3\delta }$. Since
    the set $P$ contains no more than $4k^{2s_2-2s_1}$ elements
    and
    $s_2=2s_1$, the total length of $\omega_{1}$ does not exceed $8c_0k^{-2-2s_1-3\delta
    }$. Therefore, the length of $\Omega _1$ does not exceed
    $20\pi c_0k^{-1-2s_1-3\delta }$. Considering that  $\chi
    _1^*(\lambda )$ has a length $2\pi k \bigl(1+o(1)\bigr)$, we obtain
    Statement 3.

\subsection{Non-resonant part of the isoenergetic set of
$H_{\alpha }^{(2)}$}\label{S:3.4}

Let $S_2(\lambda )$ be an isoenergetic set of the operator
$H_{\alpha}^{(2)}$:
    $S_2(\lambda)=\{\tau \in K_2 : \exists n \in \N:\
    \lambda_n^{(2)}(\alpha,\tau)=\lambda  \}$,
    here $\{\lambda_n^{(2)}(\alpha,\tau)\}_{n=1}^{\infty }$ is the
    spectrum of $H_{\alpha}^{(2)}(\tau)$.
Now we construct a non-resonance subset $\chi _2^*(\lambda )$ of
$S_2(\lambda
    )$. It
    corresponds to  non-resonance eigenvalues $\lambda
    _{\tilde j}^{(2)}(\tau )$ given by the perturbation series (\ref{3.15}).
    We start with a definition of $\cal D_1(\lambda )_{nonres}$.
    Recall that $\chi _2\subset \cal{K}_2\chi_1^*(\lambda )$ (see
    the Geometric Lemma) and $\chi_1^*(\lambda
    )=\cal{K}_1\cal D_1(\lambda )$, see (\ref{2.81}). Hence, $\chi _2\subset \cal{K}_2\cal D_1(\lambda
    )$. Let $\cal D_1(\lambda )_{nonres}$ be the preimage of $\chi
    _2$ in $\cal D_1(\lambda )$:
    \begin{equation}\cal D_1(\lambda )_{nonres}=\{\vec \varkappa \in \cal D_1(\lambda
    ):\cal{K}_2\vec \varkappa \in \chi _2\}. \label{pr}
    \end{equation}
\begin{lemma}\label{L:3.7.1/2}
The formula $\cal{K}_2\cal{D}_1(\lambda )_{nonres}=\chi_2$
establishes
 one-to-one correspondence between
$\cal{D}_1(\lambda )_{nonres}$ and $\chi_2$.
\end{lemma}
\begin{proof}

Suppose there is a pair $\vec \varkappa _1, \vec \varkappa _2\in
\cal D_1(\lambda )_{nonres}$ such that $\cal{K}_2\vec \varkappa
_1= \cal{K}_2\vec \varkappa _2=\tau $, $\tau \in \chi _2$.  We
introduce also  $t_1=\cal{K}_1\vec \varkappa _1$ and
$t_2=\cal{K}_1\vec \varkappa _2$.   The definition (\ref{2.81}) of
$\chi_1^*(\lambda
    )$ implies that $t_1,t_2 \in \chi_1^*(\lambda
    )$, since $\vec \varkappa _1, \vec \varkappa _2\in
\cal D_1(\lambda )_{nonres}\subset \cal D_1(\lambda )$. Clearly,
$\cal{K}_2t_1=\cal{K}_2t_2=\tau $ and, hence, $t_1=\tau +2\pi
p_1/N_1a$, $t_2=\tau +2\pi p_2/N_1a$ for some $p_1,p_2\in P$. Now,
by Part 1a of Geometric Lemma \ref{L:3.2}, $p_1=p_2$, and,
    therefore, $t_1=t_2$. Next, by Lemma \ref{L:May10a}, $\vec \varkappa _1=\vec \varkappa
    _2$.
     \end{proof}
We define $\cal B_2 (\lambda)$ as the set of directions
corresponding to $\cal{D}_1(\lambda
 )_{nonres}$:
$$\cal B_2(\lambda)=\{\vec{\nu} \in \cal B_1(\lambda) :
    \varkappa _1(\lambda,\vec{\nu})\vec{\nu} \in \cal{D}_1(\lambda
    )_{nonres}\},$$
 where
    $\varkappa _1(\lambda ,\vec{\nu})$ is defined by Lemma
    \ref{L:2.13}, $\varkappa _1(\lambda ,\vec{\nu})\equiv \varkappa
    _1(\varphi )$.
Note that $\cal B_2 (\lambda)$ is a unit circle with holes,
centered at the origin, and $\cal B_2(\lambda ) \subset
    \cal B_1(\lambda )$.
    We denote by $\Theta _2(\lambda )$ the set of angles $\varphi
$, corresponding to $\cal B_2(\lambda )$:
\begin{equation*}
\Theta _2(\lambda )=\{\varphi \in [0,2\pi ):\ (\cos \varphi ,\sin
\varphi )\in
\cal B_2 (\lambda) \},\ \ \ \     \Theta _2\subset \Theta _1.
 \end{equation*}
Let $\vec{\varkappa } \in \cal{D}_1(\lambda )_{nonres}$. By
(\ref{pr}), $\tau \equiv {\cal K}_2\vec \varkappa \in \chi
_2(\lambda )$. According to Theorem \ref{T:3.4}, for sufficiently
large $\lambda $, there exists an eigenvalue of the operator
$H^{(2)}_{\alpha}(\tau )$, given by (\ref{3.15}). It is convenient
here to denote $\lambda_{\tilde{j}}^{(2)} (\alpha ,\tau)$ by
$\lambda^{(2)}(\alpha,\vec{\varkappa })$.
 We can do this, since, by Lemma \ref{L:3.7.1/2},
there is one-to-one correspondence between $\vec{\varkappa }\in
\cal{D}_1(\lambda )_{nonres}$ and the pair $(\tau, \tilde j)$, $
\vec \varkappa =2\pi \tilde j/a+\tau $. We rewrite (\ref{3.15}) in
the form:
    \begin{equation}
    \lambda^{(2)}(\alpha,\vec{\varkappa })
    =\lambda^{(1)}(\vec{\varkappa })+f_2(\alpha,\vec{\varkappa }),\
    \ \
    \label{3.66}
    f_2(\alpha,\vec{\varkappa  })=\sum _{r=1}^{\infty}
\alpha^{r}g_r^{(2)}(\vec{\varkappa }),
\end{equation}
here $g_r^{(2)}(\vec{\varkappa })$ is given by (\ref{3.13}). The
function $f_2(\alpha,\vec \varkappa )$ satisfies the estimates:
\begin{align}
    |f_2(\alpha,\vec \varkappa )|&\leq 12 \alpha \epsilon_1
    ^4,\label{3.67}\\
    |\nabla f_2(\alpha,\vec \varkappa )|&
    \leq 24 \alpha \epsilon_1 ^3 k^{2l-1+\delta}.\label{3.68}
    \end{align}
By Theorems \ref{T:3.4} and \ref{T:3.5a}, the formulas
(\ref{3.66}) -- (\ref{3.68}) hold even in $(\epsilon_1
k^{-2l+1-\delta})$-neighborhood of $\cal{D}_1(\lambda )_{nonres}$,
i.e., for any $\vec \varkappa =\varkappa \vec{\nu}$ such that
$\vec{\nu} \in \cal B_2(\lambda)$ and $|\varkappa -\varkappa
_1(\lambda,\vec \nu)|<\epsilon_1 k^{-2l+1-\delta}$. We define
$\cal{D}_2(\lambda)$ as a level set for
$\lambda^{(2)}(\alpha,\vec{\varkappa })$ in this neighborhood:
    \begin{equation}
    \cal{D}_2(\lambda ):=\left\{\vec{\varkappa }=\varkappa \vec \nu: \vec \nu \in
    \cal B_2,\ \
    \bigl|\varkappa  -\varkappa _1(\lambda,\vec
\nu) \bigr|<\epsilon_1 k^{-2l+1-\delta},\
\lambda^{(2)}(\alpha,\vec{\varkappa })=\lambda
\right\}.\label{May14m}
\end{equation}
Next two lemmas are to prove   that $\cal{D}_2(\lambda)$ is a
distorted circle with holes.
%
%
\begin{lemma}\label{L:3.8}
For every $\vec{\nu} \in \cal B_2$ and every $\alpha$, $0 \leq
\alpha \leq 1$, there is a unique $\varkappa =\varkappa _2
(\lambda,\vec{\nu})$ in the interval $I_2:=\bigl[\varkappa _1
(\lambda,\vec \nu)-\epsilon_1 k^{-2l+1-\delta},\varkappa _1
(\lambda,\vec \nu)+\epsilon_1 k^{-2l+1-\delta}\bigr]$ such that
    \begin{equation}\label{3.70}
    \lambda^{(2)}(\alpha,\varkappa _2 \vec{\nu})=\lambda .
    \end{equation}
Furthermore,
    \begin{equation}\label{3.71}
    |\varkappa _2(\lambda,\vec
\nu) -\varkappa _1(\lambda,\vec \nu)| \leq 2\alpha \epsilon_1
^4k^{-2l+1}.
    \end{equation}
\end{lemma}
\begin{proof}
 Let us show that there exists a $\varkappa _2 (\lambda,\vec{\nu})$
satisfying (\ref{3.70}). Since $\varkappa _1\vec \nu \in
\cal{D}_1(\lambda )$ when $\vec \nu \in \cal B_2 (\lambda)$, for
all $\varkappa $ in the interval $I_2$, there is an eigenvalue
$\lambda^{(2)}(\alpha, \varkappa \vec{\nu})$ given by
(\ref{3.66}). In particular,
   $ \lambda^{(2)}\bigl(0,\varkappa _1\vec{\nu}\bigr)=
    \lambda^{(1)}\bigl(\varkappa _1\vec{\nu}\bigr)=\lambda.$
Let
    $\nobreak {\cal L}^{(2)}(\alpha,\vec{\nu}):=
    \{\lambda=\lambda^{(2)}(\alpha,\varkappa \vec{\nu}):\varkappa \in I_2\}.$
It is an interval, since $\lambda^{(2)}(\alpha, \varkappa
\vec{\nu})$ is a continuous function of $\varkappa $. Considering
that $\lambda^{(1)}\bigl(\varkappa \vec{\nu}\bigr)$ is a
continuous function and  $\nabla \lambda^{(1)}\bigl(\varkappa
\vec{\nu}\bigr)=2l\varkappa^{2l-1}\vec \nu \bigl(1+o(1)\bigr)$, we
readily obtain:
    ${\cal L}^{(2)}(0,\vec{\nu}) \supset
    [k^{2l}-c_1\epsilon_1 k^{-\delta},k^{2l}+c_1\epsilon_1 k^{-\delta}]$, $0<c_1
    \neq c_1(k).$
Since $\lambda^{(2)}(\alpha,\varkappa \vec{\nu})$ is continuous in
$\varkappa $ and~(\ref{3.67}) holds, we have
    ${\cal L}^{(2)}(\alpha,\vec{\nu}) \supset
    [k^{2l}-c_1\epsilon_1 k^{-\delta}+c_2\epsilon_1 ^4,
    k^{2l}+c_1\epsilon_1 k^{-\delta}-c_2\epsilon_1 ^4]$,
    $0<c_2 \neq c_2(k).$
The inequality $\epsilon_1 ^4 \ll \epsilon_1 k^{-\delta}$ yields:
    $\Lambda^{(2)}(\alpha,\vec{\nu}) \supset
    [k^{2l}-c_1\epsilon_1
    k^{-\delta}/2,k^{2l}+c_1\epsilon_1
    k^{-\delta}/2].$
Thus for every $\vec{\nu} \in \cal B_2 (\lambda)$, there exists a
$\varkappa _2(\lambda ,\vec \nu ) \in I_2$ such that
$\lambda^{(2)}(\alpha,\varkappa _2 \vec{\nu})=\lambda$.

 Using (\ref{2.20**}), we obtain
    \begin{equation}\label{3.73}
    \dfrac{\partial \lambda^{(2)}(\alpha,
    \varkappa \vec{\nu})}{\partial \varkappa } = 2lk^{2l-1}\bigl( 1+o(1) \bigr),
    \end{equation}
and this implies that $\lambda^{(2)}(\alpha,\varkappa \vec{\nu})$
is monotone with respect to $\varkappa $ in $I_2$. Hence, there is
only one $\varkappa _2(\lambda ,\vec \nu )$
satisfying~(\ref{3.70}) in $I_2$.

 Now let us estimate $\left|\varkappa _2(\lambda ,\vec \nu
)-\varkappa _1(\lambda,\vec{\nu})\right|$. By~(\ref{3.66}), we
have
    $\lambda^{(2)}(\alpha,\varkappa _1\vec{\nu})=\lambda^{(1)}\bigl(\varkappa _1\vec{\nu}
    \bigr)
    +f_2\bigl(\alpha,\varkappa _1\vec{\nu}\bigr)=\lambda +
    f_2\bigl(\alpha,\varkappa _1\vec{\nu}\bigr)$.
Using (\ref{3.67}) and (\ref{3.70}), we easily obtain:
$\bigl|\lambda^{(2)}\bigl(\alpha, \varkappa _2\vec{\nu}\bigr)-
    \lambda^{(2)}\bigl(\alpha, \varkappa _1\vec{\nu}\bigr)\bigr|<12\alpha \epsilon_1
    ^4$. Applying
 the Mean Value Theorem at the left-hand side and using
 (\ref{3.73}), we arrive at the inequality: $lk^{2l-1}|\varkappa _2 -\varkappa
 _1|<12\alpha \epsilon_1
    ^4$.
    The estimate (\ref{3.71}) immediately follows.
 \end{proof}
%
%
Let us recall that $\varPhi _2$ is the second complex non-resonant
set given by  (\ref{ut1}), see Fig. \ref{F:7}. Further, we use the
notations: $\varkappa _2(\varphi )\equiv \varkappa _2(\lambda
,\vec \nu )$, $h_2(\varphi )=\varkappa _2(\varphi )-\varkappa
_1(\varphi )$, $\vec \varkappa _2(\varphi )=\varkappa _2(\varphi
)\vec \nu $.
\begin{lemma}\label{L:3.9} \begin{enumerate}
\item The set $\cal{D}_2(\lambda )$ is a distorted circle with
holes: it can be described by the formula:
\begin{equation}
\cal{D}_2(\lambda )=\bigl\{\vec \varkappa \in \R^2: \vec \varkappa
=\vec \varkappa _2(\varphi ),\ \ {\varphi } \in \Theta _2(\lambda
)\bigr\},\label{May20a} \end{equation} where
    $ \varkappa _2(\varphi )=\varkappa _1(\varphi)+h_2(\varphi)$,
$\varkappa _1(\varphi )$ is the ``radius" of $\cal{D}_1(\lambda )$
and $h_2(\varphi )$ satisfies the estimates
    \begin{equation}\label{3.75}
    |h_2|\leq 2\alpha \epsilon_1 ^4k^{-2l+1},\ \ \
    \left|\frac{\partial h_2}{\partial \varphi} \right| \leq
    4\alpha \epsilon_1^3 k^{1+\delta}.
    \end{equation}
     \item The total length of $\cal{B}_2(\lambda)$ satisfies the estimate:
\begin{equation}\label{theta2}
    L\left(\cal{B}_1\setminus \cal{B}_2\right)<4\pi k^{-2-2s_1}.
    \end{equation}
\item The function $\varkappa _2(\varphi )$ can be extended as a
holomorphic function of $\varphi $ into the complex non-resonance
set $\varPhi _2$,  estimates (\ref{3.75}) being preserved. \item
The curve $\cal{D}_2(\lambda )$ has a length which is
asymptotically close to that of $\cal{D}_1(\lambda )$ in the
following sense:
    \begin{equation}\label{3.77}
     L\Bigl(\cal{D}_2(\lambda )\Bigr)\underset{\lambda \rightarrow
     \infty}{=}L\Bigl(\cal{D}_1(\lambda )\Bigr)\Bigl(1+O\bigl(k^{-2-2s_1}\bigr)\Bigr).
     \end{equation}
     \end{enumerate}
\end{lemma}
%
%
\begin{proof}
\begin{enumerate}
\item By Lemma \ref{L:3.8},
     $|h_2|=|\varkappa _2-\varkappa _1| \leq
     2\alpha \epsilon_1 ^4k^{-2l+1}$.
Let us prove the second estimate in (\ref{3.75}). Considering that
$\lambda =\lambda^{(2)}\left(\alpha,\vec \varkappa _2(\varphi
)\right)$, $\lambda =\lambda^{(1)}\left(\vec \varkappa _1(\varphi
)\right),$ differentiating both equations with respect to $\varphi
$ and subtracting the results, we obtain:
\begin{equation}
\frac{\partial\lambda^{(2)}}{\partial \varkappa }\left(\alpha,\vec
\varkappa _2\right)\frac{\partial h_2}{\partial
\varphi}=F(\lambda, \varphi ),\label{May13b}
     \end{equation}
     \begin{align}F(\lambda, \varphi )=\label{Oct5}
     \end{align}
     $$\Bigl(\nabla \lambda^{(1)}\left(\vec \varkappa _1\right)-
     \nabla \lambda^{(2)}\left(\alpha,\vec \varkappa _2\right)
\Bigr)\frac{\partial \varkappa _1}{\partial \varphi}+ \Bigl\langle
\varkappa _1 \nabla \lambda^{(1)}\left(\vec \varkappa _1\right)-
     \varkappa _2\nabla \lambda^{(2)}\left(\alpha,\vec \varkappa _2\right),\vec
     \mu
\Bigr\rangle ,$$ where $\vec \mu = (-\sin \varphi , \cos \varphi
)$. It is easy to check (Appendix \ref{A:3}) that
\begin{equation}\bigl|F(\lambda, \varphi )\bigr|<27\alpha
k^{2l+\delta }\epsilon _1^3. \label{May13c}
\end{equation} Then, the second estimate in (\ref{3.75}) easily follows from formulae
(\ref{May13b}) and estimates  (\ref{3.73}) and (\ref{May13c}).
\item Considering that $\varTheta _1$ is the set of angles
corresponding to $\cal{D}_1(\lambda )$ (Lemma \ref{L:2.13}) and
$\varTheta _2$ is the set of angles corresponding to
$\cal{D}_1(\lambda )_{nonres}$, we obtain
\begin{equation}L\Bigl(\cal{D}_1(\lambda )\setminus \cal{D}_1(\lambda
)_{nonres}\Bigr)=\int_{\varTheta_1\setminus \varTheta_2}
     \sqrt{\varkappa _1^2(\varphi )+\varkappa _1^{\prime}(\varphi )\ ^2}\,d\varphi.
     \label{Dec7}
\end{equation}
Using (\ref{2.75}), we easily show that the integrand is close to
$k$. Hence, \begin{equation} L\Bigl(\cal{D}_1(\lambda )\setminus
\cal{D}_1(\lambda )_{nonres}\Bigr)=L\left(\varTheta_1\setminus
\varTheta_2\right)k\bigl(1+o(1)\bigr).
     \label{Dec8}
     \end{equation}
     Let us show  that
     \begin{equation}\label{3.77a>}
     L\Bigl(\cal{D}_1(\lambda )\setminus \cal{D}_1(\lambda )_{nonres}\Bigr)<2\pi k^{-1-2s_1}\bigl(1+o(1)\bigr).
     \end{equation}
     In fact,
     by Lemma  \ref{L:3.7.1/2}, $L\bigl(\cal{D}_1(\lambda)_{nonres}\bigr)=L(\chi
     _2)$. By Lemma \ref{L:May10a},  $L\bigl(\cal{D}_1(\lambda)\bigr)=L(\chi
     _1^*)$. By Corollary
     \ref{C:May24},
      $L\bigl(\chi
     _1^*\bigr)=L\bigl({\cal K}_2\chi
     _1^*\bigr)$, and, therefore, $L\bigl(\cal{D}_1(\lambda)\bigr)=L({\cal K}_2\chi
     _1^*)$. It follows:
     $L\bigl(\cal{D}_1(\lambda )\setminus \cal{D}_1(\lambda )_{nonres}\bigr)=L\bigl(\cal{D}_1(\lambda)\bigr)-L\bigl(\cal{D}_1(\lambda)_{nonres}\bigr)=
L({\cal K}_2\chi
     _1^*)-L(\chi
     _2)=L\bigl({\cal K}_2\chi
     _1^*\setminus\chi _2)$, the last equality being valid since
     $\chi _2$ is a subset of ${\cal K}_2\chi
     _1^*$. Using the estimate (\ref{3.9}) in the Geometric
     Lemma, we obtain
     \begin{equation}\label{3.77a}
     L\Bigl(\cal{D}_1(\lambda )\setminus \cal{D}_1(\lambda )_{nonres}\Bigr)<L\Bigl(\cal{D}_1(\lambda )\Bigr)k^{-2-2s_1}.
     \end{equation} Now considering that
     $L\bigl(\cal{D}_1(\lambda)\bigr)=2\pi k \bigl(1+o(1)\bigr)$ (Lemma \ref{L:2.13}), we
     arrive at the estimate (\ref{3.77a>}). Combining it with
     (\ref{Dec8}), we get (\ref{theta2}).
\item 
 By the third statement of Lemma \ref{L:2.13}, the function $\varkappa _1(\varphi )$ is
 holomorphic in $\varPhi _1$ and $\lambda ^{(1)}\bigl(\varkappa _1(\varphi )
 \bigr)=\lambda $. If $\varphi \in \Theta _2$, then
 $\lambda ^{(2)}\bigl(\varkappa _1(\varphi )\bigr)$ is
 defined by the series (\ref{3.66}), since $\varkappa _1(\varphi )
 \in \cal D_1(\lambda )_{nonres}$ for any $\varphi \in \Theta _2$.
Now we show that $\lambda ^{(2)}
 \bigl(\varkappa _1(\varphi )\bigr)$
 can be analytically extended from $\Theta _2$ to $\varPhi _1\setminus {\cal O}$.
It is enough to check that the series (\ref{3.66}) converges in
$\varPhi _1\setminus {\cal
 O}$. We need an estimate for
 $g_r^{(2)}\bigl(\varkappa _1(\varphi )
 \bigr)$, which is  similar to (\ref{3.17}). Indeed, considering  formula
  (\ref{3.13}) for $g_r^{(2)}$ and taking into
 account that
 $\|W_2\|<\epsilon _1^{4}$, we easily conclude that
 it suffices to prove the
 estimate:
 \begin{equation}
 \sup _{z\in C_2}\Bigl\|\left(\tilde H^{(1)}\bigl(\vec \varkappa _1(\varphi
 )\bigr)-z\right)^{-1}\Bigr\|<\frac{1}{\epsilon _1}.\label{July3b}
 \end{equation}
 From the definition of $\tilde H^{(1)}$, we  see:
 $$\Bigl\|\left(\tilde H^{(1)}\bigl(\vec \varkappa _1(\varphi
 )\bigr)-z\right)^{-1}\Bigr\|=\max _{p\in P}\Bigl\|\Bigl(H^{(1)}\bigl(\vec \varkappa _1(\varphi
 )+2\pi p/N_1a\bigr)-z\Bigr)^{-1}\Bigr\|. $$
 To estimate the right-hand side, we start with the case $p=0$.
 We  obtain an upper bound on
 $\|\left(H^{(1)}\bigl(\vec \varkappa _1(\varphi
 )\bigr)-z\right)^{-1}\bigr\|$ using results of Section \ref{chapt3}. Indeed,
 ${\cal K}_1\vec \varkappa _1(\varphi )$ belongs to
 $\chi _1^*(\lambda )$, which is in the complex $(k^{-1-4s_1-2\delta })$-neighborhood
 of the first non-resonance set $\chi _1(\lambda )$.
  Hence,  we can apply Lemma \ref{L:July5} from
  Appendix \ref{A:4} with $t=
 {\cal K}_1\vec \varkappa _1(\varphi )$.
By this lemma,
 $\bigl(H^{(1)}\bigl(\vec \varkappa _1(\varphi
 )\bigr)-z\bigr)^{-1}$ has a single pole $z_0$ inside the contour $C_1$.
 It is easy to see that $z_0=\lambda $,
  since $\lambda ^{(1)}\bigl(\vec
 \varkappa _1(\varphi )\bigr)=\lambda $. Again, by Lemma \ref{L:July5},
\begin{equation}
\Bigl\|(z-k^{2l})\left( H^{(1)}\bigl(\vec \varkappa _1(\varphi
 )\bigr)-z\right)^{-1}\Bigr\|<16 \label{16*}
 \end{equation}
 for all $z$ inside $C_1$. Considering $z\in C_2$, that is, $|z-k^{2l}|=\epsilon
 _1/2$, we easily obtain from (\ref{16*}) that
 \begin{equation}
 \Bigl\|\left(H^{(1)}\bigl(\vec \varkappa _1(\varphi
 )\bigr)-z\right)^{-1}\Bigr\|<\frac{32}{\epsilon _1}, \ \ \ z\in C_2.
 \label{July3c}
 \end{equation}
 Next, we consider the case $p\in P\setminus \{0\}$. We use
 Lemma \ref{L:3.14} with $\vec b=2\pi p/N_1a$, $p\neq 0$. It is possible to
 replace $k^{2l}$ in (\ref{3.2.34.1}) by any
 $z\in C_2$, since the difference between $z$ and $k^{2l}$ is small: $|z-k^{2l}|=\epsilon
 _1/2$, $\epsilon _1=o\left(b_0k^{2l-9-12s_1-6\delta }\right)$. Thus,
 \begin{equation}\Bigl\|\left(H^{(1)}\bigl(\vec \varkappa _1(\varphi
 )+2\pi p/N_1a\bigr)-z\right)^{-1}\Bigr\|<\frac{2}{\epsilon
 }_1 \label{Oct5a}
 \end{equation} for any $p\in P\setminus \{0\}$, $z\in C_2$.
 Combining (\ref{July3c}) with the last estimate, we arrive at
 (\ref{July3b}), which
means that the series for $\lambda ^{(2)}\bigl(\alpha, \vec
\varkappa _1(\varphi
 )\bigr)$ converges. It is easy to see that estimates
 (\ref{July3c}), (\ref{Oct5a}) are stable in a vicinity of $\vec \varkappa _1(\varphi
 )$,
  namely, they hold at all points $\varkappa \vec \nu $:
 $|\varkappa -\varkappa _1(\varphi )|<\epsilon
 _1k^{-2l+1-\delta}$. Hence,  the perturbation series for the
 eigenvalue converges for such $\varkappa$ too,
 the following estimates being valid:
 \begin{equation}
 \Bigl|\lambda ^{(2)}(\alpha ,\varkappa \vec \nu  )-
 \lambda ^{(1)}(\varkappa \vec \nu  )\Bigr|<12\alpha \epsilon _1^4, \label{July3d}
 \end{equation}
 \begin{equation}
 \nabla \lambda ^{(2)}(\alpha ,\varkappa \vec \nu  )=
2l k^{2l-2}\vec k +o(k^{2l-1}). \label{July3e}
 \end{equation}
  Next, we want to show that the
 equation
 $\lambda ^{(2)}(\alpha ,\varkappa \vec \nu )=\lambda $ has a solution
 $\varkappa _2(\varphi )$\footnote{In fact, it should be $\varkappa _2(\alpha ,\varphi
 )$;
 we omit $\alpha $ for shortness.}
  for
 every $\varphi \in \varPhi _1\setminus {\cal O}.$
 It easily follows from (\ref{2.66}) -- (\ref{2.67a}) that
 \begin{equation}\frac{\partial \lambda ^{(1)}(\varkappa  \vec
 \nu )}{\partial \varkappa  }=2lk^{2l-1}(1+o(1)),\ \ \ \frac{\partial ^2\lambda ^{(1)}(\varkappa  \vec
 \nu )}{\partial \varkappa  ^2}=O(k^{2l-2})\label{**}
 \end{equation}
  for any $\varphi \in \varPhi
 _1$ and $\varkappa  :|\varkappa  -k|<k^{-1-4s_1-2\delta }$. Using these
 estimates we readily obtain that $|\lambda ^{(1)}(\varkappa  \vec
 \nu )-\lambda |>k^{2l-1}\epsilon _1^4$ for all $\varkappa $ on the circle
 $|\varkappa  -\varkappa _1
 (\varphi )|=\epsilon _1^4$. Considering (\ref{July3d}) and
 applying Rouch\'{e}'s theorem, we obtain that there is a single
 solution $\varkappa _2(\varphi )$ of the equation $\lambda
 ^{(2)}(\alpha ,\varkappa  \vec \nu )=\lambda $ such that $|\varkappa _2(\varphi )
 -\varkappa _1
 (\varphi )|\leq \epsilon _1^4$. Applying
 (\ref{July3e}) and Implicit function theorem, we obtain that
 $\varkappa _2 (\varphi )$ is a holomorphic function of $\varphi
 $ in $\varPhi _1\setminus {\cal O}$ and estimates (\ref{3.75})
  hold.

 Let us make here a remark for the future.
Convergence of the series
 for the resolvent $\left( H^{(2)}\left(\vec \varkappa _1(\varphi
 )\right)-z\right)^{-1}$, $z\in C_2$, following from (\ref{July3b}), means that the
 resolvent  has a single pole $z= \lambda ^{(2)}(\vec \varkappa _1(\varphi
 )\vec \nu
 )$ inside $C_2$. Similar result holds when we replace $\vec \varkappa _1(\varphi
 )$ by $\vec \varkappa _2(\varphi
 )$, since $\vec \varkappa _1(\varphi
 )$ and $\vec \varkappa _2(\varphi
 )$ are close: $\bigl|\vec \varkappa _2(\varphi
 )-\vec \varkappa _1(\varphi
 )|=o(\epsilon _1)$. Considering that $\lambda ^{(2)}(\vec \varkappa
 _2(\varphi )
 )=\lambda $, we obtain that $(z-\lambda )\left( H^{(2)}\left(\vec \varkappa _2(\varphi
 )\right)-z\right)^{-1}$ is holomorphic inside $C_2$ and the estimate
 similar to (\ref{16*}) holds:
 \begin{equation}
 \Bigl\|(z-\lambda )\left(H^{(2)}\bigl(\vec \varkappa _2(\varphi
 )\bigr)-z\right)^{-1}\Bigr\|<32. \label{16**}
 \end{equation}

\item First, we show that
\begin{equation}\label{3.77a+}
     L\bigl(\cal{D}_2(\lambda )\bigr)\underset{\lambda \rightarrow
     \infty}{=}L\bigl(\cal{D}_1(\lambda )_{nonres}\bigr)\Bigl(1+O(k^{-2-2s_1})\Bigr).
     \end{equation}
     Second, we check that
     \begin{equation}\label{3.77a?}
     L\bigl(\cal{D}_1(\lambda )_{nonres}\bigr)\underset{\lambda \rightarrow
     \infty}{=}L\bigl(\cal{D}_1(\lambda )\bigr)\Bigl(1+O(k^{-2-2s_1})\Bigr).
     \end{equation}
     From these two estimates (\ref{3.77}) follows. In fact,
 considering that
    $$
     L\bigl(\cal{D}_2(\lambda )\bigr)=\int_{\varTheta_2}
     \sqrt{\varkappa _2^2+\varkappa _2^{\prime}\ ^2}\,d\varphi ,\ \ \
     L\bigl(\cal{D}_1(\lambda )_{nonres}\bigr)=\int_{\varTheta_2}
     \sqrt{\varkappa _1^2+\varkappa _1^{\prime}\ ^2}\,d\varphi,
     $$
     and taking into account that $h_2(\varphi )=
     \varkappa _2(\varphi )-\varkappa _1(\varphi )$ satisfies
     (\ref{3.75}), we easily get (\ref{3.77a+}).
     Formula (\ref{3.77a?}) easily follows from (\ref{3.77a}).
     \end{enumerate}
 \end{proof}
%
Now define the non-resonance set $\chi_2^*(\lambda )$ in
$S_2(\lambda )$ by
    \begin{equation}\label{3.77.7}
    \chi_2^*(\lambda ):=\cal{K}_2\cal{D}_2(\lambda ).
    \end{equation}
    \begin{lemma} \label{Apr4a}
    The set $\chi_2^*(\lambda )$ belongs to the
    $\left(2\alpha \epsilon _1^4k^{-2l+1}\right)$-neighborhood of $\chi_2(\lambda
    )$ in $K_2$. If $\tau \in \chi_2^*(\lambda )$, then the operator
    $H^{(2)}_{\alpha }(\tau )$ has a simple eigenvalue
   equal to $\lambda $. This
    eigenvalue is given by the perturbation series (\ref{3.15}),
     where $p\in
P,j\in \Z^2$ are described as in Geometric Lemma \ref{L:3.2}, part
2.
\end{lemma}

\begin{proof} By Lemma \ref{L:3.9}, $\cal D_2(\lambda )$ is in the $\left(
2\alpha \epsilon _1^4k^{-2l+1}\right)$ neighborhood of $\cal
D_1(\lambda )_{nonres}$. Considering that $\chi_2^*(\lambda )=\cal
K_2\cal D_2(\lambda )$ and $\chi_2(\lambda )=\cal K_2\cal
D_1(\lambda )_{nonres}$ (see (\ref{pr})), we immediately obtain
that $\chi_2^*(\lambda )$ is in the
    $(2\alpha \epsilon _1^4k^{-2l+1})$-neighborhood of $\chi_2(\lambda
    )$. The size of this neighborhood is
less than $\epsilon _1k^{-2l+1-\delta }$, hence Theorem 3.1 holds
in it, i.e., for any $\tau \in \chi_2^*(\lambda )$ there is a
single eigenvalue of $H^{(2)}_{\alpha }(\tau )$ in the interval
$\varepsilon_2 (k,\delta )$. Since $\chi_2^*(\lambda )\subset
S_2(\lambda )$, this eigenvalue is equal to $\lambda $. By the
theorem, the eigenvalue is given by the series (\ref{3.15}),
     where $p\in
P,j\in \Z^2$ are described as in Geometric Lemma \ref{L:3.2}, part
2.  \end{proof}

    \begin{lemma}\label{L:May10a*} Formula (\ref{3.77.7}) establishes
    one-to-one correspondence between $\chi_2^*(\lambda )$ and $\cal{D}_2(\lambda
    )$.\end{lemma}
    \begin{remark} \label{R:May14}
    From geometric point of view this means that $\chi_2^*(\lambda
    )$ does not have self-intersections.\end{remark}
    \begin{proof} Suppose there is a pair $\vec \varkappa , \vec \varkappa
    ^*
    \in \cal{D}_2(\lambda)$ such that $\cal K_2\vec \varkappa =\cal
    K_2\vec \varkappa
    ^*
    =\tau $, $\tau \in \chi_2^*(\lambda )$. By the definition (\ref{May14m}) of
    $\cal{D}_2(\lambda
    )$, we have $\lambda^{(2)}(\alpha,\vec{\varkappa })=\lambda^{(2)}(\alpha,\vec{\varkappa
    }^*)=
    \lambda $, i.e., the eigenvalue $\lambda $ of
    $H^{(2)}_{\alpha }(\tau )$ is not simple. This contradicts to the
    previous lemma.
     \end{proof}

%
%
%
%
%
%
%
%
%
%
%

\subsection{Preparation for the Next Approximation} \label{S:3.6}
\subsubsection{Contracted set ${\cal O}_s(\vec b)$}\label{S:3.6'}
 Let us recall that ${\cal O}(\vec b)$ is the union of disks of
 the radius
 $r=k^{-4-6s_1-3\delta }$ centered at the zeros of the unperturbed determinant
 $\det\bigl(I+A_0(\vec y_0(\varphi))\bigr)$, see Definition \ref{D:gamma}. We proved that the ``perturbed" determinant
$\det\bigl(I+A_1(\vec y(\varphi)\bigr)$
has the same number of zeros inside  each $\Gamma (\vec b)$ as the
unperturbed one, when $b_0$ is big enough.
To prepare the next approximation, we contract the set $\Gamma
(\vec b)$ around  zeros of the perturbed determinant.
First, we consider the disks with the radius
$r^{\prime}=rk^{-2-4s_2-\delta }$ centered at each zero $\varphi
_{j,0}$ of the perturbed determinant $\det \bigl(I+A_1(\vec
y(\varphi )\bigr)$ in the set ${\cal O}(\vec b)$. Obviously each
of these disks is in ${\cal O}(\vec b)$, since $r'<<r$ and the
distance between zeros of the perturbed and unperturbed
determinants is smaller than $r/2$ (Lemma \ref{L:*1*}, Part 2).
Next, we take the union
 of these smaller disks and denote it by ${\cal O}_s(\vec b)$, here the index $s$ stands for
``small". Clearly, ${\cal O}_s(\vec b)\subset {\cal O}(\vec b)$.
 We denote by $\Gamma _s(\vec
b)$  a connected component of these new disks which intersects
with $\varPhi _1$ and by $\gamma _s(\vec b)$  its boundary,
$\gamma _s=\partial \Gamma _s$. Let us remind that in all our
previous considerations $\vec b$ satisfies the conditons $b_0 >
k^{-2l+9+12s_1+7 \delta }$.\footnote{Indeed, we had two
restrictions on $b_0$: $b_0 > k^{7+8s_1-2\gamma_0+6\delta}$ (Lemma
\ref{L:3.13}) and $b_0 > k^{-2l+9+12s_1+7\delta}$ (Lemmas
\ref{L:3.14}--\ref{L:3.17}). Recalling that
$\gamma_0=2l-2-4s_1-2\delta$, we get
$7+8s_1-2\gamma_0+6\delta<-2l+9+12s_1+7\delta$, we can combine two
restrictions into
    $b_0 > k^{-2l+9+12s_1+7\delta}$.
    }
\begin{lemma}\label{L:July9b}\begin{enumerate}
\item The set ${\cal O}_s(\vec b)$ contains the same number of
disks as  ${\cal O}(\vec b)$. This number  does not exceed
$J=c_0k^{2+2s_1}$. \item For any $\varphi \in \varPhi_1\setminus
{\cal O}_s(\vec b)$,
    \begin{equation}\label{3.7.2.2}
    \Bigl\|\Bigl(H^{(1)}(\vec y(\varphi))-k^{2l}\Bigr)^{-1}\Bigr\|
    <
    cb_0^{-1}k^{-2l+9+12s_1+6\delta}(2rJ/r^{\prime})^J,
    \end{equation}
    \begin{equation}\label{3.7.2.4}
    \Bigl\|\Bigl(H^{(1)}(\vec y(\varphi))-k^{2l}\Bigr)^{-1}\Bigr\|_1
    <
    cb_0^{-1}k^{-2l+11+14s_1+6\delta}(2rJ/r^{\prime})^J.
    \end{equation}
    \end{enumerate}
\end{lemma}
\begin{corollary}\label{C:july9a}
If $b_0 > k^{-2l+9+12s_1+7 \delta }$,
\begin{equation}\label{3.7.2.2*}
    \Bigl\|\Bigl(H^{(1)}(\vec y(\varphi))-k^{2l}\Bigr)^{-1}\Bigr\|
    < \frac{1}{\epsilon _1},
    \end{equation}
    \begin{equation}\label{3.7.2.4*}
    \Bigl\|\Bigl(H^{(1)}(\vec y(\varphi))-k^{2l}\Bigr)^{-1}\Bigr\|_1
    < \frac{c_0k^{2+2s_1}}{\epsilon _1}.
    \end{equation}
    \end{corollary}
    \begin{proof}[Corollary \ref{C:july9a}] We use the condition on $\eta $: $\eta
    >
2+64/(2l-11)=2+2/s_1$ and the formulae for $\epsilon
_1=e^{-\frac{1}{4}k^{\eta
    s_1}}$, $r/r'=k^{2+4s_2+\delta }$. Simple computation shows that  the
    right-hand side in (\ref{3.7.2.2}) does not exceed $1/\epsilon
    _1$. Hence, the corollary holds.
     \end{proof}
\begin{proof}[Lemma \ref{L:July9b}] By construction, the set ${\cal O}_s(\vec b)$ contains the same number of disks
as  ${\cal O}(\vec b)$. Since the number of disks in ${\cal
O}(\vec b)$ does not exceed  $J=c_0k^{2+2s_1}$ (see the remark
after Definition \ref{D:gamma}), the same is true for ${\cal
O}_s(\vec b)$.

If $\varphi \in \varPhi _1\setminus {\cal O}(\vec b)$, then the
estimates just follow from Lemma \ref{L:3.14}. Thus, we have to
consider ${\cal O}\setminus {\cal O}_s$. Note that $(H^{(1)}(\vec
y(\varphi))-k^{2l})^{-1}\det \left(\dfrac{H^{(1)}(\vec
y(\varphi))-k^{2l}} {H_0(\vec y(\varphi))+k^{2l}}\right)$ is
holomorphic in every connected component $\Gamma (\vec b)$ of
${\cal O}(\vec b)$. We denote zeros of $\det
\left(\dfrac{H^{(1)}(\vec y(\varphi))-k^{2l}} {H_0(\vec
y(\varphi))+k^{2l}}\right)$ in $\Gamma $  by $\varphi_{j,0},\
j=1,\cdots,J^*$, $J^*\leq J$. Obviously, $(H^{(1)}(\vec
y(\varphi))-k^{2l})^{-1}\prod_{j=1}^{J^*}(\varphi-\varphi_{j,0})$
is holomorphic in $\Gamma$. For any $\varphi$ on $\gamma$,
    $$\Bigl\|(H^{(1)}(\vec
    y(\varphi))-k^{2l})^{-1}\prod_{j=1}^{J^*}(\varphi-\varphi_{j,0})\Bigr\|
    \leq \bigl\|(H^{(1)}(\vec y(\varphi))-k^{2l})^{-1}\bigr\|
    \prod_{j=1}^{J^*}|\varphi-\varphi_{j,0}|.$$
Considering that $|\varphi-\varphi_{j,0}| \leq 2rJ^{*}$ for any
$\varphi \in \gamma$ and the estimate (\ref{3.2.34.1}), we get
    \begin{equation}\label{3.7.2.3}
    \|(H^{(1)}(\vec
    y(\varphi))-k^{2l})^{-1}\prod_{j=1}^{J^*}(\varphi-\varphi_{j,0})\|
    < cb_0^{-1}k^{-2l+9+12s_1+6\delta}(2rJ^*)^{J^*}.
    \end{equation}
By the maximum principle the inequality (\ref{3.7.2.3}) holds for
any $\varphi \in \Gamma $. Note that any $\varphi \in \Gamma
\setminus \Gamma _s $ satisfies the inequality
$|\varphi-\varphi_{j,0}| \geq r^{\prime}$ and, hence,
$$\Bigl\|\Bigl(H^{(1)}(\vec y(\varphi))-k^{2l}\Bigr)^{-1}\Bigr\| <
    cb_0^{-1}k^{-2l+9+12s_1+6\delta}(2rJ^*/r^{\prime})^{J^*}$$
    holds for $\varphi \in \Gamma \setminus \Gamma _s $.
Considering that $r/r'>1$ and $J^*\leq J$, we obtain
(\ref{3.7.2.2}).
 To prove
(\ref{3.7.2.4}), we use the following Hilbert relation:
    \begin{multline}\label{July4a}
    \Bigl(H^{(1)}(\vec y(\varphi))-k^{2l}\Bigr)^{-1}=\\
    \Bigl(H_0(\vec y(\varphi))+k^{2l}\Bigr)^{-1}
    +\Bigl(H_0(\vec y(\varphi))+k^{2l}\Bigr)^{-1}(-W_1+2k^{2l})
    \Bigl(H^{(1)}(\vec y(\varphi))-k^{2l}\Bigr)^{-1}.
    \end{multline}
    Summarizing the terms of a diagonal operator, we easily get
    \begin{equation}\Bigl\|\Bigl(H_0(\vec y(\varphi))+k^{2l}\Bigr)^{-1}\Bigr\|_1<c
    k^{-2l+2+2s_1}.\label{Nov16}
    \end{equation}
     Substituting (\ref{3.7.2.2}) and the last
    estimate into (\ref{July4a}), we obtain (\ref{3.7.2.4}).
 \end{proof}
\subsubsection{The set ${\cal O}_s(\vec b)$ for small $b _0$}\label{SS:3.7.1}

Everything  we  considered so far is valid for $\vec b$ obeying
the inequality $b_0 > k^{-2l+9+12s_1+7\delta}$, here $b_0$ is the
distance from $\vec b$ to the nearest vertex of $K_1$.
 We showed in Section  \ref{St3} that $\vec b=2\pi
p/N_1a$, $p\in P\setminus \{0\}$, satisfies the requirement.
However, in the next section and later, $b_0$ will be taken
smaller, since the reciprocal lattice is getting finer with each
step. To prepare for this, let us consider $\vec b$ being close to
a vertex of $K_1$:
\begin{equation}
0<b_0  \leq k^{-2l+9+12s_1+7\delta}. \label{july9d}
\end{equation}
We show that for such $\vec b$ the resolvent $\bigl(H^{(1)}(\vec
y(\varphi ))-k^{2l}\bigr)^{-1}$
has no more than two poles $\varphi ^{\pm}$ in $\varPhi _1$ and
even its small neighborhood. We surround these poles by two
contours $\gamma ^{\pm}$ and obtain estimates for
$\bigl(H^{(1)}(\vec y(\varphi ))-k^{2l}\bigr)^{-1}$ when
 $\varphi $ is outside $\gamma ^{\pm}$.

In fact, it easily follows from (\ref{july9d}) that:
$b_0=o(k^{-1-4s_1-2\delta})$ as $k \to \infty$. Let $\tilde
\varPhi _1$ be the $(\frac{1}{2}k^{-2-4s_1-2\delta})$ neighborhood
of $\varPhi _1$.
Suppose $|\vec b |=b_0$, i.e.,  the closest vertex of $K_1$ for
$\vec b$ is $(0,0)$.  The perturbation series (\ref{2.66})
converge for both $\lambda ^{(1)}\left(\vec \varkappa _1(\varphi
)\right)$ and $\lambda ^{(1)}\left(\vec y(\varphi )\right)$ when
$\varphi \in \tilde \varPhi _1$, and both  functions are
holomorphic  in $\tilde \varPhi _1$, because $\vec \varkappa
_1(\varphi )$ and $\vec y(\varphi )$ are in the complex
$(2k^{-1-4s_1-2\delta})$-neighborhood of $\cal{D}_0(\lambda
)_{nonres}$ for such $\varphi $ (Lemma \ref{L:2.13}). Note that
$\lambda ^{(1)}\left(\vec \varkappa _1(\varphi )\right)=k^{2l}$
for all $\varphi \in \tilde \varPhi _1$. We base our further
considerations on these perturbation series expansions. For $\vec
b$ being close to a vertex $\vec e$ other than $(0,0)$, we take
$\vec y(\varphi )=\vec \varkappa _1(\varphi )+\vec b-\vec e$.

We define $\varphi _b\in [0,2\pi )$ by the formula $\vec
b=b_0(\cos \varphi _b, \sin \varphi _b)$ when  $|\vec b |=b_0$,
and by the analogous formula $\vec b-\vec e=b_0(\cos \varphi _b,
\sin \varphi _b)$ when $\vec b$ is close to a vertex $\vec e$
other than $(0,0)$.


\begin{lemma}\label{L:3.7.1.1}
If $\vec b$ satisfies (\ref{july9d}) and
$|\epsilon_0|<b_0k^{2l-3-4s_1-3\delta}$, then the equation
    \begin{equation}\label{3.7.1.2}
    \lambda ^{(1)}\left(\vec y (\varphi)\right)=k^{2l}+\epsilon_0
    \end{equation}
 has no more than two
solutions, $\varphi^{\pm}_{\epsilon _0}$, in $\tilde \varPhi _1$.
They satisfy the inequality
    \begin{equation}
    \left|\varphi^{\pm}_{\epsilon _0}-(\varphi_b \pm \pi
    /2)\right|<\frac{1}{8}k^{-2-4s_1-2\delta}. \label{july5a}
    \end{equation}
\end{lemma}
\begin{proof}  Suppose $W_1=0$ and $|\vec b|=b_0$, i.e., the closest vertex of $K_1$ for $\vec
b$ is $(0,0)$. Then the equation (\ref{3.7.1.2}) has the form
$|k\vec \nu +\vec b|^{2l}=k^{2l}+\epsilon_0$. It is easy to show
that it has two solutions $\varphi^{\pm}_{\epsilon _0}$ satisfying
(\ref{july5a}). Applying perturbative arguments and Rouch\'{e}'s
    theorem, we prove the lemma for nonzero $W_1$. A detailed proof is in Appendix \ref{A:5}. In the case
    when $\vec b$ is close to a vertex other than $(0,0)$, the
    considerations are the same up to a parallel shift.
 \end{proof}
\begin{lemma}\label{L:july5a} Suppose $\vec b$ satisfies (\ref{july9d}) and $\varphi \in \tilde \varPhi
_1$ obeys the inequality analogous to (\ref{july5a}):
$\left|\varphi-(\varphi_b \pm \pi
    /2)\right|<k^{-2-4s_1-2\delta}$. Then,
    \begin{equation}\label{3.7.1.6.1/2}
    \frac{\partial}{\partial \varphi}\lambda^{(1)}\left(\vec
    y(\varphi)\right)=_{k\to \infty}\pm 2lb_0k^{2l-1}\bigl(1+o(1)\bigr).
    \end{equation}
\end{lemma}
\begin{proof}
Let $W_1=0$ and $|\vec b|=b_0$. Then $\lambda^{(1)}(\vec
    y(\varphi))=|k\vec \nu +\vec b|^{2l}$ and
    $$\frac{\partial}{\partial \varphi}|k\vec \nu +\vec b|^{2l}=
    2l|k\vec \nu +\vec b|^{2l-2} \langle k\vec \nu +\vec b,k\vec \mu \rangle, \ \ \vec \mu
    =(-\sin \varphi, \cos \varphi ).$$
    For $\varphi $ close to $\varphi_b \pm \pi
    /2$, we have $\langle \vec b,\vec \mu
\rangle =\pm b_0(1+o(1))$. Considering also that $\langle \vec
\mu, \vec \nu \rangle =0$, we obtain $\frac{\partial}{\partial
\varphi}|k\vec \nu +b|^{2l}=\pm 2lb_0k^{2l-1}(1+o(1))$. Applying
perturbative arguments, we get a similar formula for nonzero
$W_1$. For a detailed proof see Appendix \ref{A:6}. In the case
    when $\vec b$ is close to a vertex other than $(0,0)$, the
    considerations are the same up to a parallel shift.
 \end{proof}

\begin{definition} \label{SN}
Let $\Gamma^{\pm}_s(\vec b)$ be  open disks centered at
$\varphi_0^{\pm}\in \tilde \varPhi _1$
\footnote{$\varphi_0^{\pm}=\varphi_{\epsilon _0}^{\pm}$, when
$\epsilon _0 =0$.} with the radius $r'$ introduced in the previous
subsection,
 $r'=rk^{-2-4s_2-\delta }$; $\gamma^{\pm}_s(\vec
b)$ be their boundary circles and ${\cal O}_s(\vec
b)=\Gamma^{+}_s\cup \Gamma^{-}_s$.
\end{definition}
%
%
%
%
%
\begin{lemma}\label{L:3.7.1}
For any $\varphi$ in $\varPhi _1\setminus {\cal O}_s(\vec b)$,
    \begin{equation}\label{3.7.1.10}
    |\lambda^{(1)}(\vec
    y(\varphi))-k^{2l}|\geq b_0k^{2l-1-\delta}r'.
    \end{equation}
\end{lemma}
\begin{proof} Suppose (\ref{3.7.1.10}) does not hold for some $\varphi$ in $\varPhi _1\setminus {\cal O}_s(\vec b)$. This means that
$\varphi$ satisfies  equation (\ref{3.7.1.2}) with some
$\varepsilon _0$: $|\varepsilon _0|
<b_0k^{2l-1}r'<b_0k^{2l-3-4s_1-3\delta}$. By Lemma
\ref{L:3.7.1.1}, $\varphi $ obeys  (\ref{july5a}). Thus $\varphi$
could be either $\varphi_{\epsilon _0}^{+}$ or $\varphi_{\epsilon
_0}^{-}$. Without loss of generocity, assume $\varphi
=\varphi_{\epsilon _0}^{+}$. Applying (\ref{3.7.1.6.1/2}) in the
segment between $\varphi$ and $\varphi_{0}^{+}$, we obtain:
$\varepsilon _0=\lambda^{(1)}(\vec
    y(\varphi))-k^{2l} = \pm 2lb_0k^{2l-1}\bigl(1+o(1)\bigr)\left(\varphi
    -\varphi_0^{+}\right)$. It immediately follows that $\left|\varphi
    -\varphi_0^{+}\right|<r'$, i.e., $\varphi \in \Gamma_s^+ \subset O_s(\vec b)$, which contradicts the assumption $\varphi \in \varPhi _1\setminus {\cal O}_s(\vec
    b)$.
    \end{proof}

\begin{lemma}\label{L:july9c}
For any $\varphi \in \varPhi _1\setminus {\cal O}_s(\vec b)$,
    \begin{equation}\label{3.7.1.10.1/2}
    \Bigl\|\Bigl(H^{(1)}(\vec
    y(\varphi))-k^{2l}\Bigr)^{-1}\Bigr\|<\frac{16}{b_0r'k^{2l-1-\delta}},
    \end{equation}
    \begin{equation}\label{3.7.1.13}
    \Bigl\|\Bigl(H^{(1)}(\vec y(\varphi))-k^{2l}\Bigr)^{-1}\Bigr\|_1<
    \frac{16c_0k^{2+2s_1}}{b_0r'k^{2l-1-\delta}}
    .
    \end{equation}
\end{lemma}
\begin{corollary}\label{C:new} If $\epsilon _1k^{-2l+1-2\delta
}<b_0<k^{-2l+9+12s_1+7\delta}$ and $\varphi \in \varPhi
_1\setminus {\cal O}_s(\vec b)$, then
    \begin{equation}\label{3.7.1.10.1/2++}
    \Bigl\|\Bigl(H^{(1)}(\vec
    y(\varphi))-k^{2l}\Bigr)^{-1}\Bigr\|<\frac{1}{\epsilon _1^2},
    \end{equation}
    \begin{equation}\label{3.7.1.13++}
    \Bigl\|\Bigl(H^{(1)}(\vec y(\varphi))-k^{2l}\Bigr)^{-1}\Bigr\|_1<
    \frac{c_0k^{2+2s_1}}{\epsilon _1^2}
    .
    \end{equation}
\end{corollary} Corollary follows from the condition on $b_0$ and the
estimate $r'k^{-3\delta}>16\epsilon _1$, which is obviously valid
for sufficiently large $k$.

\begin{proof}
Since $\vec y(\varphi)$ is in the
$(2k^{-1-4s_1-2\delta})-$neighborhood of $\cal D_0(\lambda
)_{nonres}$, we can apply Lemma  \ref{L:July5} from Appendix
\ref{A:4}. By this lemma,
$$\left\|\Bigl(\lambda^{(1)}(\vec y(\varphi))-k^{2l}\Bigr)
    \Bigl(H^{(1)}(\vec y(\varphi))-k^{2l}\Bigr)^{-1}\right\| \leq 16.$$
Using  (\ref{3.7.1.10}),  we easily get (\ref{3.7.1.10.1/2}).
Using (\ref{3.7.1.10.1/2}) and considering  (\ref{July4a}),
(\ref{Nov16}), we obtain (\ref{3.7.1.13}).  \end{proof}

\section{The Third Approximation} \label{chapt5}

\setcounter{equation}{0}

\subsection{The Operator $H_{\alpha}^{(3)}$}

Choosing $s_3=2s_2$, we define the third operator
$H_{\alpha}^{(3)}$ by the formula:
 \begin{equation}\label{4.1}
     H_{\alpha }^{(3)}=H^{(2)}+\alpha W_3,\quad      (0\leq \alpha \leq
     1),\qquad
     W_3=\sum_{r=M_2+1}^{M_3}V_r, \notag
     \end{equation}
 where $M_3$ is chosen
in such a way that $2^{M_3}\approx k^{s_3}$. Obviously, the
periods of $W_3$ are $2^{M_3-1} (b_1,0)$ and $2^{M_3-1} (0,b_2)$.
We  write them in the form: $N_2N_1(a_1,0)$ and $N_2N_1(0,a_2)$,
here $N_2=2^{M_3-M_2}$, $\frac{1}{4}k^{s_3-s_2}<N_2<
4k^{s_3-s_2}$. Note that
    \begin{align*}\|W_3\|_{\infty} \leq \sum_{r=M_2+1}^{M_3}
    \|V_r\|_{\infty} \leq \sum_{r=M_2+1}^{M_3}\exp(-2^{\eta r})
    <\exp(-k^{\eta s_2}).\end{align*}
\subsection{Multiple Periods of $W_2(x)$}
The operator,
%
    $H^{(2)}=H_1 + W_2(x)$,
%
has the periods $N_1a_1, N_1a_2$. The corresponding family of
operators, $\{H^{(2)}(\tau)\}_{\tau \in K_2}$, acts in $L_2(Q_2)$,
where $Q_2=[0,N_1a_1] \times [0,N_1a_2]$ and
 $K_2=[0, 2\pi/N_1a_1)\times [0, 2\pi/N_1a_2)$.  Since now on we
denote  quasimomentum $t$ from the first step by $t^{(1)}$,
quasimomentum $\tau $ from the second  step by $t^{(2)}$.
Correspondingly, quasimomentum for $H_{\alpha }^{(3)}$ we denote
by $t^{(3)}$. Eigenvalues of $H^{(2)}\left(t^{(2)}\right)$ are
denoted by $\lambda_n^{(2)}\left(t^{(2)}\right)$, $n \in \N$ and
its spectrum by $\Lambda ^{(2)}\left(t^{(2)}\right)$.

Next, let us consider  $W_2(x)$ as a periodic function with the
periods $N_2N_1a_1$, $N_2N_1a_2$. When changing the periods, the
family of operators
$\bigl\{H^{(2)}\left(t^{(2)}\right)\bigr\}_{t^{(2)} \in K_2}$ is
replaced by the family of operators $\bigl\{\tilde
H^{(2)}\left(t^{(3)}\right)\bigr\}_{ t^{(3)}\in K_3}$, acting in
$L_2(Q_3)$, where $Q_3=[0,N_2N_1a_1] \times [0,N_2N_1a_2]$ and
$K_3=[0, 2\pi/N_2N_1a_1)\times [0,2\pi/N_2N_1a_2)$. We denote
eigenvalues of $\tilde H^{(2)}\left(t^{(3)}\right)$ by
$\tilde{\lambda}_n^{(2)}\left(t^{(3)}\right)$, $n \in \N$, and its
spectrum by $\tilde{\Lambda}^{(2)}\left(t^{(3)}\right)$.  We
denote now by $P^{(1)}$ the set $P$, introduced by (\ref{May10a}),
its elements being  $p^{(1)}$. By Bloch theory (see
e.g.\cite{RS}), for any $t^{(3)} \in K_3$,
    \begin{equation}\label{4.3}
    \tilde{\Lambda}^{(2)}\left(t^{(3)}\right)=\bigcup_{p^{(2)}\in P^{(2)}}
    \Lambda^{(2)}\left(t^{(3)}+2\pi p^{(2)}/N_2N_1a\right),
    \end{equation}
where $P^{(2)}=\left\{p^{(2)}=\bigl(p^{(2)}_1,p^{(2)}_2\bigr) \in
\Z^2 : 0 \leq p^{(2)}_1 \leq N_2-1,\ 0 \leq p^{(2)}_2 \leq
N_2-1\right\}$, $2\pi p^{(2)}/N_2N_1a=\left(2\pi p_1^{(2)}/a_1,
2\pi p_2^{(2)}/a_2\right)/N_2N_1$. An isoenergetic set
$\tilde{S}_2(\lambda _0)\subset K_3$ of the operator $\tilde
H^{(2)}$ is defined by the formula:
    \begin{align*}
    \tilde{S}_2(\lambda )&=\left\{t^{(3)} \in K_3: \exists n \in \N: \
    \tilde{\lambda}_n^{(2)}\bigl(t^{(3)}\bigr)=\lambda \right\}\\
    &=\left\{t^{(3)} \in K_3: \exists n \in \N,\ p^{(2)} \in P^{(2)}:
    \lambda_n^{(2)}\left(t^{(3)}+2\pi p^{(2)}/N_2N_1a\right)=\lambda \right\}.
    \end{align*}
Obviously,
    $\tilde{S}_2=\cal{K}_3 S_2, $ where $\cal{K}_3$ is the
    parallel shift into $K_3$, that is,
    $\cal{K}_3:\R^2 \rightarrow K_3$, $\cal{K}_3\left(t^{(3)}+2\pi
    m/N_2N_1a\right)=t^{(3)}$, $ m \in \Z^2$, $t^{(3)} \in K_3.$
    We denote  index $j$, introduced in Part 1 of
    Geometric
    Lemma \ref{L:2.1}, by $j^{(1)}$ and $\tilde j $, introduced in
    (\ref{3.15}), by $j^{(2)}$, $j^{(2)}=j^{(1)}+p^{(1)}/N_1.$


\subsection{Perturbation Formulae}

 The results of this section are analogous to those in the previous step, the index 2 just being replaced by 3.  Let us start with establishing a
 lower bound for $k$.
 Since $\eta s_1>2+2s_1$, there is a number $k_*>e$ such that
\begin{equation}
 \hat C(1+s_1)k^{2+2s_1}\ln k <k^{\eta s_1}, \hat C=400l(c_0+1)^2, \label{k}
 \end{equation}
  for any $k>k_*$. Assume also, that $k_*$ is sufficiently
 large to ensure validity of all estimates in the first two steps
 for any $k>k_*$.
\begin{lemma}[Geometric Lemma]\label{L:4.2}
For 
a sufficiently large $\lambda $, $\lambda
>k_*^{2l}$, there exists a non-resonance
set $\chi _3(\lambda, \delta) \subset \cal{K}_3\chi_2^*$
 such that:
\begin{enumerate}
\item For any point $t^{(3)}\in \chi _3$, the following conditions
hold:
\begin{enumerate}
    \item There exists a unique $p^{(2)} \in P^{(2)}$
    such that $t^{(3)} + 2\pi p^{(2)}/N_2N_1a \in \chi_2^*$.
    \item The following relation holds:
    $$ \lambda _{{j}^{(2)}}^{(2)}\left(t^{(3)} + 2\pi
    p^{(2)}/N_2N_1a\right)=k^{2l},$$
      where $\lambda _{{j}^{(2)}}^{(2)}\left(t^{(3)} + 2\pi p^{(2)}/N_2N_1a\right)$
      is
      given  by the perturbation series (\ref{3.15}) with $\alpha =1$ and
      ${j}^{(2)}={j}+p/N_1$, here ${j}$ and
      $p$
       are
       defined by the point $\tau = t^{(3)} + 2\pi p^{(2)}/N_2N_1a \ $ as it is described in
        Part 2
      of the Geometric Lemma for the previous step.
    \item The eigenvalue $\lambda _{{j}^{(2)}}^{(2)}\bigl(t^{(3)} + 2\pi
    p^{(2)}/N_2N_1a\bigr)$ is
    a simple eigenvalue of $\tilde H^{(2)}\left(t^{(3)}\right)$ and its distance to all other eigenvalues $\lambda _n^{(2)}\left(t^{(3)} + 2\pi
    \hat{p}^{(2)}/N_2N_1a\right)$ of $\tilde H^{(2)}\bigl(t^{(3)}\bigr)$ is greater than
    $\epsilon_2 =e^{-\frac{1}{4}k^{\eta s_2}}$:
    \begin{equation}\label{4.6}
    \Bigl|\lambda _{j^{(2)}}^{(2)}\left(t^{(3)} + 2\pi p^{(2)}/N_2N_1a\right)-
    \lambda _n^{(2)}\left(t^{(3)} + 2\pi
    \hat{p}^{(2)}/N_2N_1a\right)\Bigl|>\epsilon_2.
    \end{equation}
\end{enumerate}
\item For any $t^{(3)}$ in the $(\epsilon_2 k^{-2l+1-\delta
})$-neighborhood in $\C ^2$ of  $\chi_3$, there exists a unique
$p^{(2)} \in P^{(2)}$ such that $t^{(3)} +2\pi p^{(2)}/N_2N_1a$ is
in the  $(\epsilon_2 k^{-2l+1-\delta })$-neighborhood in $\C ^2$
of $\chi_2^*$ and
    \begin{equation}\label{4.7}
    \Bigl| \lambda_{j^{(2)}}^{(2)}\left(t^{(3)} +2\pi p^{(2)}/N_2N_1a\right)-k^{2l}\Bigr| < \epsilon_2 k^{-\delta
    },
    \end{equation}${j}^{(2)}={j}+p/N_1$, here ${j}$ and
      $p$
       are
       defined by the point $\tau = t^{(3)} + 2\pi p^{(2)}/N_2N_1a \ $
        as it is described in
        Part 2
      of the Geometric Lemma for the previous step.
\item The third nonresonance set $\chi_3$ has an asymptotically
full measure on $\chi_2^*$ in the following sense:
\begin{equation}\label{4.9}
\frac{L\left(\cal{K}_3\chi_2^* \setminus \chi
_3)\right)}{L\left(\chi_2^*\right)}< k^{-4-2s_1-2s_2}.
\end{equation}
\end{enumerate}
%
\end{lemma}
 The proof of the lemma is analogous to that for Geometric Lemma in the second step.
 It will be presented in Section \ref{S:4.2}.

\begin{corollary}\label{C:4.3}
If $t^{(3)}$ belongs to the complex $(\epsilon_2
k^{-2l+1-\delta})-$neighborhood of $\chi_3(\lambda,\delta)$, then
for any $z$ lying on the circle $C_3=\{ z: |z-k^{2l}|
=\epsilon_2/2 \} $, the following inequalities hold:
\begin{align}
\Bigl\|\bigl(\tilde H^{(2)}(t^{(3)}\bigr)-z)^{-1}\Bigr\| & <
\frac{4}{\epsilon_2},\\ \|(\tilde H^{(2)}(t^{(3)})-z)^{-1}\|_1 & <
\frac{4c_0k^{2+2s_3}}{\epsilon_2}. \label{4.11}
\end{align}
\end{corollary}


%
%
\begin{remark} Every point $2\pi q/N_2N_1a $ ($q\in
\Z ^2$) of the dual lattice  for periods $N_2N_1a_1$, $N_2N_1a_2$
can be uniquely represented in the form: $2\pi q/N_2N_1a =2\pi
m/N_1a+2\pi p^{(2)}/N_2N_1a$, where $m\in \Z^2$, $p^{(2)}\in
P^{(2)}$. Note that $2\pi m/N_1a$ is a point of a dual lattice for
periods $N_1a_1$, $N_1a_2$ and $p^{(2)}\in P^{(2)}$ is responsible
for refining the lattice. By Remark \ref{R:11}, $2\pi q/N_2N_1a $
also can be uniquely represented as $2\pi q/N_2N_1a =2\pi j/a+2\pi
p^{(1)}/N_1a+2\pi p^{(2)}/N_2N_1a$, here $j\in \Z^2$, $p^{(1)}\in
P^{(1)}$, $p^{(2)}\in P^{(2)}$.

 Let us
consider a normalized eigenfunction $\psi_n\left(t^{(2)},x\right)$
of $H^{(2)}\left(t^{(2)}\right)$ in $L_2(Q_2)$. We extend it
quasiperiodically to $L_2(Q_3)$, renormalize and denote the new
function by $\tilde{\psi}_n\left(t^{(3)},x\right)$, $t^{(3)}={\cal
K}_3t^{(2)}$. The Fourier representations of
$\psi_n\left(t^{(2)},x\right)$ in $L_2(Q_2)$ and
$\tilde{\psi}_n(t^{(3)},x)$ in $L_2(Q_3)$ are simply related. If
we denote Fourier coefficients of $\psi_n\left(t^{(2)},x\right)$
with respect to the basis
    $|Q_2|^{-1/2}e^{i\left(2\pi m/N_1a+t^{(2)},x\right)}$, $m \in \Z^2,$
in $L_2(Q_2)$  by $C_{nm}$, then, obviously, the Fourier
coefficients $\tilde{C}_{nq}$ of
$\tilde{\psi}_n\left(t^{(3)},x\right)$ with respect to the basis
    $|Q_3|^{-1/2}e^{i\left(2\pi q/N_2N_1a+t^{(3)},x\right)}$, $q \in
    \Z^2$,
in $L_2(Q_3)$ are given by the formula
    \begin{equation*}
    \tilde{C}_{nq}=
        \begin{cases}
        C_{nm},  &\text{if $q=mN_2+p^{(2)}$;}\\
        0,       &\text{otherwise},
        \end{cases}
    \end{equation*}
    $p^{(2)}$ being defined from the relation
    $t^{(2)}=t^{(3)}+2\pi p^{(2)}/N_1N_2a$.
Correspondingly, matrices of the projections on $\psi_n(\tau,x)$
and $\tilde{\psi}_n(t^{(3)},x)$ with respect to the above bases
are simply related:
    \begin{equation*}
    (\tilde{E}_n)_{q\hat{q}}=
        \begin{cases}
        (E_n)_{m\hat{m}},  &\text{if $q=mN_2+p^{(2)},\ \hat{q}=\hat{m}N_2+p^{(2)}$;}\\
        0,       &\text{otherwise},
        \end{cases}
    \end{equation*}
$\tilde{E}_n$ and $E_n$ being projections in $L_2(Q_3)$ and
$L_2(Q_2)$, respectively.
\end{remark}

\medskip

 We define functions $g_r^{(3)}(k,t^{(3)})$ and
operator-valued functions $G_r^{(3)}(k,t^{(3)})$, $r=1, 2, \cdots
,$ as follows:
\begin{equation*}
 g_r^{(3)}\bigl(k,t^{(3)}\bigr)=\frac{(-1)^r}{2\pi ir}\mbox{Tr}\oint _{C_3}
 \left(\Bigl(\tilde H^{(2)}\bigl(t^{(3)}\bigr)-z\Bigr)^{-1}W_3\right)^rdz,
 \end{equation*}
\begin{equation*}
G_r^{(3)}\bigl(k,t^{(3)}\bigr)=\frac{(-1)^{r+1}}{2\pi i}\oint
_{C_3}\left(\Bigl(\tilde
H^{(2)}\bigl(t^{(3)}\bigr)-z\Bigr)^{-1}W_3\right)^r \Bigl(\tilde
H^{(2)}\bigl(t^{(3)}\bigr)-z\Bigr)^{-1}dz.
 \end{equation*}
%
%
%
%
\begin{theorem}\label{T:4.4}
 Suppose $t^{(3)}$ belongs to the
$(\epsilon_2 k^{-2l+1-\delta })$-neighborhood in $K_3$ of the
third nonresonance set $\chi _3(\lambda ,\delta )$, $0<7\delta
<2l-11-16s_1$. Then, for sufficiently large $\lambda $, $\lambda
>k_*^{2l}$ and for all $\alpha $, $0
\leq \alpha \leq 1$, there exists a unique eigenvalue of the
operator $H_{\alpha}^{(3)}\left(t^{(3)}\right)$
 in
the interval $\varepsilon_3 (k):= (k^{2l}-\epsilon_2 /2,
k^{2l}+\epsilon_2 /2)$, $\epsilon_2=e^{-\frac{1}{4}k^{\eta s_2}}$.
It is given by the series:
\begin{equation}\label{4.15}
\lambda_{{j^{(3)}}}^{(3)} \bigl(\alpha
,t^{(3)}\bigr)=\lambda_{j^{(2)}}^{(2)}\left( t^{(3)}+2\pi
p^{(2)}/N_2N_1a \right)+\sum _{r=1}^{\infty }\alpha ^r
g_r^{(3)}\bigl(k,t^{(3)}\bigr),
\end{equation}
converging absolutely in the disk $|\alpha|  \leq 1$, where
${j^{(3)}}:=j^{(2)}+p^{(2)}/N_2N_1$, $p^{(2)}$, $j^{(2)}$ being
described in Geometric Lemma  \ref{L:4.2}. The spectral
projection, corresponding to $\lambda_{{j^{(3)}}}^{(3)} (\alpha
,t^{(3)})$, is given by the series:
\begin{equation}\label{4.16}
E_{{j^{(3)}}}^{(3)} \bigl(\alpha
,t^{(3)}\bigr)=\tilde{E}_{j^{(2)}}^{(2)}\left(t^{(3)}+2\pi
p^{(2)}/N_2N_1a \right)+\sum _{r=1}^{\infty }\alpha
^rG_r^{(3)}\bigl(k,t^{(3)}\bigr),
\end{equation}
which converges in the trace class $\mathbf{S_1}$ uniformly with
respect to $\alpha $ in the disk  $| \alpha | \leq 1$.

The following estimates hold for coefficients
$g_r^{(3)}(k,t^{(3)})$,  $G_r^{(3)}(k,t^{(3)})$:
%
\begin{equation}\label{4.17}
\Bigl\| g_r^{(3)}\bigl(k,t^{(3)}\bigr)
\Bigr|<\frac{3\epsilon_2}{2}\left(4\epsilon_2^3\right)^r, \ \ \ \
\
 \Bigl\| G_r^{(3)}\bigl(k,t^{(3)}\bigr)\Bigr\| _1< 6r \left(4 \epsilon_2^3\right)^r.
\end{equation}
\end{theorem}
\begin{corollary}\label{C:4.5}
 The
following estimates hold for the perturbed eigenvalue and its
spectral projection:
\begin{equation}\label{4.19}
\Bigl| \lambda_{{j^{(3)}}}^{(3)} \bigl(\alpha
,t^{(3)}\bigr)-\lambda_{j^{(2)}}^{(2)} \bigl(t^{(3)}+2\pi
p^{(2)}/N_2N_1a\bigr)\Bigr| \leq 12 \alpha \epsilon_2 ^4,
\end{equation}
\begin{equation}\label{4.20}
\Bigl\|E_{{j^{(3)}}}^{(3)} \bigl(\alpha
,t^{(3)}\bigr)-\tilde{E}_{j^{(2)}}^{(2)}\bigl(t^{(3)}+2\pi
p^{(2)}/N_2N_1a\bigr) \Bigr\|_1\leq 48 \alpha \epsilon_2^3 .
\end{equation}
\end{corollary}
The series~(\ref{4.15}),~(\ref{4.16}) can be extended as
holomorphic functions of $t^{(3)}$ in the complex
$\left(\dfrac{1}{2}\epsilon _2k^{-2l+1-\delta
}\right)$-neighborhood of $\chi _3$; they can be differentiated
any number of times with respect to $t^{(3)}$ and retain their
asymptotic character. The results analogous to Lemma \ref{T:3.6}
-- Lemma \ref{L:3.5.1/2} hold.

\subsection{Proof of the Geometric Lemma}\label{S:4.2}

We define the third nonresonance set, $\chi_3(\lambda )\subset
\cal{K}_3\chi_2^*(\lambda )$ as follows: $\chi_3(\lambda
)=\cal{K}_3\chi_2^*(\lambda ) \setminus
    {\Omega}_2(\lambda )$,
    \begin{multline}\label{4.2.1}
{\Omega}_2(\lambda )=\Bigl\{t^{(3)} \in \cal{K}_3\chi_2^*: \exists
n,\hat{n} \in
    \N,\ p^{(2)},\hat{p}^{(2)} \in P^{(2)},\ \  p^{(2)}\neq \hat
    p^{(2)}: \\ \lambda_{n}^{(2)}\left(t^{(3)}+2\pi p^{(2)}/N_2N_1a\right)=\lambda ,\ \ \  t^{(3)}+2\pi p^{(2)}/N_2N_1a \in
\chi_2^*(\lambda
    ),\\
     \Bigl|\lambda_{n}^{(2)}\left(t^{(3)}+2\pi p^{(2)}/N_2N_1a\right)-
     \lambda_{\hat{n}}^{(2)}\left(t^{(3)}+2\pi
    \hat{p}^{(2)}/N_2N_1a\right)\Bigr|\leq \epsilon_2  \Bigr\}.
    \end{multline}
    Proofs of Statements 1 and 2 of Geometric Lemma are identical to
those for the second step up to the shift of indices by 1.

From now on we denote a vector $\vec b\in K_1$, introduced in the
second step, by $\vec b^{(1)}$. Naturally, we denote the set
${\cal O}(\vec b)$ constructed in the previous step (Section
\ref{CRS}) by ${\cal O}^{(1)}(\vec b^{(1)})$. Correspondingly,
$\Gamma (\vec b)$ is $\Gamma ^{(1)}(\vec b^{(1)})$, ${\cal
O}_*={\cal O}_*^{(1)}$ (see (\ref{ut1})), and ${\cal O}_s(\vec b)$
is ${\cal O}_s^{(1)}(\vec b^{(1)})$. Note that ${\cal
O}_s^{(1)}(\vec b^{(1)})$  is defined for an arbitrarily small
$\vec b^{(1)}$, see Sections \ref{S:3.6'} and
    \ref{SS:3.7.1}. Let us recall that the complex non-resonant set
    $\varPhi _2$ is given by (\ref{ut1}).
\begin{definition}\label{D:Aug15}
   By analogy with the definition of $\omega _1^*(\lambda )$
(Section \ref{CRS}),
     we consider a complex resonance set $\omega
_2^*(\lambda )$, which is the set of $\varphi \in \varPhi _2$
satisfying
\begin{equation}
\det \left(\dfrac{H^{(2)}\bigl(\vec
y^{(2)}(\varphi)\bigr)-k^{2l}-\epsilon }{H_0\bigl(\vec
y^{(2)}(\varphi)\bigr)+k^{2l}}\right)=0,\ \ \ \vec
y^{(2)}(\varphi)=\vec \varkappa _2(\varphi )+\vec b^{(2)},\ \ \vec
b^{(2)}=\frac{2\pi p^{(2)}}{N_2N_1a}, \label{May18c+1}
\end{equation}
for some $p^{(2)}\in P^{(2)}\setminus \{0\}$ and  $|\epsilon
|<\epsilon _2$.  Let ${\omega}_2(\lambda )$ be the set of $\varphi
\in
    \Theta _2$ corresponding to ${\Omega}_2(\lambda )$.
    Considering as in the proof of Lemma \ref{L:Sept25}, it is easy to show that $\omega _2=\omega _2^*\cap \Theta _2$.
    \end{definition}

 By analogy with (\ref{3.2.3.1}), let us consider an arbitrary
 $\vec b^{(2)}\in K_2$ and its distance $b_0^{(2)}$ to the closest vertex of $K_2$:
   \begin{equation*}
b_0^{(2)}=\min _{m=(0,0), (0,1), (1,0), (1,1)}\left|\vec
b^{(2)}-2\pi m/N_1a\right|. \label{july9e}
\end{equation*}
     Now we construct ${\cal O}^{(2)}\left(\vec b^{(2)}\right)\subset \varPhi
     _2$, assuming $b_0^{(2)}>0$. Indeed,
    for each $\vec b^{(2)}+2\pi
p^{(1)}/N_1a$, $p^{(1)}\in P^{(1)}$,  we take the set ${\cal
    O}_s^{(1)}\bigl(\vec b^{(2)}+2\pi
p^{(1)}/N_1a \bigr)$ described in Section
    \ref{S:3.6}
    and consider a
    union:
    \begin{equation}
    {\cal O}^{(2)}\left(\vec b^{(2)}\right)=\cup _{p^{(1)}\in P^{(1)}}{\cal
    O}^{(1)}_s\left(\vec b^{(2)}+2\pi
p^{(1)}/N_1a \right).\label{july8a}
    \end{equation}
    A connected component of ${\cal O}^{(2)}\left(\vec
    b^{(2)}\right)$ we denote by $\Gamma ^{(2)}\left(\vec
    b^{(2)}\right)$ and its boundary by $\gamma ^{(2)}\left(\vec
    b ^{(2)}\right)$.
    \begin{remark} \label{R:july}
    The set ${\cal O}^{(2)}\left(\vec b^{(2)}\right)$
    consists of  disks of the radius $r'$ around all zeros of
    the determinant $\det \Bigl(I+\tilde A_1\bigl(\vec
    y^{(1)}(\varphi )\bigr)\Bigr)$, here and below:
   \begin{equation}
    \vec y^{(1)}(\varphi )=\vec
    \varkappa _1(\varphi )+\vec
    b^{(2)},\label{Nov26}
    \end{equation}
     $$I+\tilde
A_1\bigl(\vec y^{(1)} \bigr)=\Bigl(\tilde H^{(1)}\bigl(\vec
y^{(1)}\bigr)-\lambda\Bigr) \Bigl(\tilde H_0\bigl(\vec
y^{(1)}\bigr)+\lambda\Bigr)^{-1}.$$
\end{remark} This remark
easily follows from the fact that each ${\cal
    O}^{(1)}_s\bigl(\vec b^{(2)}+2\pi
p^{(1)}/N_1a \bigr)$ is built around  zeros of $\det\Bigl(I+
A_1\left(\vec \varkappa _1(\varphi )+\vec b^{(2)}+2\pi
p^{(1)}/N_1a
 \right)\Bigr)$, and $\tilde A_1\bigl(\vec
    y^{(1)}(\varphi )\bigr)$ is a direct sum of the operators
    $A_1\left(\vec \varkappa ^{(1)}+\vec b^{(2)}+2\pi
    p^{(1)}/N_1a\right)$, $p^{(1)}\in P^{(1)}$.

    \begin{remark} \label{R:Nov} Let us estimate the number of disks in ${\cal O}^{(2)}\left(\vec b^{(2)}\right)$ and the size of a connected component
    $\Gamma ^{(2)}\left(\vec
    b ^{(2)}\right)$. Indeed, the set ${\cal O}^{(2)}\left(\vec b^{(2)}\right)$ contains no more than
    $4c_0k^{2+2s_2}$ disks, since each ${\cal O}_s^{(1)}$ consists of no more than
    $c_0k^{2+2s_1}$ disks (Lemma \ref{L:July9b}, Definition \ref{SN}), and $P^{(1)}$ contains no more than
    $4k^{2(s_2-s_1)}$ elements. The set  ${\cal O}^{(2)}\left(\vec b^{(2)}\right)$ is formed by disks
of the radius $r^{\prime}$, here $r'=rk^{-2-4s_2-\delta }$, $r$
being the radius of disks constituting ${\cal O}_*^{(1)}$,
$r=k^{-4-6s_1-3\delta }$. Hence,  the size of each connected
component of ${\cal O}^{(2)}\left(\vec
    b^{(2)}\right)$ does not exceed $8c_0r'k^{2+2s_2}=8c_0rk^{-\delta -2s_2}$. Thus, the size of $\Gamma ^{(2)}\left(\vec
    b ^{(2)}\right)$ is much smaller
    than the radius of disks constituting ${\cal O}_*^{(1)}$. Here, even if a component $\Gamma ^{(2)}\left(\vec
    b ^{(2)}\right)$ is not strictly inside $\varPhi _2$, it has the same properties as a component inside
 $\varPhi _2$. Further, we consider $\Gamma ^{(2)}\left(\vec
    b ^{(2)}\right)\subset \varPhi _2$, ${\cal O}^{(2)}\left(\vec b^{(2)}\right)\subset \varPhi _2$.
    \end{remark}
     By analogy with (\ref{ut1}), we introduce new notations:
    \begin{equation}
    {\cal O}^{(2)}_*=\cup _{p^{(2)}\in P^{(2)}\setminus \{0\}}{\cal
    O}^{(2)}\bigl(2\pi p^{(2)}/N_2N_1a\bigr), \ \ \ \ \varPhi
    _3=\varPhi _2\setminus {\cal O}^{(2)}_*.
     \label{ut1*}
    \end{equation}
We will show (Corollary \ref{C:ut*}) that  $\omega _2^*\subset
{\cal
    O}^{(2)}_*$ and $\omega _2 \subset {\cal
    O}^{(2)}_*\cap \Theta _2.$ From now on we call ${\cal
    O}^{(2)}_*$ the second (complex) resonance set in $\varPhi
    _2$ and ${\cal
    O}^{(2)}_*\cap \Theta _2$ the second resonance set in
    $\Theta _2$. From (\ref{ut1*}) and (\ref{july8a}) we see:
    \begin{equation}
    {\cal O}^{(2)}_*=\cup _{p^{(1)}\in P^{(1)}, p^{(2)}\in P^{(2)}\setminus \{0\}}{\cal
    O}^{(1)}_s\Bigl(2\pi p^{(1)}/N_1a+2\pi p^{(2)}/N_2N_1a\Bigr),
     \label{ut1*a}
    \end{equation}
    each ${\cal
    O}^{(1)}_s\Bigl(2\pi p^{(1)}/N_1a+2\pi p^{(2)}/N_2N_1a\Bigr)$
    being a union of disks with centers at zeros of the
    determinants
    $\det \Bigl(I+A_1\Bigl(\vec \varkappa ^{(1)}+2\pi p^{(1)}/N_1a+2\pi
    p^{(2)}/N_2N_1a\Bigr)\Bigr)$, the size of disks being $r'$.
    \begin{remark} \label{R:X}
    The set ${\cal
O}^{(2)}_*$ consists of  no more than $16c_0k^{2+2s_3}$ disks of
the radius $r'$, since
    each ${\cal
    O}^{(1)}_s$ is formed by no more than $c_0k^{2+2s_1}$ disks and
    $P^{(1)}$,  $P^{(2)}$ contains no more than $4k^{2(s_2-s_1)}$ and
$4k^{2(s_3-s_2)}$ elements, respectively. \end{remark} Obviously,
$O_*^{(2)}$ contains more disks than ${\cal
    O}^{(1)}_*$, however, disks in $O_*^{(2)}$ are of a smaller size than
    ones
    in ${\cal
    O}^{(1)}_*$. Disks in ${\cal
    O}^{(1)}_*$ are centered at zeros of unperturbed
    determinants, while smaller disks in
 $O_*^{(2)}$ are centered about zeros of the perturbed determinant
 (see Remark \ref{R:july}). The shift of centers is of principle
 importance: we can reduce
 the size of disks, since they are more precisely targeted. If $W_1=0$, then $O_*^{(2)}$ is just a union of disks
 centered at quasi-intersections of the circle of radius $k$,
 centered at the origin, with circles of the same radius $k$,
 centered at points $2\pi j/a+2\pi p^{(1)}/N_1a+2\pi
    p^{(2)}/N_2N_1a$, $p^{(2)}\neq 0$, of the dual lattice for
    periods $N_2N_1a_1, N_2N_1a_2$.

\begin{figure}
\centering \psfrag{Phi_3}{$\Phi_3$}
\includegraphics[totalheight=.2\textheight]{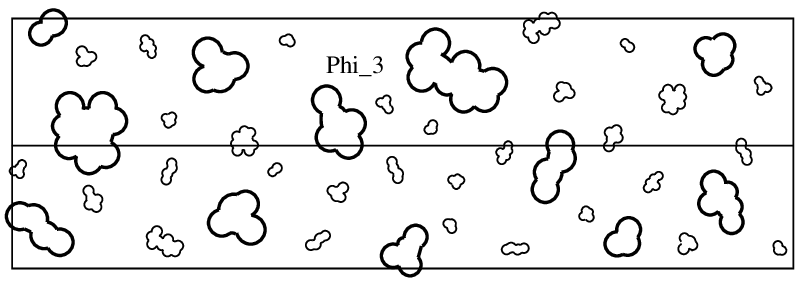}
\caption{The set $\varPhi
    _3$.}\label{F:8}
\end{figure}

To obtain $\varPhi _3$ we delete ${\cal
    O}^{(2)}_*$ from  $\varPhi _2$. Thus, the set $\varPhi
_3$ has a
    structure of Swiss cheese (Fig. \ref{F:8}); we will add more holes of a smaller
    size at each  step of approximation.


 By analogy with
(\ref{july9a}), we define the operator $A_2\left( \vec
y^{(2)}(\varphi )\right)$ by the formula:
$$I+A_2\bigl(\vec
y^{(2)}(\varphi ) \bigr)=\Bigl(H^{(2)}\bigl(\vec y^{(2)}(\varphi
)\bigr)-\lambda\Bigr) \Bigl(\tilde H_0\bigl(\vec y^{(2)}(\varphi
)\bigr)+\lambda\Bigr)^{-1}, $$ $\vec y^{(2)}(\varphi)$ being given
in Definition \ref{D:Aug15}, $\tilde H_0\left(\vec y^{(2)}(\varphi
)\right)$ acting in $L_2(Q_2)$.
The goal is to show that   $\det\Bigl(I+\tilde A_1\bigl(\vec
y^{(1)}(\varphi)\bigr)\Bigr)$ and $\det\Bigl(I+A_2\bigl(\vec
y^{(2)}(\varphi)\bigr)\Bigr)$ have the same number of zeros inside
each $\Gamma ^{(2)}\bigl(\vec
    b ^{(2)}\bigr)$ and that the resolvents for $\tilde H^{(1)}\bigl(\vec
y^{(1)}(\varphi)\bigr)$ and $H^{(2)}\bigl(\vec
y^{(2)}(\varphi)\bigr)$ satisfy similar estimates when $\varphi
\in \varPhi _1\setminus {\cal O}^{(2)}\bigl(\vec b^{(2)}\bigr)$.
\begin{lemma}\label{L:july9a}
If $\vec b^{(2)}\in K_2$  and $b_0^{(2)}>\epsilon
_1k^{-2l+1-2\delta }$, then
\begin{enumerate}
\item For any $\varphi \in \varPhi _1\setminus {\cal
O}^{(2)}\left(\vec b^{(2)}\right)$:
\begin{equation}\label{3.7.2.2**}
   \Bigl \|\Bigl(\tilde H^{(1)}\bigl(\vec y^{(1)}(\varphi)\bigr)-k^{2l}\Bigr)^{-1}\Bigr\|
    <\frac{1}{\epsilon _1^2}
    \end{equation}
    \begin{equation}\label{3.7.2.4**}
    \Bigl\|\Bigl(\tilde H^{(1)}\bigl(\vec y^{(1)}(\varphi)\bigr)-k^{2l}\Bigr)^{-1}\Bigr\|_1 <
    \frac{4c_0k^{2+2s_2}}{\epsilon _1^2}
    \end{equation}
    \item The  number of zeros of $\det\Bigl(I+\tilde A_1\left(\vec
    y^{(1)}(\varphi)\right)\Bigr)$  in $\varPhi _1$ does not exceed $4c_0k^{2+2s_2}$.
    \end{enumerate}
    \end{lemma}
    \begin{corollary}\label{C:*1}
    There are no zeros of $\det\Bigl(I+\tilde A_1\left(\vec
    y^{(1)}(\varphi)\right)\Bigr)$ in $\varPhi _1\setminus {\cal
    O}^{(2)}\left(\vec b^{(2)}\right)$.
    \end{corollary}
\begin{proof}
    \begin{enumerate}
    \item
    Considering the relation between $H^{(1)}$ and $\tilde H^{(1)}$, we
    easily see
    $$\Bigl\|\Bigl(\tilde H^{(1)}\bigl(\vec
    y^{(1)}(\varphi)\bigr)-k^{2l}\Bigr)^{-1}\Bigr\|=
    \max _{p^{(1)}\in P^{(1)}}\Bigl \|\Bigl( H^{(1)}\left(\vec
    y^{(1)}(\varphi)+2\pi
    p^{(1)}/N_1a\right)-k^{2l}\Bigr)^{-1}\Bigr\|.$$
    The condition  $\varphi \in \varPhi _1\setminus {\cal
O}^{(2)}\left(\vec b^{(2)}\right)$ yields $\varphi \in \varPhi
_1\setminus {\cal O}^{(1)}_s\left(\vec b^{(2)}+2\pi
    p^{(1)}/N_1a\right)$ for every $p^{(1)}\in P^{(1)}$. If $\vec b^{(2)}+2\pi
    p^{(1)}/N_1a $ satisfies  $b_0\geq k^{-2l+9+12s_1+7\delta}$, where $b_0$ is the distance
    from
    $\vec b^{(2)}+2\pi
    p^{(1)}/N_1a $ to the nearest vertex of $K_1$, then we
     use Lemma \ref{L:July9b} and Corollary
    \ref{C:july9a}:
    \begin{equation}
    \Bigl \|\Bigl( H^{(1)}\left(\vec
    y^{(1)}(\varphi)+2\pi
    p^{(1)}/N_1a\right)-k^{2l}\Bigr)^{-1}\Bigr\|<\frac{1}{\epsilon
    _1}<\frac{1}{\epsilon
    _1^2}.\label{Denis2}
    \end{equation}
     Suppose $\vec b^{(2)}+2\pi
    p^{(1)}/N_1a $ satisfies $b_0<k^{-2l+9+12s_1+7\delta}$. It is easy to see that  $b_0>\epsilon _1k^{-2l+1-2\delta}$, since
    $b_0^{(2)}>\epsilon _1k^{-2l+1-2\delta}$ by the hypothesis of the lemma. Using Corollary
     \ref{C:new} to obtain an  estimate similar to (\ref{Denis2}).
Thus,  (\ref{3.7.2.2**}) is proved.
    The estimate (\ref{3.7.2.4**}) can be checked in the analogous
    way from  (\ref{3.7.2.4*}) and (\ref{3.7.1.13++}), when we take into account that $P^{(1)}$ contains less than
    $4k^{2s_2-2s_1}$ elements.
\item The determinant of $I+\tilde A_1\left(\vec
y^{(1)}(\varphi)\right)$ is equal to zero if and only if at least
one of the determinants $\det \Bigl(I+ A_1\left(\vec
y^{(1)}(\varphi)+2\pi p^{(1)}/N_1a\right)\Bigr)$, $p^{(1)}\in
P^{(1)}$, is zero. Since each of them has no more than
$c_0k^{2+2s_1}$ zeros (Corollary \ref{C:May23}, Definition
\ref{SN}) and $P^{(1)}$ contains no more than $4k^{2(s_2-s_1)}$
elements, the total number of zeros of $\det \Bigl(I+\tilde
A_1\left(\vec y^{(1)}(\varphi)\right)\Bigr)$ does not exceed
$4c_0k^{2+2s_2}$.
\end{enumerate}
     \end{proof}
The following five lemmas are the analogs of Lemmas \ref{L:3.13}
-- \ref{L:3.17}.
\begin{lemma} \label{L:s1} If $\vec b^{(2)}\in K_2$  and $b_0^{(2)}>\epsilon _1k^{-2l+1-2\delta}$, then:
\begin{enumerate}
\item The number of zeros of $\det\Bigl(I+\tilde A_1\bigl(\vec
y^{(2)}(\varphi)\bigr)\Bigr)$ inside $\Gamma ^{(2)}\left(\vec
b^{(2)}\right)$ is the same as that of $\det\Bigl(I+\tilde
A_1\bigl(\vec y^{(1)}(\varphi)\bigr)\Bigr).$ \item For any
$\varphi \in \varPhi _2\setminus {\cal O}^{(2)}\left(\vec
b^{(2)}\right)$, the following inequalities hold:
    \begin{equation}
    \Bigl\|\Bigl(\tilde H^{(1)}\left(\vec y^{(2)}(\varphi)\right)-k^{2l}\Bigr)^{-1}\Bigr\|
    < \frac{2}{\epsilon_1^2}, \label{*2}
    \end{equation}
    \begin{equation}
    \Bigl\|\Bigl(\tilde H^{(1)}(\vec y^{(2)}(\varphi))-k^{2l}\Bigr)^{-1}\Bigr\|_1
    < \frac{8c_0k^{2+2s_2}}{\epsilon_1^2}.\label{*3}
    \end{equation}
    \end{enumerate}
\end{lemma}
\begin{corollary} \label{C:May22a*}
There are no zeros of $\det \Bigl(I+\tilde A_1\left(\vec
y^{(2)}(\varphi)\right)\Bigr)$ in $\varPhi _2\setminus {\cal
O}^{(2)}\left(\vec b^{(2)}\right)$.
\end{corollary}
\begin{proof}
\begin{enumerate}
\item Direct calculation gives: $$\det \Bigl(I+\tilde
A_1\bigl(\vec
    y^{(2)}\bigr)\Bigr)\Big/\det \Bigl(I+\tilde A_1\bigl(\vec
    y^{(1)}\bigr)\Bigr)=
    \det (I+T).$$
    $$T=\Bigl(W_1-2\lambda\Bigr)\Bigl(\tilde{H}_0\bigl(\vec y^{(2)}\bigr)+\lambda\Bigr)^{-1}
                  \Bigl(\tilde{H}_0\bigl(\vec y^{(1)}\bigr)-
                  \tilde{H}_0\bigl(\vec y^{(2)}\bigr)\Bigr)
\Bigl(\tilde H^{(1)}\bigl(\vec y^{(1)}\bigr)-\lambda\Bigr)^{-1}.$$
The estimate (\ref{3.75}) yields: $\bigl|\vec
y^{(2)}(\varphi)-\vec y^{(1)}(\varphi)\bigr|<2\alpha \epsilon_1
^4k^{-2l+1}$. Using the last estimate we readily show that
    \begin{equation}
    \Bigr\|\Bigl(\tilde{H}_0\bigl(\vec y^{(2)}\bigr)+\lambda\Bigr)^{-1}\Bigl(\tilde{H}_0\bigl(\vec y^{(1)}\bigr)
              -\tilde{H}_0\bigl(\vec y^{(2)}\bigr)\Bigr)\Bigr\|
        < 2l\alpha \epsilon_1^4 k^{-2l}.\label{*1}
        \end{equation}
Considering (\ref{3.7.2.4**}), we obtain
    \begin{gather}\label{01}
    \|T\|_1
    \leq 4k^{2l}
    \left(4l\epsilon_1^4 k^{-2l}\right)\left(4c_0k^{2+2s_2}/\epsilon_1^2 \right) \leq
    64lc_0
    \epsilon_1^2 k^{2+2s_2}<1.
    \end{gather}
Hence, $|\det(I+T)-1|<1$ for any $\varphi \in \gamma
^{(2)}\left(\vec b^{(2)}\right)$.
Applying  Rouch\'{e}'s Theorem, we complete the proof. \item Let
us consider the Hilbert relation:
    $$ \Bigl(\tilde H^{(1)}\bigl(\vec y^{(2)}(\varphi)\bigr)-\lambda\Bigr)^{-1}
     =\Bigl(\tilde H^{(1)}\bigl(\vec y^{(1)}(\varphi)\bigr)-\lambda\Bigr)^{-1}
     +T_*
    \Bigl(\tilde H^{(1)}\bigl(\vec
    y^{(2)}(\varphi)\bigr)-\lambda\Bigr)^{-1},$$ $$
 T_*=
    \Bigl(\tilde H^{(1)}\bigl(\vec y^{(1)}(\varphi)\bigr)-\lambda\Bigr)^{-1}
    \Bigl(\tilde{H}_0\bigl(\vec y^{(1)}(\varphi)\bigr)-\tilde{H}_0\bigl(\vec
    y^{(2)}(\varphi)\bigr)\Bigr).$$
    To estimate $\|T_*\|$,
we represent $T_*$ in the form $T_*=(I+T_1)T_2$, here
    $$T_1=
    \Bigl(\tilde H^{(1)}\bigl(\vec y^{(1)}(\varphi)\bigr)-\lambda\Bigr)^{-1}
    \Bigl(-W_1+2\lambda\Bigr),$$ $$
    T_2=\Bigl(\tilde{H}_0\bigl(\vec y^{(1)}(\varphi)\bigr)+\lambda\Bigr)^{-1}
    \Bigl(\tilde{H}_0\bigl(\vec y^{(1)}(\varphi)\bigr)-\tilde{H}_0\bigl(\vec
    y^{(2)}(\varphi)\bigr)\Bigr).$$
Using (\ref{3.7.2.2**}), we get $\| T_1\|\leq
4\lambda\epsilon_1^{-2}$.
The norm of $T_2$ can be bounded as (\ref{*1}). Multiplying the
estimates for $\|T_1\|$, $\|T_2\|$, we obtain: $\|T_*\|<8 \alpha l
\varepsilon _1^2<1/2$. Substituting the last estimate into the
Hilbert relation and considering (\ref{3.7.2.2**}), we arrive at
(\ref{*2}).
 Using (\ref{3.7.2.4**}), we prove (\ref{*3}) in the analogous way.
\end{enumerate}
     \end{proof}
%
%
\begin{lemma} \label{L:ti} If $\vec b^{(2)}\in K_2$ and $b_0^{(2)}>\epsilon _1k^{-2l+1-2\delta }$, then
 the following estimates hold for any $\varphi \in \varPhi _2\setminus {\cal O}^{(2)}\left(\vec
b^{(2)}\right)$:
    \begin{equation}
    \Bigl\|\Bigl(H^{(2)}\bigl(\vec y^{(2)}(\varphi)\bigr)-\lambda\Bigr)^{-1}\Bigr\|
    < \frac{4}{\epsilon_1^2},\label{*4}
    \end{equation}
    \begin{equation}
    \Bigl\|\Bigl(H^{(2)}\bigl(\vec y^{(2)}(\varphi)\bigr)-\lambda\Bigr)^{-1}\Bigr\|_1
    < \frac{16c_0k^{2+2s_2}}{\epsilon_1^2}.\label{*5}
    \end{equation}
\end{lemma}
\begin{corollary} \label{C:Nov12} There are no zeros of $\det\Bigl(I+A_2\left(\vec y^{(2)} (\varphi)\right)\Bigr)$
 in $\varPhi _2\setminus {\cal O}^{(2)}\left(\vec
b^{(2)}\right)$.
\end{corollary}
\begin{proof} The proof is analogous to that of Lemma \ref{L:3.14}; we use
Hilbert relation, the estimates (\ref{*2}), (\ref{*3}) and the
inequality $\|W_2\|<\epsilon _1^4$, which immediately follows from
(\ref{W2}) and the definition of $\epsilon _1$.      \end{proof}
\begin{lemma} \label{L:ti1} If  $\vec b^{(2)}\in K_2$  and $b_0^{(2)}>\epsilon _1k^{-2l+1-2\delta }$, then
 the number of zeros of
$\det\left(I+A_2(\vec y^{(2)} (\varphi))\right)$ inside $\Gamma
^{(2)}(\vec b^{(2)})$ is the same as that of
$\det\left(I+\tilde{A}_1(\vec y^{(2)} (\varphi))\right)$.
\end{lemma}
\begin{proof} Using (\ref{*3}), we obtain that for any  $\varphi \in \gamma
^{(2)}$:
    \begin{align*}
    &\left| \det \left(\frac{H^{(2)}\left(\vec y^{(2)} (\varphi)\right)-k^{2l}}
    {\tilde H^{(1)}\left(\vec y^{(2)}(\varphi)\right)-k^{2l}}\right) -\det I \right|
    =\left| \det \left(I+\frac{W_2}
    {\tilde H^{(1)}\left(\vec y^{(2)}(\varphi)\right)-k^{2l}}\right) -\det I \right|
    \\
    &\leq 2 \|W_2\|\Bigl\|\Bigl(\tilde H^{(1)}\bigl(\vec y^{(2)}(\varphi)\bigr)-k^{2l}\Bigr)^{-1}\Bigr\|_1
    \leq 2 \epsilon_1^4 \cdot
    \frac{4c_0k^{2+2s_2}}{\epsilon_1^2}=8c_0\epsilon_1^2k^{2+2s_2}
    \ll 1.
    \end{align*}
By Rouch\'{e}'s Theorem, we complete the proof.
     \end{proof}
\begin{lemma} \label{L:ti12} \begin{enumerate}
\item If  $\vec b^{(2)}\in K_2$  and $b_0^{(2)}>\epsilon
_1k^{-2l+1-2\delta }$, then
 the number of zeros of
$\det\left(I+A_2(\vec y^{(2)} (\varphi))\right)$ in any $\Gamma
^{(2)}(\vec b^{(2)})$ is the same as that of
$\det\left(I+\tilde{A}_1(\vec y^{(1)} (\varphi))\right)$. \item
The distance from a zero of $\det\left(I+A_2(\vec y^{(2)}
(\varphi))\right)$ to the nearest zero of \\
$\det\left(I+\tilde{A}_1(\vec y^{(1)} (\varphi))\right)$ does not
exceed $r^{(2)}/2$.
\end{enumerate}
\end{lemma}
\begin{corollary}
\label{C:**} The  number of zeros of $\det\left(I+A_2(\vec y^{(2)}
(\varphi))\right)$
 does not exceed
$4c_0k^{2+2s_2}$.
\end{corollary}
 The corollary is the combination
of the first part of the lemma with the second part of Lemma
\ref{L:july9a}.
\begin{proof} [Lemma \ref{L:ti1}] The first statement immediately
follows from Lemmas \ref{L:s1} (Part 1) and \ref{L:ti1}. Next,
note that Lemmas \ref{L:july9a} -- \ref{L:ti1} hold not only for
${\cal O}^{(2)}(\vec b^{(2)})$, but also for a set $\tilde {\cal
O}^{(2)}(\vec b^{(2)})$, constructed from disks of twice smaller
radius $r^{(2)}/2$ with the same centers, since all estimates in
the lemmas preserved under such change up to some insignificant
constants. This proves that all zeros of $\det\left(I+A_2(\vec
y^{(2)} (\varphi))\right)$ in ${\cal O}^{(2)}(\vec b^{(2)})$ are,
in fact, in the smaller set $\tilde {\cal O}^{(2)}(\vec b^{(2)})$,
i.e., the second statement of the lemma holds.
     \end{proof}
\begin{lemma}\label{L:3.17*}
If $\vec b^{(2)}\in K_2$  and $b_0 ^{(2)}\geq \epsilon
_1k^{-2l+1-2\delta }$, $|\epsilon |< \epsilon_1^4$, then
\begin{enumerate} \item The determinant of $\ \dfrac{H^{(2)}(\vec
y^{(2)}(\varphi))-k^{2l}-\epsilon }{H_0(\vec
y^{(2)}(\varphi))+k^{2l}}$
 has no zeros in $\varPhi _2\setminus {\cal O}^{(2)}\left(\vec b^{(2)}\right)$.
\item The number of zeros of  $\det\left(\dfrac{H^{(2)}(\vec
y^{(2)}(\varphi))-k^{2l}-\epsilon }{H_0(\vec
y^{(2)}(\varphi))+k^{2l}}\right)$ in $\Gamma ^{(2)}\left(\vec
b^{(2)}\right)$ is the same as that of
$\det\left(\dfrac{H^{(2)}(\vec y^{(2)}(\varphi))-k^{2l}}{H_0(\vec
y^{(2)}(\varphi))+k^{2l}}\right)$.
\end{enumerate}
\end{lemma}
 \begin{corollary}\label{C:ut*}
The following relation hold: $\omega _{2}^*\subset {\cal
O}^{(2)}_*$.
\end{corollary}
\begin{proof}[Lemma \ref{L:3.17*} and Corollary \ref{C:ut*}] Proof of the lemma is completely analogous to that of the
previous two lemmas up to replacement of the estimate
$\|W_2\|<\epsilon _1^4$ by $\|W_2+\epsilon \|<2\epsilon _1^4$.
Corollary follows from the lemma and Definition \ref{D:Aug15}. In
fact, $\omega _2^*(\lambda )=\cup _{p^{(2)}\in P^{(2)}\setminus
\{0\}} \omega_{2,p^{(2)}}$, where $\omega_{2,p^{(2)}}$ is the set
of zeros of the determinant (\ref{May18c+1}) for a fixed
$p^{(2)}$. By Lemma \ref{L:3.17*} (Part 1),
$\omega_{2,p^{(2)}}\subset {\cal O}^{(2)}\left(\vec
b^{(2)}\right)$, $\vec b^{(2)}=\frac{2\pi p^{(2)}}{N_2N_1a}$,
since $\epsilon _2< \epsilon _1^4$. Note that such $\vec b^{(2)}$
satisfies the condition of the lemma, since $N_2N_1a_{1,2} \leq
2k^{s_3}$. Using the definition (\ref{ut1*}) of ${\cal O}^{(2)}_*$
and the formula $\omega _2^*(\lambda )=\cup _{p^{(2)}\in
P^{(2)}\setminus \{0\}} \omega_{2,p^{(2)}}$ yields the corollary.
     \end{proof}
From now on we denote the radius $r$ of the disks in ${\cal
O}^{(1)}_*$ by $r^{(1)}$ and the radius $r'$ of the disks in
${\cal O}^{(2)}$ by $r^{(2)}$.

\begin{lemma} \label{L:St3} (proof of Statement 3) Under conditions of Geometric lemma \ref{L:4.2} , estimate
(\ref{4.9}) holds.\end{lemma}

\begin{proof}
 By Corollary \ref{C:ut*},
    $\omega_{2}^*\subset {\cal O}^{(2)}_*$. Hence, $\omega_{2}\subset {\cal O}^{(2)}_*\cap \Theta
    _2$, the set $\omega_{2}$ being given in Definition \ref{D:Aug15}.
     Considering that
    that ${\cal O}^{(2)}_*$ is formed by no more than
    $16c_0k^{2+2s_3}$  disks of the radius
    $r^{(2)}$ (see Remark \ref{R:X}),  we obtain that the total length
    of $\omega_{2}$ does not exceed $32c_0k^{2+2s_3
    }r^{(2)}$. Taking into account that $L(\Omega _2)\approx  k L(\omega _2)$ and
    $r^{(2)}=r^{(1)}k^{-2-4s_2-\delta }=k^{-6-6s_1-4s_2-4\delta }$,
    $s_3=2s_2=4s_1$,
we obtain that the the length of $\Omega _2$ does not exceed
    $64c_0k^{-3-2s_1-2s_2-4\delta }$. Considering that  $\chi
    _2^*(\lambda )$ has a length  $2\pi k (1+o(1))$ we arrive at
    Statement 3.     \end{proof}

    \subsection{Nonresonant part of the isoenergetic set of
$H_{\alpha }^{(3)}$}\label{S:3.4*}

This section is analogous to Section \ref{S:3.4} for the second
step. Indeed, let $S_3(\lambda )$ be an isoenergetic set of the
operator $H_{\alpha}^{(3)}$:
    $S_3(\lambda)=\{t^{(3)} \in K_3 : \exists n \in \N:
    \lambda_n^{(3)}(\alpha,t^{(3)})=\lambda  \}$,
    here $\{\lambda_n^{(3)}(\alpha,t^{(3)})\}_{n=1}^{\infty }$ is the
    spectrum of $H_{\alpha}^{(3)}(t^{(3)})$.
Now we construct a non-resonance subset $\chi _3^*(\lambda )$ of
$S_3(\lambda
    )$. It
    corresponds to  non-resonance eigenvalues $\lambda
    _{j^{(3)}}^{(3)}(t^{(3)} )$ given by the perturbation series (\ref{4.15}).
    We start with a definition of $\cal{D}_2(\lambda )_{nonres}$.
    Recall that $\chi _3\subset \cal{K}_3\chi_2^*(\lambda )$ (see
     Geometric Lemma \ref{L:4.2}) and $\chi_2^*(\lambda
    )=\cal{K}_2\cal{D}_2(\lambda )$, see (\ref{3.77.7}). Hence, $\chi _3\subset \cal{K}_3\cal{D}_2(\lambda
    )$. Let $\cal{D}_2(\lambda )_{nonres}$ be the preimage of $\chi
    _3$ in $\cal{D}_2(\lambda )$:
    \begin{equation}
    \cal{D}_2(\lambda )_{nonres}=\{\vec \varkappa \in \cal{D}_2(\lambda
    ):\cal{K}_3\vec \varkappa \in \chi _3\}. \label{pp*}
    \end{equation}
The following lemma is analogous to Lemma \ref{L:3.7.1/2} in the
second step.
\begin{lemma}\label{L:3.7.1/2*}
The formula $\cal{K}_3\cal{D}_2(\lambda )_{nonres}=\chi_3$
establishes
 one-to-one correspondence between
$\cal{D}_2(\lambda )_{nonres}$ and $\chi_3$.
\end{lemma}

\begin{proof} The proof is analogous to that of
Lemma \ref{L:3.7.1/2} up to the shift of indices by 1, i.e., $\chi
_2\to \chi _3$, $\chi_1^*(\lambda )\to \chi_2^*(\lambda )$, $\tau
=t^{(2)}\to  t^{(3)}$; we use  formula (\ref{3.77.7}) instead of
(\ref{2.81}), Part 1a of the Geometric Lemma for the third step
instead of Part 1a of the Geometric Lemma for the second step, and
Lemma  \ref{L:May10a*} instead of  \ref{L:May10a}.
     \end{proof}

We define $\cal{B}_3(\lambda )$ as the set of directions
corresponding to $\cal{D}_2(\lambda
 )_{nonres}$:
$$\cal{B}_3(\lambda)=\{\vec{\nu} \in \cal B_2(\lambda):
    \varkappa _{2}(\lambda,\vec{\nu})\vec{\nu} \in \cal{D}_2(\lambda
    )_{nonres}\}$$
 where
    $\varkappa _{2}(\lambda ,\vec{\nu})$ is defined by Lemma \ref{L:3.9}, $\varkappa _{2}(\lambda ,\vec{\nu})\equiv \varkappa _{2}(\varphi )$.
Note that $\cal{B}_3(\lambda )$ is a unit circle with holes
centered at the origin and $\cal{B}_3(\lambda ) \subset
    \cal{B}_2(\lambda )$.
    We denote by $\Theta _3(\lambda )$ the set of angles $\varphi
$ corresponding to $\cal{B}_3(\lambda )$:
\begin{equation}
\Theta _3(\lambda )=\{\varphi \in [0,2\pi ):\ (\cos \varphi ,\sin
\varphi )\in \cal{B}_3(\lambda )\} \label{May17a*},\ \ \ \
\Theta _3\subset \Theta _2.
 \end{equation}
We define $\cal{D}_3(\lambda)$ as a level set for
$\lambda^{(3)}(\alpha,\vec{\varkappa })$ in the $\left(\epsilon
_2k^{-2l+1-\delta }\right)$-neighborhood of $\cal{D}_2(\lambda
)_{nonres}$:
    \begin{equation*}
    \cal{D}_3(\lambda ):=\left\{\vec{\varkappa  }=\varkappa  \vec \nu: \vec \nu \in
    \cal{B}_3(\lambda ),\ \
    \bigl| \varkappa  -\varkappa _2(\lambda,\vec
\nu) \bigr|<\epsilon_2k^{-2l+1-\delta},\
\lambda^{(3)}(\alpha,\vec{\varkappa  })=\lambda
\right\}.\label{May14m*}
\end{equation*}
Next two lemmas are to prove   that $\cal{D}_3(\lambda )$ is a
distorted circle with holes. Their formulations and proofs are
analogous to those of Lemmas \ref{L:3.8} and \ref{L:3.9}.
%
%
\begin{lemma}\label{L:3.8*}
For every $\vec{\nu} \in \cal{B}_3(\lambda )$ and every $\alpha$,
$0 \leq \alpha \leq 1$, there is a unique $\varkappa   =\varkappa
_3 (\lambda,\vec{\nu})$ in the interval $I_3:=\bigl[\varkappa _2
(\lambda,\vec \nu)-\epsilon_2 k^{-2l+1-\delta},\varkappa _2
(\lambda,\vec \nu)+\epsilon_2 k^{-2l+1-\delta}\bigr]$ such that
    \begin{equation}\label{3.701}
    \lambda^{(3)}(\alpha,\varkappa _3 \vec{\nu})=\lambda .
    \end{equation}
Furthermore,
    \begin{equation}\label{3.71m}
    |\varkappa _3 (\lambda,\vec
\nu)-\varkappa _2(\lambda,\vec \nu)| \leq 2\alpha \epsilon_2
^4k^{-2l+1}.
    \end{equation}
\end{lemma}
Let us recall that $\varPhi _3$ is the third complex non-resonant
set given by  (\ref{ut1*}), see Fig. \ref{F:8}. Further we use the
notations $\varkappa _3(\varphi )\equiv \varkappa _3
(\lambda,\vec{\nu})$, $h_3(\varphi )\equiv \varkappa _3(\varphi
)-\varkappa _2(\varphi )$, $\vec \varkappa _3(\varphi )=\varkappa
_3(\varphi )\vec \nu $.
\begin{lemma}\label{L:3.9*} \begin{enumerate}
\item The set $\cal{D}_3(\lambda )$ is a distorted circle with
holes: it can be described by the formula:
\begin{equation}
\cal{D}_3(\lambda )=\bigl\{\vec \varkappa \in \R^2: \vec \varkappa
=\vec \varkappa _3(\varphi ),\ \ \varphi \in \Theta _3(\lambda
)\bigr\},\label{May20a1}
\end{equation} where
    $ \varkappa _3(\varphi )=
    \varkappa _2(\varphi)+h_3(\varphi ), $
$\varkappa _2(\varphi )$ is the ``radius" of $\cal{D}_2(\lambda )$
and $h_3(\varphi )$ satisfies the estimates
    \begin{equation}\label{3.75*}
    |h_3|<2\alpha \epsilon_2 ^4k^{-2l+1},\ \ \
    \left|\frac{\partial h_3}{\partial \varphi} \right| \leq
    4\alpha \epsilon_2^3 k^{1+\delta}.
    \end{equation}
    \item The total length of $\cal{B}_3(\lambda)$ satisfies the estimate:
\begin{equation}\label{theta2*s}
    L\left(\cal{B}_2\setminus \cal{B}_3\right)<4\pi k^{-4-2s_1-2s_2}.
    \end{equation}
\item The function $\varkappa _3(\varphi )$ can be extended as a
holomorphic function of $\varphi $ to the complex non-resonce set
$\varPhi _3$, estimates (\ref{3.75*}) being preserved. \item The
curve $\cal{D}_3(\lambda )$ has a length which is asymptotically
close to that of $\cal{D}_2(\lambda )$ in the following sense:
    \begin{equation}\label{3.77m}
     L\Bigl(\cal{D}_3(\lambda )\Bigr)\underset{\lambda \rightarrow
     \infty}{=}L\Bigl(\cal{D}_2(\lambda )\Bigr)\Bigl(1+O(k^{-4-2s_1-2s_2})\Bigr).
     \end{equation}
     \end{enumerate}
\end{lemma}
\begin{proof} The proof is analogous to that of Lemma \ref{L:3.9}.
Note only that, in Part 2, when proving convergence  of the series
for the resolvent $\left(H^{(3)}(\vec \varkappa _2(\varphi
 ))-z\right)^{-1}$, we use the estimate
 \begin{equation}
 \sup _{z\in C_3}\Bigl\|\left(\tilde H^{(2)}(\vec \varkappa _2(\varphi
 ))-z\right)^{-1}\Bigr\|<\frac{64}{\epsilon _2},\label{July3b2}
 \end{equation}
 analogous to (\ref{July3b}), the operator  $\tilde H^{(2)}$ acting in $L_2(Q_3)$. The estimate (\ref{July3b2}) follows from (\ref{*4}) for
 $\vec b=2\pi p^{(2)}/N_2N_1a $,
 $p^{(2)}\neq 0$, and estimate (\ref{16**}), which takes care about the case $p^{(2)}=0$. As
 a side result of these considerations, we obtain an estimate
 analogous to (\ref{16**}) for the new resolvent
 $\left(H^{(3)}(\vec \varkappa _3(\varphi
 ))-z\right)^{-1}$.
      \end{proof}
We define the non-resonance set, $\chi_3^*(\lambda )$ in
$S_3(\lambda )$ by the formula analogous to (\ref{3.77.7}):
    \begin{equation}\label{3.77.7*}
    \chi_3^*(\lambda ):=\cal{K}_3\cal{D}_3(\lambda ).
    \end{equation}
    The
following lemmas are analogous to Lemmas \ref{Apr4a} and
\ref{L:May10a*}.
    \begin{lemma} \label{Apr4a*}
    The set $\chi_3^*(\lambda )$ belongs to the
    $\left(2\alpha \epsilon _2^4k^{-2l+1}\right)$-neighborhood of $\chi_3(\lambda
    )$ in $K_3$. If $t^{(3)}\in \chi_3^*(\lambda )$, then the operator
    $H^{(3)}_{\alpha }(t^{(3)})$ has a simple eigenvalue
   equal to $\lambda $. This
    eigenvalue is given by the perturbation series (\ref{4.15}).
\end{lemma}

    \begin{lemma}\label{L:May10a**} Formula (\ref{3.77.7*}) establishes
    one-to-one correspondence between $\chi_3^*(\lambda )$ and $\cal{D}_3(\lambda
    )$.\end{lemma}
    \begin{remark} \label{R:May14*}
    From geometric point of view this means that $\chi_3^*(\lambda
    )$ does not have self-intersections.\end{remark}

\subsection{Preparation for the Next Approximation} \label{S:3.6*}
\subsubsection{Contracted set ${\cal O}^{(2)}_s\left(\vec b^{(2)}\right)$}\label{S:3.6'm}
For any $\vec b^{(2)}$ such that $b_0^{(2)}>\epsilon
_1k^{-2l+1-2\delta }$, we have constructed  the set ${\cal
O}^{(2)}(\vec b^{(2)}) \subset \varPhi _2$, which surrounds zeros
of the determinant $\det\Bigl(I+\tilde A_1\left(\vec y
^{(1)}(\varphi)\right)\Bigr)$, $\vec y ^{(1)}(\varphi)=\vec
\varkappa _1(\varphi)+\vec b^{(2)}$,
 see Remark \ref{R:july}. The set consists of disks of
 the radius
 $r^{(2)}=r^{(1)}k^{-2-4s_2-\delta}$,
 $r^{(1)}=k^{-4-6s_1-3\delta}$ and contains no more than
 $4c_0k^{2+2s_2}$ disks, see Remark \ref{R:Nov}.
We proved  that the  new determinant $\det\Bigl(I+A_2\bigl(\vec
y^{(2)}(\varphi)\bigr)\Bigr)$, $\vec y^{(2)}(\varphi)=\vec
\varkappa _2(\varphi)+\vec b ^{(2)}$,
has the same number of zeros inside  each $\Gamma ^{(2)}(\vec
b^{(2)})$ as   $\det\Bigl(I+\tilde A_1\bigl(\vec
y^{(1)}(\varphi)\bigr)\Bigr)$ (Lemma  \ref{L:ti12}, Part 1) and
both have no zeros in $\in \varPhi _2\setminus {\cal O}^{(2)}(\vec
b^{(2)})$ (Corollaries \ref{C:*1}, \ref{C:Nov12}).
To prepare the next approximation, we contract the set ${\cal
O}^{(2)}_s(\vec b^{(2)})$ around the zeros of the  new determinant
$\det\Bigl(I+A_2\bigl(\vec y^{(2)}(\varphi)\bigr)\Bigr)$. We
consider the disks with the radius
$r^{(3)}=r^{(2)}k^{-2-4s_3-\delta }$ centered at each zero
$\varphi _{j,0}^{(2)}$ of the  determinant $\det
\Bigl(I+A_2\bigl(\vec y^{(2)}(\varphi )\bigr)\Bigr)$. Obviously
each of these disks is in ${\cal O}^{(2)}\bigl(\vec
b^{(2)}\bigr)$, since $r^{(3)}<<r^{(2)}$ and the distance between
zeros of the old and new determinants is smaller than $r^{(2)}/2$
(Lemma \ref{L:ti12}, Part 2). Next, we take the union
 of these smaller disks and denote it by ${\cal O}_s^{(2)}\bigl(\vec b^{(2)}
 \bigr)$.
 We denote by $\Gamma _s^{(2)}\bigl(\vec
b\bigr)$  a connected component of these new disks, which
intersects with $\varPhi_2$, and by $\gamma _s^{(2)}\bigl(\vec
b^{(2)}\bigr)$ its boundary, $\gamma _s^{(2)}=\partial \Gamma
_s^{(2)}$.
\begin{lemma}\label{L:July9b*}
\begin{enumerate}
\item The set ${\cal O}_s^{(2)}\bigl(\vec b^{(2)}\bigr)$ contains
the same number of disks as   ${\cal O}^{(2)}\bigl(\vec
b^{(2)}\bigr)$. This number  does not exceed $4c_0k^{2+2s_2}$.
\item For any $\varphi \in \varPhi _2\setminus {\cal
O}_s^{(2)}\bigl(\vec b^{(2)}\bigr)$,
    \begin{equation}\label{3.7.2.2*m}
    \Bigl\|\Bigl(H^{(2)}\bigl(\vec y^{(2)}(\varphi)\bigr)-k^{2l}\Bigr)^{-1}\Bigr\| \leq
    \frac{4}{\epsilon _1^2}\left(\frac{2r^{(2)}J^{(2)}}{r^{(3)}}\right)^{J^{(2)}},
    \ \ \ J^{(2)}=4c_0k^{2+2s_2},
    \end{equation}
    \begin{equation}\label{3.7.2.4*m}
    \Bigl\|\Bigl(H^{(2)}\bigl(\vec y^{(2)}(\varphi)\bigr)-k^{2l}\Bigr)^{-1}\Bigr\|_1 \leq
    \frac{4J^{(2)}}{\epsilon _1^2}\left(\frac{2r^{(2)}J^{(2)}}{r^{(3)}}\right)^{J^{(2)}}.
    \end{equation}
\end{enumerate}
\end{lemma}
The proof is analogous to that of Lemma \ref{L:July9b}, Lemma
 \ref{L:ti} being used instead of Lemma \ref{L:3.14}.
Considering the expressions for $r^{(2)}, r^{(3)},
J^{(2)},\epsilon _2$, we easily obtain that the right-hand part of
(\ref{3.7.2.2*m}) does not exceed $1/\epsilon _2$. Hence, the
following corollary holds:
\begin{corollary}\label{C:july9a1}
If $b_0^{(2)} > \epsilon _1 k^{-2l+1-2\delta }$, then
\begin{equation}\label{3.7.2.2**=}
    \Bigl\|\Bigl(H^{(2)}\bigl(\vec y^{(2)}(\varphi)\bigr)-k^{2l}\Bigr)^{-1}\Bigr\|
    <\frac{1}{\epsilon _2},
    \end{equation}
    \begin{equation}\label{3.7.2.4*=}
    \Bigl\|\Bigl(H^{(2)}\bigl(\vec y^{(2)}(\varphi)\bigr)-k^{2l}\Bigr)^{-1}\Bigr\|_1
    < \frac{4c_0k^{2+2s_2}}{\epsilon _2}.
    \end{equation}
\end{corollary}

\subsubsection{The set ${\cal O_s^{(2)}}\left(\vec b^{(2)}\right)$ for small $\vec b^{(2)}$}
\label{SS:3.7.1m}

Everything  we  considered so far is valid if
    $b_0 ^{(2)}>\epsilon _1k^{-2l+1-2\delta }$.
 However, in the next chapter
and later, $b_0^{(2)}$ is taken smaller, since the reciprocal
lattice is getting finer with each step. To prepare for this, let
us consider $\vec b^{(2)}$ being close to a vertex of $K_2$:
$0<b_0 ^{(2)}\leq \epsilon _1k^{-2l+1-2\delta }$.
We  show  that for such $\vec b^{(2)}$, the resolvent
$\Bigl(H^{(2)}\bigl(\vec y^{(2)}(\varphi
)\bigr)-k^{2l}\Bigr)^{-1}$
has no more than two poles  in $\varPhi _2$ and even its small
neighborhood. We  surround these poles by two contours $\gamma
^{\pm \ (2)}_s$ and prove an estimate for the norm of
$\Bigl(H^{(2)}\left(\vec y^{(2)}(\varphi
)\right)-k^{2l}\Bigr)^{-1}$ when $\varphi $ is outside these
contours.

 Let $\tilde \varPhi _2$ be
the $\frac{r^{(1)}}{2}$-neighborhood of $\varPhi _2$. It is easy
to see that $\tilde \varPhi _2\subset \tilde \varPhi _1$. Suppose
$|\vec b^{(2)} |=b_0^{(2)}$, i.e., the closest vertex of $K_2$ for
$\vec b^{(2)}$ is $(0,0)$. We consider the functions $\lambda
^{(1)}(\vec y ^{(1)}(\varphi ))$ and $\lambda ^{(2)}(\vec y
^{(2)}(\varphi ))$ defined  by perturbation series (\ref{2.66})
and (\ref{3.66}) for $\varphi \in \tilde \varPhi _2$. The
convergence of these series can be easily justified. In fact, by
Lemmas \ref{L:2.13} and \ref{L:3.9}, $\vec \varkappa _1(\varphi )$
and $\vec \varkappa _2(\varphi )$ are holomorphic functions of
$\varphi $ in $\varPhi _2$. It is easy to show that the lemmas
hold even in $\tilde \varPhi _2$, since all estimates involved are
preserved in $\tilde \varPhi _2$.  The perturbation series
(\ref{2.66}) and (\ref{3.66}) converge for $\lambda ^{(1)}(\vec
\varkappa _1(\varphi ))$ and $\lambda ^{(2)}(\vec \varkappa
_2(\varphi ))$, respectively, because $\vec \varkappa _1(\varphi
)$ and $\vec \varkappa _2(\varphi )$ are in small neighborhoods of
$\cal{D}_0(\lambda )_{nonres}$ and $\cal{D}_1(\lambda )_{nonres}$
(see ({2.75}), ({3.75})).  Since the estimates involved are stable
with respect to a change of $\vec \varkappa _{1,2}$ not exceeding
$\epsilon _1k^{-2l+1-\delta }$, the perturbation series for
$\lambda ^{(1)}\bigl(\vec y^{(1)}(\varphi )\bigr)$ and $\lambda
^{(2)}\bigl(\vec y ^{(2)}(\varphi )\bigr)$ also converge, both
functions being holomorphic in $\tilde \varPhi _2$.
 We base our further
considerations on these perturbation series expansions. For $\vec
b^{(2)}$ being close to a vertex $\vec e$ other than $(0,0)$, we
take $\vec y^{(2)}(\varphi )=\vec \varkappa _2(\varphi )+\vec
b^{(2)}-\vec e$.

From now on, we denote the solutions $\varphi ^{\pm }_{\epsilon
_0}$ of the equation $\lambda ^{(1)}\bigl(\vec y^{(1)}(\varphi
)\bigr)=k^{2l}+\varepsilon _0$, introduced in Lemma
\ref{L:3.7.1.1}, by $\varphi ^{\pm (1)}_{\epsilon _0}$. As in the
previous subsection, we choose $r^{(3)}=r^{(2)}k^{-2-4s_3-\delta
}$ here.
\begin{lemma}\label{L:3.7.1.1m}
If $0<b_0 ^{(2)}\leq \epsilon _1k^{-2l+1-2\delta }$ and $
|\epsilon_0|<b_0^{(2)}k^{2l-1-\delta}r^{(3)}$, then the equation
    \begin{equation}\label{3.7.1.2*}
   \lambda ^{(2)}\left(\vec y ^{(2)}(\varphi)\right)=k^{2l}+\epsilon_0
   \end{equation}
 has no more than two
solutions $\varphi^{\pm\ (2)}_{\epsilon _0}$ in $\tilde \varPhi
_2$. For any $\varphi^{\pm\ (2)}_{\epsilon _0}$ there is
$\varphi^{\pm (1)}_{0}\in \tilde \varPhi _1$ such that
    \begin{equation}
    \left|\varphi^{\pm (2)}_{\epsilon _0}-\varphi^{\pm (1)}_{0}\right|
    <r^{(3)}/4, \label{july5a*}
    \end{equation}
    here and below $\varphi^{\pm (1)}_{0}$ is $\varphi ^{\pm (1)}_{\epsilon
    _0}$ for $\epsilon _0=0$.
\end{lemma}
\begin{proof}  Let $|\vec b^{(2)}|=b_0^{(2)}$. First, we expand $\lambda ^{(2)}(\vec y ^{(2)}(\varphi))$
and $\lambda ^{(1)}(\vec y ^{(1)}(\varphi))$  near the point $\vec
b^{(2)}=0$ and  consider that $\lambda ^{(2)}(\vec \varkappa
_2(\varphi))=\lambda ^{(1)}(\vec \varkappa _1(\varphi))=\lambda$.
Then, using (\ref{3.68}), the estimates of the type (\ref{2.20a}),
(\ref{2.20**a}) and (\ref{3.71}), we  check that
\begin{equation}\left|\lambda ^{(2)}(\vec y ^{(2)}(\varphi))-\lambda
^{(1)}(\vec y ^{(1)}(\varphi))\right|<b_0^{(2)}\epsilon _1
\label{Nov16a} \end{equation} in $\tilde \varPhi _2$ and even in
its $r^{(3)}$-neighborhood, the neighborhood being a subset of
$\tilde \varPhi _1$. Suppose (\ref{3.7.1.2*}) holds for some
$\varphi \in \tilde \varPhi _2$.
By (\ref{Nov16a}), $\lambda ^{(1)}(\vec y
^{(1)}(\varphi))=k^{2l}+\varepsilon _0'$, $\varepsilon
_0'=O\left(b_0^{(2)}k^{2l-1-\delta}r^{(3)}\right)$. Hence,
$\varphi $ satisfies conditions of Lemmas \ref{L:3.7.1.1} and
\ref{L:july5a}. Surrounding $\varphi $ by a circle $C$ of the
radius $r^{(3)}/4$, we see that $|\lambda ^{(1)}(\vec y
^{(1)}(\varphi))-k^{2l}-\varepsilon
_0'|=\frac{1}{2}lk^{2l-1}b_0^{(2)}r^{(3)}\left(1+o(1)\right)>|\varepsilon
_0'|$ on this circle. Applying Rouch\'{e}'s theorem, we obtain
that there is a solution of $\lambda ^{(1)}(\vec y
^{(1)}(\varphi))=k^{2l}$ inside this circle. Thus, any solution of
(\ref{3.7.1.2*}) is in the circle of the radius $r^{(3)}/4$
surrounding $\varphi^{\pm (1)}_{0}$, the point $\varphi^{\pm
(1)}_{0}$ being in the $(r^{(3)}/4)$-neighborhood of $\tilde
\varPhi _2$. It remains to check that (\ref{3.7.1.2*}) has no more
than two solutions and no more than one in a vicinity of each
$\varphi^{\pm (1)}_{0}$. We construct the disk of the radius
$r^{(3)}/4$ centered at $\varphi^{\pm (1)}_{0}$, described above,
and note that $|\lambda ^{(1)}(\vec y
^{(1)}(\varphi))-k^{2l}|=\frac{1}{2}lk^{2l-1}b_0^{(2)}r^{(3)}\left(1+o(1)\right)$
on the circle. Using (\ref{Nov16a}) and
 Rouch\'{e}'s Theorem, we obtain that there is only one solution of
 (\ref{3.7.1.2*}) in the disk.
If $\vec b$ is close to a vertex other than $(0,0)$, the
    considerations are the same up to a parallel shift.
     \end{proof}
\begin{lemma}\label{L:july5a*} Suppose $0<b_0 ^{(2)}\leq \epsilon _1k^{-2l+1-2\delta }$  and $\varphi \in \tilde \varPhi
_2$ obeys the inequality analogous to (\ref{july5a*}):
$\left|\varphi -\varphi^{\pm (1)}_{0}\right|
    <r^{(3)}$.
 Then,
    \begin{equation}\label{3.7.1.6.1/2+}
    \frac{\partial}{\partial \varphi}\lambda^{(2)}\left(\vec
    y^{(2)}(\varphi)\right)=_{k\to
    \infty}\pm 2lb_0^{(2)}k^{2l-1}\bigl(1+o(1)\bigr).
    \end{equation}

\end{lemma}
The proof is completely analogous to that of Lemma \ref{L:july5a}.


\begin{definition}\label{SN1}
Let $\Gamma^{\pm (2)}_s\left(\vec b^{(2)}\right)$ be the open
disks centered at $\varphi_0^{\pm (2)}$ with radius $r^{(3)}$;
$\gamma^{\pm (2)}_s\bigl(\vec b^{(2)}\bigr)$ be their boundary
circles and ${\cal O}_s^{(2)}\bigl(\vec
b^{(2)}\bigr)=\Gamma^{+(2)}_s\cup
\Gamma^{-(2)}_s$.\end{definition}
%
%
%
%
%
\begin{lemma}\label{L:3.7.1*}
For any $\varphi$ in $\varPhi _2\setminus {\cal O}_s^{(2)}(\vec
b^{(2)})$,
    \begin{equation}\label{3.7.1.10*}
    |\lambda^{(2)}(\vec
    y^{(2)}(\varphi))-k^{2l}|\geq b_0^{(2)}k^{2l-1-\delta}r^{(3)}.
    \end{equation}
\end{lemma}
 The proof
is analogous to that of Lemma \ref{L:3.7.1}.

\begin{lemma}\label{L:july9c*}
For any $\varphi \in \varPhi _2\setminus {\cal
O}_s^{(2)}\left(\vec b^{(2)}\right)$,
    \begin{equation}\label{3.7.1.10.1/2*}
    \Bigl\|\Bigl(H^{(2)}\left(\vec
    y^{(2)}(\varphi)\right)-k^{2l}\Bigr)^{-1}\Bigr\|<\frac{16}{b_0^{(2)}r^{(3)}k^{2l-1-\delta}},
    \end{equation}
    \begin{equation}\label{3.7.1.13m}
    \Bigl\|\Bigl(H^{(2)}\left(\vec y^{(2)}(\varphi)\right)-k^{2l}\Bigr)^{-1}\Bigr\|_1<
    \frac{64c_0k^{2+2s_2}}{b_0^{(2)}r^{(3)}k^{2l-1-\delta }}.
    \end{equation}
\end{lemma}
\begin{corollary} \label{C:new1}
If $\epsilon _2k^{-2l+1-2\delta }<b_0^{(2)}\leq  \epsilon
_1k^{-2l+1-2\delta }$, and $\varphi \in \varPhi _2\setminus {\cal
O}_s^{(2)}\left(\vec b^{(2)}\right)$, then
    \begin{equation}\label{3.7.1.10.1/2**}
    \Bigl\|\Bigl(H^{(2)}\left(\vec
    y^{(2)}(\varphi)\right)-k^{2l}\Bigr)^{-1}\Bigr\|<\frac{1}{\epsilon _2^2},
    \end{equation}
    \begin{equation}\label{3.7.1.13**}
    \Bigl\|\Bigl(H^{(2)}\left(\vec y^{(2)}(\varphi)\right)-k^{2l}\Bigr)^{-1}\Bigr\|_1<
    \frac{4c_0k^{2+2s_2}}{\epsilon _2^2}.
    \end{equation}
    \end{corollary}
    The corollary follows from the condition on $b_0^{(2)}$, for
    the formula $r^{(3)}=r^{(2)}k^{-2-4s_3-\delta }=k^{-8-6s_1-4s_2-4s_3-5\delta
    }$ and estimate (\ref{k}).
    The proof of the lemma is analogous to that of Lemma
\ref{L:july9c}.

\section{The $n$-th Step of Approximation. Swiss Cheese Method.}
\setcounter{equation}{0}
\subsection{Perturbation Formulae}
\label{Section 5} On the $n$-th step, $n\geq 4$, we choose
$s_{n}=2s_{n-1}$ define the operator $H_{\alpha}^{(n)}$ by the
formula:
 \begin{equation}\label{4.1m}
     H_{\alpha }^{(n)}=H^{(n-1)}+\alpha W_{n},\quad      (0\leq \alpha \leq
     1),\qquad
     W_{n}=\sum_{r=M_{n-1}+1}^{M_n}V_r, \notag
     \end{equation}
where $M_{n}$ is choosen in such a way that $2^{M_{n}}\approx
k^{s_{n}}$. Obviously, the periods of $W_{n}$ are $2^{M_{n}-1}
(b_1,0)$ and $2^{M_{n}-1} (0,b_2)$. We  write them in the form:
$N_{n-1}\cdot...\cdot N_1(a_1,0)$ and $N_{n-1}\cdot...\cdot
N_1(0,a_2)$, here  $N_{n-1}=2^{M_{n}-M_{n-1}}$,
$\frac{1}{4}k^{s_{n}-s_{n-1}}<N_{n-1}< 4k^{s_n-s_{n-1}}$. Note
that
    $\|W_n\|_{\infty} \leq \sum_{r=M_{n-1}+1}^{M_n}
    \|V_r\|_{\infty} <
    \exp(-k^{\eta s_{n-1}}).$

    The $n$-th step is analogous to the third step up to replacement
    the indices 3 by $n$, 2 by $n-1$, the product $N_2N_1$ by
    $N_{n-1}\cdot ...\cdot N_1$, etc.

    We note that  $k$, satisfying (\ref{k}), obeys the analogous
    condition for the $n$-th step:
    \begin{equation}
 \hat C(1+s_{n-2})k^{2+2s_{n-2}}\ln k <k^{\eta s_{n-2}}\label{k*}
 \end{equation}
with the same constant $\hat C$. The inequality (\ref{k*}) can be
obtained from (\ref{k}) by  induction. This is an important fact:
it means that the lower bound for $k$ does not grow with $n$,
i.e., all steps hold uniformly in $k$ for $k>k_*$, $k_*$ being
introduced by (\ref{k}). Further we assume $k>k_*$.

The formulation of the geometric lemma for $n$-th step is the same
as that for Step 3 up to a shift of indices, we skip it here for
shortness. Note only that in the lemma we use the set
$\chi_{n-1}^*(\lambda)$ to define $\chi_{n}(\lambda)$. In fact,
we started with the
    definition of $\chi _1(\lambda )$ and then use it to define $\chi _1^*(\lambda
    )$ (Step 1). Considering $\chi _1^*(\lambda
    )$, we constructed $\chi _2(\lambda
    )$  (Lemma \ref{L:3.2}) and later used it to define $\chi _2^*(\lambda
    )$ (Section \ref{S:3.4}). Using $\chi _2^*(\lambda
    )$, we introduced $\chi _3(\lambda )$ (Lemma \ref{L:4.2}) and, then $\chi _3^*(\lambda
    )$ (Section \ref{S:3.4*}).
    Thus, the process goes like $\chi _1\to \chi _1^*\to \chi
    _2\to \chi _2^*\to \chi _3\to \chi _3^*$.
    Here we consider the  set $\chi_{n-1}^*(\lambda)$  defined by
(\ref{3.77.7*}) for $n=4$ and by (\ref{3.77.7*+}) for $n>4$.
\footnote{Strictly speaking we assume that there is a subset
$\chi_{n-1}^*(\lambda)$ of the isoenergetic surface
$S_{n-1}(\lambda)$ of $H^{(n-1)}$ such that perturbation series of
the type (\ref{4.15}), (\ref{4.16}) converges for $t^{(n-1)}\in
\chi_{n-1}^*(\lambda)$ and $\chi_{n-1}^*(\lambda)$ has properties
described in Section \ref{S:3.4*} up to replacement of $3$ by
$n-1$. In particular, we assume that
$\chi_{n-1}^*(\lambda)=\cal{K}_{n-1}\cal{D}_{n-1}(\lambda )$,
where $\cal{D}_{n-1}(\lambda)$ satisfies the analog of Lemma
\ref{L:3.9*} and that the analogs of
 Lemmas \ref{Apr4a*} and \ref{L:May10a**} hold too. Here, $\cal{K}_{n-1}$ is
 the parallel shift into $K_{n-1}$. Further in this section we
 describe the next set
 $\chi_{n}^*(\lambda)$ which has analogous properties.}
 The estimate (\ref{4.9}) for $n$-th step takes the form:
\begin{equation}\label{4.9*}
\frac{L\left(\cal{K}_n\chi_{n-1}^*\setminus \chi
_n)\right)}{L\left(\chi_{n-1}^*\right)}<k^{-S_n},\ \ \ \ S_n=2\sum
_{i=1}^{n-1}(1+s_i).
\end{equation}
It is easy to see that $S_n=2(n-1)+\left(2^{n}-2\right)s_1$ and
$S_n\approx 2^{n}s_1\approx s_n$ for large $n$. The formulation of
the main results (perturbation formulae) for $n$-th step is the
same as for the third step up to the shift of indices. The formula
for the resonance set $\Omega _{n-1}$ and non-resonance set $\chi
_n$ are analogous to those for $\Omega _2$, $\chi _2$ (see
(\ref{4.2.1})).
\subsection{Proof of Geometric Lemma} Proofs of the first and second
statements of Geometric Lemma are simple and  completely analogous
to those in the second step and third step. We skip them here for
shortness. Proof of the  third statement is completely analogous
to that for the third step. For the purpose of rigor we present  a
short version of it.

 In the
second and third steps we defined the sets ${\cal
O}^{(1)}\left(\vec b^{(1)}\right)$ and ${\cal O}^{(2)}\left(\vec
b^{(2)}\right)$ (see Definition \ref{D:gamma} and formula
(\ref{july8a})). Considering (\ref{july8a}), we define ${\cal
O}^{(n-1)}\left(\vec b^{(n-1)}\right)$,  $\vec b^{(n-1)}\in
K_{n-1}$,  $n\geq 4$, by the analogous formula
    \begin{equation}
    {\cal O}^{(n-1)}\left(\vec b^{(n-1)}\right)=\cup _{p^{(n-2)}\in P^{(n-2)}}{\cal
    O}^{(n-2)}_s\left(\vec b^{(n-1)}+2\pi
p^{(n-2)}/\hat N_{n-2}a \right), \label{july8a*}
    \end{equation}
    here and below
    $\hat N_{n-2}\equiv
N_{n-2}\cdot ....\cdot N_1,$
$$P^{(m)}=\{p^{(m)}=\left(p_1^{(m)},p_2^{(m)}\right),\ 0\leq
p_1^{(m)} \leq N_m-1, 0 \leq p_2^{(m)} \leq N_m-1\};$$ the set
${\cal
    O}^{(m)}_s(\vec b^{(m)})$, $m\geq 1$, is a collection of disks of the radius
    $r^{(m+1)}=
    r^{(m)}k^{-2-4s_{m+1}-\delta }$ around zeros of the determinant
    $\det\Bigl(I+A_{m}\left(\vec y^{(m)}(\varphi )\right)\Bigr)$ in $\varPhi
    _m$, here $\vec y^{(m)}(\varphi )=\vec \varkappa _m(\varphi )+\vec
    b^{(m)}$, $\vec \varkappa _m(\varphi )$ being defined by Lemmas \ref{L:2.13} $(m=1)$, \ref{L:3.9} $(m=2)$, \ref{L:3.9*}
    $(m=3)$
    and
    \ref{L:3.9*+} $(m\geq 4)$.
     Let
\begin{equation}
    {\cal O}^{(n-1)}_*=\cup _{p^{(n-1)}\in P^{(n-1)}\setminus \{0\}}{\cal
    O}^{(n-1)}\Bigl(2\pi p^{(n-1)}/\hat N_{n-1}a\Bigr), \ \ \ \ \varPhi
    _n=\varPhi _{n-1}\setminus {\cal O}^{(n-1)}_*.
     \label{ut1**}
    \end{equation}
   We denote by $\Gamma ^{(n-1)}\left(\vec b^{(n-1)}\right)$ a connected component of ${\cal O}^{(n-1)}\left(\vec b^{(n-1)}\right)$. Note that the complex non-resonance set $\varPhi _m$ is defined by
    the  recurrent formula analogous to
 (\ref{ut1*}).
    \begin{lemma} \label{L:new2}
    The set ${\cal
    O}^{(m)}_s\left(\vec b^{(m)} \right)$, $\vec b^{(m)}\in K_m$ contains no more than
    $4^{m-1}c_ok^{2+2s_{m}}$ disks.
    \end{lemma}
    \begin{corollary}\label{C:c1}
    The set ${\cal O}^{(n-1)}\left(\vec b^{(n-1)}\right)$ contains no more than
    $4^{n-2}c_ok^{2+2s_{n-1}}$ disks.
    \end{corollary}
    \begin{corollary} \label{C:c2}
    The set ${\cal O}^{(n-1)}_*$ contains no more than
    $4^{n-1}c_ok^{2+2s_{n}}$ disks.
    \end{corollary}
    The lemma is proved by induction procedure. For $m=2$ it holds
    by Part 1 of Lemma \ref{L:July9b*} and by Definition \ref{SN1}. For $m>2$
    it follows from Corollary \ref{C:obvi} and Definition \ref{D:sat}.
    Rigorously
    speaking, we assume for now that the lemma holds for $m\leq n-2$. At the end
    of this section, we  obtain the analog of the lemma for the
    next step ($n\to n+1$), this will complete the induction
    procedure. Corollaries
    \ref{C:c1} and \ref{C:c2} are based on the fact that $P^{(n-1)}$
    contains no more than $4k^{2(s_{n}-s_{n-1})}$ elements and a
    similar estimate holds for $P^{(n-2)}$.

 Obviously, $\varPhi _{n}$ has a structure of Swiss cheese,
    more and more holes of smaller and smaller radii appear at each
    step of approximation (Fig. \ref{F:8}). Note that the disks are more
    and more precisely ``targeted" at each step of approximation.
    At the n-th step the disks of ${\cal O}^{(n-1)}_*$ are centered around the zeros of the
    determinants
     $$\det\Bigl(I+A_{n-2}\left(\vec \varkappa _{n-2}(\varphi )+2\pi p^{(n-2)}/\hat N_{n-2}a
     +2\pi p^{(n-1)}/\hat N_{n-1}a\right)\Bigr),$$ where
     $p^{(n-2)}\in P^{(n-2)}$,
      $p^{(n-1)}\in P^{(n-1)}$, the
     corresponding operator $H^{(n-2)}$ being closer and closer to
     the operator $H$ at each step of approximation. If $W_1=W_2=...=W_{n-2}=0$, then ${\cal O}^{(n-1)}_*$ is just a
    union of circles centered at quasi-intersections of the circle
    of radius $k$, centered at the origin, with circles of the same
    radius centered at points $2\pi j/a+2\pi
    p^{(1)}/N_1a+....+2\pi p^{(n-1)}/\hat N_{n-1}a$, which are points of the dual lattice corresponding to
    the periods $\hat N_{n-1}a_1,\hat N_{n-1}a_2$.

The following is a proof of third statement of Geometric Lemma for
$n$-th step, namely, estimate (\ref{4.9*}). It is completely
analogous to that in the third step, we only have to provide that
condition (\ref{k*}) is sufficient for all necessary estimates.
    We start with an analog of Lemma \ref{L:july9a}: we shift
    indices
    $1\to n-2, 2\to n-1$ and replace the coefficient $4c_0$ in the
    estimate (\ref{3.7.2.4**}) by $4^{n-2}c_0$. The operator $\tilde H^{(n-2)}$ is $H^{(n-2)}$ extended from $Q_{n-2}$ to $Q_{n-1}$. By analogy with
    (\ref{Nov26}), $\vec y^{(n-2)}(\varphi)=\vec \varkappa _{n-2}(\varphi)+\vec b^{(n-1)}$, the function being defined in $\varPhi _{n-2}$. Thus, we have the
    following lemma and a corollary:
    \begin{lemma}\label{L:july9a+}
If $\vec b^{(n-1)}\in K_{n-1}$  and $b_0^{(n-1)}>\epsilon
_{n-2}k^{-2l+1-2\delta }$, then
\begin{enumerate}
\item For any $\varphi \in \varPhi _{n-2}\setminus {\cal
O}^{(n-1)}\left(\vec b^{(n-1)}\right)$:
\begin{equation}\label{3.7.2.2**a}
   \Bigl \|\Bigl(\tilde H^{(n-2)}\left(\vec y^{(n-2)}(\varphi)\right)-k^{2l}\Bigr)^{-1}\Bigr\|
    < \frac{1}{\epsilon _{n-2}^2}
    \end{equation}
    \begin{equation}\label{3.7.2.4**a}
    \Bigl\|\Bigl(\tilde H^{(n-2)}\left(\vec y^{(n-2)}(\varphi)\right)-k^{2l}\Bigr)^{-1}\Bigr\|_1
    <
    \frac{4^{n-2}c_0k^{2+2s_{n-1}}}{\epsilon _{n-2}^2}
    \end{equation}
   \item The  number of zeros of $\det\Bigl(I+\tilde A_{n-2}\bigl(\vec
   y^{(n-2)}(\varphi)\bigr)\Bigr)$  in $\varPhi _{n-2}$ does not exceed $4^{n-2}c_0k^{2+2s_{n-1}}$.
    \end{enumerate}
    \end{lemma}
    \begin{corollary}\label{C:*1+}
    There are no zeros of $\det\Bigl(I+\tilde A_{n-2}\bigl(\vec
    y^{(n-2)}(\varphi)\bigr)\Bigr)$ in  \\ $\varPhi _{n-2}\setminus {\cal
    O}^{(n-1)}\left(\vec b^{(n-1)}\right)$.
    \end{corollary}
\begin{proof} Proof is analogous to that of Lemma \ref{L:july9a}. We  use
Corollary \ref{C:july9a1} for $n=4$ and Corollary \ref{C:+} for
$n>4$, as well as Corollary \ref{C:new1} ($n=4$) and Corollary
\ref{C:new1+} ($n>4$).\footnote{Rigorously speaking, we have to
assume that the lemma holds for $n$-th step. For $(n+1)$-th step
it follows from   Corollaries \ref{C:+} and \ref{C:new1+}.} Part 2
of the lemma follows from the definition of ${\cal
O}^{(n-1)}\left(\vec b^{(n-1)}\right)$ and Corollary
\ref{C:c1}.\end{proof}

Next, we repeat Lemmas \ref{L:s1} -- \ref{L:St3} up to replacement
of indices. Proofs are  completely analogous to those for Step 3.
In
 Lemma \ref{L:St3} we show that the total length of
$\Omega _{n-1}$ does not exceed $k^{-S_n}$, here $\Omega _{n-1}$
is the analog of $\Omega _1$, $\Omega _2$, see (\ref{3.2.1}),
(\ref{4.2.1}). Let $\vec y^{(n-1)}(\varphi)=\vec \varkappa
_{n-1}(\varphi)+\vec b^{(n-1)}$, the function being defined in
$\varPhi _{n-1}$.

\begin{lemma} \label{L:s+} If $\vec b^{(n-1)}\in K_{n-1}$  and $b_0^{(n-1)}>\epsilon _{n-2}k^{-2l+1-2\delta}$, then:
\begin{enumerate}
\item The number of zeros of $\det\Bigl(I+\tilde A_{n-2}\bigl(\vec
y^{(n-1)}(\varphi)\bigr)\Bigr)$ in any $\Gamma ^{(n-1)}\left(\vec
b^{(n-1)}\right)$ is the same as that of $\det\Bigl(I+\tilde
A_{n-2}\bigl(\vec y^{(n-2)}(\varphi)\bigr)\Bigr)$. \item For any
$\varphi \in \varPhi _{n-1}\setminus {\cal O}^{(n-1)}\left(\vec
b^{(n-1)}\right)$, the following inequalities hold:
    \begin{equation}
    \Bigl\|\Bigl(\tilde{H}^{(n-2)}\bigl(\vec y^{(n-1)}(\varphi)\bigr)-k^{2l}
    \Bigr)^{-1}\Bigr\|
    < \frac{2}{\epsilon_{n-2}^2}, \label{*2+}
    \end{equation}
    \begin{equation}
    \Bigl\|\Bigl(\tilde{H}^{(n-2)}\bigl(\vec y^{(n-1)}(\varphi)\bigr)-k^{2l}\Bigr)^{-1}\Bigr\|_1
    <\frac{2\cdot 4^{n-2}c_0k^{2+2s_{n-1}}}{\epsilon_{n-2}^2}.\label{*3+}
    \end{equation}
    \end{enumerate}
\end{lemma}
\begin{corollary} \label{C:May22a*+}
There are no zeros of $\det \Bigl(I+\tilde A_{n-2}\bigl(\vec
y^{(n-1)}(\varphi)\bigr)\Bigr)$ in\\ $\varPhi _{n-1}\setminus
{\cal O}^{(n-1)}\left(\vec b^{(n-1)}\right)$.
\end{corollary}

\begin{proof}
 The proof is analogous to that of Lemma \ref{L:s1}. In Part 1 we use the
estimate $\bigl|\vec y^{(n-1)}(\varphi)-\vec
y^{(n-2)}(\varphi)\bigr|<2\alpha \epsilon_{n-2} ^4k^{-2l+1}$,
which follows from assumed properties of $\chi _{n-1}^*$, see
(\ref{3.75*}) with $3\to n-1$.   We apply (\ref{3.7.2.4**a})
instead of (\ref{3.7.2.4**}) and produce an analog of (\ref{01}),
the estimate $16\cdot 4^{n-2}c_0\epsilon _{n-2}^2k^{2+2s_{n-1}}<1$
following from (\ref{k*}). In Part 2 we use the obvious estimate
$\epsilon _{n-2}^2<1/4$.
\end{proof}


\begin{lemma} If $\vec b^{(n-1)}\in K_{n-1}$  and $b_0^{(n-1)}>\epsilon _{n-2}k^{-2l+1-2\delta }$, then
 the following estimates
hold for any $\varphi \in \varPhi _{n-1}\setminus {\cal
O}^{(n-1)}\left(\vec b^{(n-1)}\right)$:
    \begin{equation}
    \Bigl\|\Bigl(H^{(n-1)}\left(\vec y^{(n-1)}(\varphi)\right)-k^{2l}\Bigr)^{-1}\Bigr\|
    <\frac{4}{\epsilon_{n-2}^2},\label{*4m}
    \end{equation}
    \begin{equation}
    \Bigl\|\Bigl(H^{(n-1)}\left(\vec y^{(n-1)}(\varphi)\right)-k^{2l}\Bigr)^{-1}\Bigr\|_1
    <\frac{4^{n-1}c_0k^{2+2s_{n-1}}}{\epsilon_{n-2}^2}.\label{*5m}
    \end{equation}
\end{lemma}
\begin{corollary} \label{C:Nov29} There are no zeros of $\det\Bigl(I+A_{n-1}\left(\vec y^{(n-1)} (\varphi)\right)\Bigr)$
 in\\ $\varPhi _{n-1}\setminus {\cal
O}^{(n-1)}\left(\vec b^{(n-1)}\right)$.
\end{corollary}

\begin{lemma} If $\vec b^{(n-1)}\in K_{n-1}$  and $b_0^{(n-1)}>\epsilon _{n-2}k^{-2l+1-2\delta }$, then
 the number of zeros of
$\det\Bigl(I+A_{n-1}\bigl(\vec y^{(n-1)} (\varphi)\bigr)\Bigr)$ in
any $\Gamma ^{(n-1)}\left(\vec b^{(n-1)}\right)$ is the same as
that of $\det\Bigl(I+\tilde{A}_{n-2}\bigl(\vec y^{(n-1)}
(\varphi)\bigr)\Bigr)$.
\end{lemma}
\begin{lemma} \label{L:ti12} \begin{enumerate}
\item If  $\vec b^{(n-1)}\in K_{n-1}$  and $b_0^{(n-1)}>\epsilon
_{n-2}k^{-2l+1-2\delta }$, then
 the number of zeros of
$\det\Bigl(I+A_{n-1}\bigl(\vec y^{(n-1)} (\varphi)\bigr)\Bigr)$ in
any $\Gamma ^{(n-1)}(\vec b^{(n-1)})$ is the same as that of
$\det\Bigl(I+\tilde{A}_{n-2}\bigl(\vec y^{(n-2)}
(\varphi)\bigr)\Bigr)$. \item The distance from a zero of
$\det\Bigl(I+A_{n-1}\bigl(\vec y^{(n-1)} (\varphi)\bigr)\Bigr)$
to the nearest zero of $\det\Bigl(I+\tilde{A}_{n-2}\bigl(\vec
y^{(n-2)} (\varphi)\bigr)\Bigr)$ does not exceed $r^{(n-1)}/2$.
\end{enumerate}
\end{lemma}
\begin{corollary}
\label{C:**} The  number of zeros of
$\det\left(I+A_{n-1}\left(\vec y^{(n-1)} (\varphi)\right)\right)$
 does not exceed
$4^{n-2}c_0k^{2+2s_{n-1}}$.
\end{corollary}
 The corollary is the combination
of the first part of the lemma with the second part of Lemma
\ref{L:july9a+}.
The following is an analog of Definition \ref{D:Aug15}.
\begin{definition}\label{D:Aug15+}
 A complex resonance set $\omega
_{n-1}^*(\lambda )$ is the set of $\varphi \in \varPhi _{n-1}$
satisfying
\begin{equation}
\det \left(\dfrac{H^{(n-1)}\bigl(\vec
y^{(n-1)}(\varphi)\bigr)-k^{2l}-\epsilon }{H_0\bigl(\vec
y^{(n-1)}(\varphi)\bigr)+k^{2l}}\right)=0,\ \ \ \vec
y^{(n-1)}(\varphi)=\vec \varkappa _{n-1}(\varphi )+\vec
b^{(n-1)},\label{May18c+1s}
\end{equation}
for some $\vec b^{(n-1)}=2\pi p^{(n-1)}/\hat N_{n-1}a$,
$p^{(n-1)}\in P^{(n-1)}\setminus \{0\}$, and  $|\epsilon
|<\epsilon _{n-1}$.  We denote by ${\omega}_{n-1}(\lambda )$  the
set of $\varphi \in
    \Theta _{n-1}$ corresponding to ${\Omega}_{n-1}(\lambda )$.
    Considering as in the proof of Lemma \ref{L:Sept25}, it is easy to show that $\omega _{n-1}=\omega _{n-1}^*\cap \Theta _{n-1}$.
    \end{definition}

\begin{lemma}\label{L:3.17**}
If $\vec b^{(n-1)}\in K_{n-1}$  and $b_0 ^{(n-1)}\geq \epsilon
_{n-2}k^{-2l+1-2\delta}$, $|\epsilon |< \epsilon_{n-2}^4$, then
\begin{enumerate} \item The determinant of $\ \dfrac{H^{(n-1)}\bigl(\vec
y^{(n-1)}(\varphi)\bigr)-k^{2l}-\epsilon }{H_0\bigl(\vec
y^{(n-1)}(\varphi)\bigr)+k^{2l}}$
 has no zeros in \\ $\varPhi _{n-1}\setminus {\cal O}^{(n-1)}\left(\vec b^{(n-1)}\right)$.
 \item
The number of zeros of  $\det\left(\dfrac{H^{(n-1)}\bigl(\vec
y^{(n-1)}(\varphi)\bigr)-k^{2l}-\epsilon }{H_0\bigl(\vec
y^{(n-1)}(\varphi)\bigr)+k^{2l}}\right)$ in any\\ $\Gamma
^{(n-1)}\left(\vec b^{(n-1)}\right)$ is the same as that of
$\det\left(\dfrac{H^{(n-1)}\bigl(\vec
y^{(n-1)}(\varphi)\bigr)-k^{2l}}{H_0\bigl(\vec
y^{(n-1)}(\varphi)\bigr)+k^{2l}}\right)$.
\end{enumerate}
\end{lemma}
\begin{corollary}\label{C:ut*+} The following relation holds:
$\omega _{n-1}^*\subset {\cal O}^{(n-1)}_*.$
\end{corollary}
\begin{lemma} \label{L:St3+} The estimate
(\ref{4.9*}) holds.\end{lemma}

\begin{proof}
 By Corollary \ref{C:ut*+},
    $\omega_{n-1}^*\subset {\cal O}^{(n-1)}_*$. Hence, $\omega_{n-1}\subset {\cal O}^{(n-1)}_*\cap \Theta
    _{n-1}$.
    By Corollary \ref{C:c2},
     ${\cal O}^{(n-1)}_*$ is formed by no more than
    $4^{n-1}c_0k^{2+2s_{n}}$  disks of the radius
    $r^{(n-1)}=r^{(n-2)}k^{-2-4s_{n-1}-\delta }$. Hence, the total length of
    $\omega_{n-1}$ does not exceed  $2\cdot
    4^{n-1}c_0r^{(n-2)}k^{-\delta }$.
    Considering the recurrent formula for $r^{(m+1)}$, $r^{(m+1)}=
    r^{(m)}k^{-2-4s_{m+1}-\delta }$,
we easily get that $4^{n}c_0r^{(n-2)}<k^{-S_n}$, $n>1$.
     Therefore, the length of $\Omega _{n-1}$ does not exceed
    $\frac{k}{2}k^{-S_n}$.  Considering that  $\chi
    _{n-1}^*(\lambda )$ has a length $2\pi k \bigl(1+o(1)\bigr)$, we obtain
    (\ref{4.9*}).
\end{proof}

\subsection{Nonresonant part of the isoenergetic set of
$H_{\alpha }^{(n)}$}

Now we construct a nonresonance subset  $\chi _n^*(\lambda )$ of
the isoenergetic surface $S_n(\lambda
    )$ of  $H_{\alpha }^{(n)}$ in $K_n$, $S_n(\lambda
    )\subset K_n.$ It
    corresponds to  nonresonance eigenvalues  given by  perturbation series.
     The sets $\chi _1^*(\lambda )$, $\chi _2^*(\lambda
    )$,
    $\chi _3^*(\lambda )$ are defined in the previous steps as well
    as the non-resonance sets $\chi _1(\lambda )$, $\chi _2(\lambda
    )$,
    $\chi _3(\lambda )$. Let us recall that we started with the
    definition of $\chi _1(\lambda )$ and then use it to define $\chi _1^*(\lambda
    )$. Considering $\chi _1^*(\lambda
    )$, we constructed $\chi _2(\lambda
    )$ (Step 2).  Next,  we defined $\chi _2^*(\lambda
    )$, using $\chi _2(\lambda
    )$. Using $\chi _2^*(\lambda
    )$, we introduced $\chi _3(\lambda )$ and, then $\chi _3^*(\lambda
    )$.
    Thus, the process looks like $\chi _1\to \chi _1^*\to \chi
    _2\to \chi _2^*\to \chi _3\to \chi _3^*$. The geometric lemma in
    this section gives us $\chi _4$ and every next $\chi _n$ if $\chi _{n-1}^*$ is defined.
    To ensure the reccurent procedure  we show now how to define
     $\chi _n^*(\lambda )$ using  $\chi _n(\lambda
    )$.

    As in previous sections, we start with the definition of $\cal{D}_{n-1}(\lambda )_{nonres}$.
   First note that $\chi _{n-1}^*$ is defined by
   the formula $\chi _{n-1}^*=\cal{K}_{n-1}\cal{D}_{n-1}$, where $\cal{D}_{n-1}$
   is a distorted circle with holes, and the shift gives one-to-one
   correspondence between $\chi _{n-1}^*$ and $\cal{D}_{n-1}$
   (it is proven for $n=2,3,4$ and we we assume this here for $n>4$ to provide an induction procedure).
    By analogy with previous steps, we define  $\cal{D}_{n-1}(\lambda )_{nonres}$ as
     the preimage of $\chi
    _n$ in $\cal{D}_{n-1}(\lambda )$:
    \begin{equation}
    \cal{D}_{n-1}(\lambda )_{nonres}=\{\vec \varkappa \in \cal{D}_{n-1}(\lambda
    ):\cal{K}_n\vec \varkappa \in \chi _n\}, \label{pp*m}
    \end{equation}
     $\cal{D}_{n-1}(\lambda )_{nonres}$ being  defined
    since $\chi _n \subset \chi _{n-1}^*.$
We define $\cal{B}_n$ as the set of directions corresponding to
$\cal{D}_{n-1}(\lambda
 )_{nonres}$:
$$\cal{B}_n(\lambda)=\{\vec{\nu} \in S_1 :
    \varkappa _{n-1}(\lambda,\vec{\nu})\vec{\nu} \in \cal{D}_{n-1}(\lambda
    )_{nonres}\},$$
 where
    $\varkappa _{n-1}(\lambda ,\vec{\nu})$ is the ``radius" of the distorted
    circle
    $\cal{D}_{n-1}(\lambda
    )$
    defined by the analog of
    Lemma \ref{L:3.9*} for step $n-1$.
Note that $\cal{B}_n$ is a unit circle with holes centered at the
origin and $\cal{B}_n(\lambda ) \subset
    \cal{B}_{n-1}(\lambda )$.
    We denote by $\Theta _n$ the set of angles $\varphi
$ in polar coordinates, corresponding to $\cal{B}_n$:
\begin{equation}
\Theta _n=\{\varphi \in [0,2\pi ):\ (\cos \varphi ,\sin \varphi
)\in \cal{B}_n\} \label{May17a*+},\ \ \ \     \Theta _n\subset
\Theta _{n-1}.
 \end{equation}
We define $\cal{D}_n(\lambda)$ as a level set for
$\lambda^{(n)}(\alpha,\vec{\varkappa})$ in this neighborhood:
    $$\cal{D}_n(\lambda )=\left\{\vec{\varkappa }=\varkappa \vec \nu: \vec \nu \in
    \cal{B}_n,
    \bigl| \varkappa -\varkappa _{n-1}(\lambda,\vec
\nu)\bigr|<\epsilon_{n-1}k^{-2l+1-\delta},
\lambda^{(n)}(\alpha,\vec{\varkappa })=\lambda \right\}.$$
Considering as in the previous step, we prove the analogs of
Lemmas \ref{L:3.8*} and \ref{L:3.9*}. For shortness, we provide
here only the second lemma. By analogy with previous sections, we
shorten notations here:  $\varkappa _{n-1}(\varphi )\equiv
\varkappa _{n-1}(\lambda ,\vec{\nu})$,  $\vec \varkappa
_{n-1}(\varphi )\equiv \varkappa _{n-1}(\lambda
,\vec{\nu})\vec{\nu}$.
\begin{lemma}\label{L:3.9*+} \begin{enumerate}
\item The set $\cal{D}_n(\lambda )$ is a distorted circle with
holes: it can be described by the formula:
\begin{equation}
\cal{D}_n(\lambda )=\bigl\{\vec \varkappa  \in \R^2: \vec
\varkappa
 =\vec \varkappa _{n}(\varphi ),\ \ \varphi
\in \Theta _n(\lambda)\bigr\},\label{May20a+}
\end{equation} where
    $ \varkappa _{n}(\varphi )=
    \varkappa _{n-1}(\varphi )+h_n(\varphi), $
and $h_n(\varphi )$ satisfies the estimates
    \begin{equation}\label{3.75*+}
    |h_n|<2\alpha \epsilon_{n-1} ^4k^{-2l+1},\ \ \
    \left|\frac{\partial h_n}{\partial \varphi} \right| \leq
    4\alpha \epsilon_{n-1}^3 k^{1+\delta}.
    \end{equation}
     \item The total length of $\cal{B}_n(\lambda)$ satisfies the estimate:
\begin{equation}\label{theta2*}
    L\left(\cal{B}_{n-1}\setminus \cal{B}_{n}\right)<4\pi k^{-S_n}.
    \end{equation}
\item Function $\varkappa _{n}(\varphi )$ can be extended as a
holomorphic function of $\varphi $ to the complex non-resonance
set $\varPhi _{n}$, estimates (\ref{3.75*+}) being preserved.
\item The curve $\cal{D}_n(\lambda )$ has a length which is
asymptotically close to that of $\cal{D}_{n-1}(\lambda )$ in the
following sense:
    \begin{equation}\label{3.77mm}
     L\Bigl(\cal{D}_n(\lambda )\Bigr)\underset{\lambda \rightarrow
     \infty}{=}L\Bigl(\cal{D}_{n-1}(\lambda )\Bigr)\Bigl(1+O(k^{-S_n})\Bigr).
     \end{equation}
     \end{enumerate}
\end{lemma}

Now define the nonresonance set, $\chi_n^*(\lambda )$ in
$S_n(\lambda )$ by the formula analogous to (\ref{3.77.7*}).
Indeed,
    \begin{equation}\label{3.77.7*+}
    \chi_n^*(\lambda ):=\cal{K}_n\cal{D}_n(\lambda ).
    \end{equation}
The following lemmas are analogous to Lemmas \ref{Apr4a*} and
\ref{L:May10a**}.

    \begin{lemma}\label{Apr4a*+}
    The set $\chi_n^*(\lambda )$ belongs to the
    $\left(2\alpha \epsilon _{n-1}^4k^{-2l+1}\right)$-neighborhood of $\chi_n(\lambda
    )$ in $K_n$. If $t^{(n)}\in \chi_n^*(\lambda )$, then the operator
    $H^{(n)}_{\alpha }(t^{(n)})$ has a simple eigenvalue
   equal to $\lambda $. This
    eigenvalue is given by the perturbation series analogous to (\ref{4.15}).
\end{lemma}

    \begin{lemma}\label{L:May10a**+} Formula (\ref{3.77.7*+}) establishes
    one-to-one correspondence between $\chi_n^*(\lambda )$ and $\cal{D}_n(\lambda
    )$.\end{lemma}
    \subsection{Preparation for the Next Approximation} \label{S:3.6*m}
\subsubsection{Contracted set ${\cal O}^{(n-1)}_s\bigl(\vec b^{(n-1)}\bigr)$}
 We constructed  the set ${\cal
O}^{(n-1)}\bigl(\vec b^{(n-1)}\bigr)$, which  surrounds zeros of
the determinants
$$\det\Bigl(I+ A_{n-2}\left(\vec \varkappa _{n-2}(\varphi)+\vec
b^{(n-1)}+2\pi p^{(n-2)}/\hat N_{n-2}a \right)\Bigr),\ \
p^{(n-2)}\in P^{(n-2)},$$ or, shorter, zeros of the determinant
$\det\Bigl(I+\tilde A_{n-2}\left( \vec y
^{(n-2)}(\varphi)\right)\Bigr)$, here $\vec y
^{(n-2)}(\varphi)=\vec \varkappa _{n-2}(\varphi)+\vec b^{(n-1)}$,
$\tilde A_{n-2}$ corresponds to $\tilde H^{(n-2)}$. The set
consists of disks of
 the radius
 $r^{(n-1)}$ and contains no more than
 $4^{n-2}c_0k^{2+2s_{n-1}}$ disks.
We proved that the  new determinant $\det\Bigl(I+A_{n-1}\bigl(\vec
y^{(n-1)}(\varphi)\bigr)\Bigr)$, $\vec y^{(n-1)}(\varphi)=\vec
\varkappa _{n-1}(\varphi)+\vec b ^{(n-1)}$,
has the same number of zeros  as  the old one inside  each $\Gamma
^{(n-1)}\bigl(\vec b^{(n-1)}\bigr)$ (Lemma \ref{L:ti12}); and both
have no zeros in $\in \varPhi _{n-1}\setminus {\cal
O}^{(n-1)}\bigl(\vec b^{(n-1)}\bigr)$ (Corollaries \ref{C:*1+},
\ref{C:Nov29}).
To prepare the next approximation, we contract the set ${\cal
O}^{(n-1)}_s\bigl(\vec b^{(n-1)}\bigr)$ around the zeros of the
new determinant. We consider the disks with the radius
$r^{(n)}=r^{(n-1)}k^{-2-4s_{n}-\delta }$ centered at
zeros of the new determinant. Obviously each of these disks is in
${\cal O}^{(n-1)}\bigl(\vec b^{(n-1)}\bigr)$, since
$r^{(n)}<<r^{(n-1)}$ and the distance between zeros of the old and
new determinants is smaller than $r^{(n-1)}/2$ (Lemma
\ref{L:ti12}). Next, we take the union
 of these smaller disks and denote it by ${\cal O}_s^{(n-1)}\bigl(\vec b^{(n-1)}
 \bigr)$.
Obviously, the following lemma holds. \begin{lemma} \label{L:obvi}
The set ${\cal O}_s^{(n-1)}\bigl(\vec b^{(n-1)}\bigr)$ belongs to
${\cal O}^{(n-1)}\bigl(\vec b^{(n-1)}\bigr)$ and  consists of the
same number of disks as ${\cal O}^{(n-1)}\bigl(\vec
b^{(n-1)}\bigr)$.
\end{lemma} \begin{corollary} \label{C:obvi} The set ${\cal
O}_s^{(n-1)}\bigl(\vec b^{(n-1)}\bigr)$ contains no more than
$4^{n-2}c_0k^{2+2s_{n-1}}$ disks.
\end{corollary}
The corollary follows from the lemma and Corollary \ref{C:c1}.

\begin{lemma}\label{L:July9b*+}
For any $\varphi \in \varPhi _{n-1}\setminus {\cal
O}_s^{(n-1)}\bigl(\vec b^{(n-1)}\bigr)$,
    \begin{equation}\label{3.7.2.2*+}
    \Bigl\|\Bigl(H^{(n-1)}\bigl(\vec y^{(n-1)}(\varphi)\bigr)-k^{2l}\Bigr)^{-1}\Bigr\| \leq
\frac{4}{\epsilon ^2
_{n-2}}\left(\frac{2r^{(n-1)}J^{(n-1)}}{r^{(n)}}\right)
    ^{J^{(n-1)}},
    \end{equation}
    \begin{equation}\label{3.7.2.4*+}
    \Bigl\|\Bigl(H^{(n-1)}\bigl(\vec y^{(n-1)}(\varphi)\bigr)-k^{2l}\Bigr)^{-1}\Bigr\|_1 \leq
    \frac{4J^{(n-1)}}{\epsilon _{n-2}^2}
    \left(\frac{2r^{(n-1)}J^{(n-1)}}{r^{(n)}}\right)^{J^{(n-1)}},
    \end{equation}
      $$ J^{(n-1)}=4^{n-2}c_0k^{2+2s_{n-1}},\ \ \epsilon
_{n-2}=e^{-\frac{1}{4}k^{\eta s_{n-2}}}.$$
\end{lemma}
The proof is analogous to that of Lemma \ref{L:July9b}. Using the
recurrent formula for $r^{(n)}$, condition (\ref{k*}) and a simple
induction procedure, we easily obtain that the right-hand part of
(\ref{3.7.2.2*+}) does not exceed $1/\epsilon _{n-1}$. Hence, the
following is true:
\begin{corollary}\label{C:+}
\begin{equation}\label{3.7.2.2**+}
    \Bigl\|\Bigl(H^{(n-1)}\bigl(\vec y^{(n-1)}(\varphi)\bigr)-k^{2l}\Bigr)^{-1}\Bigr\|
    <\frac{1}{\epsilon _{n-1}}
    \end{equation}
    \begin{equation}\label{3.7.2.4**+}
    \Bigl\|\Bigl(H^{(n-1)}\bigl(\vec y^{(n-1)}(\varphi)\bigr)-k^{2l}\Bigr)^{-1}\Bigr\|_1
    \leq \frac{ J^{(n-1)}}{\epsilon _{n-1}}.
    \end{equation}
    \end{corollary}
\subsubsection{The set ${\cal O}_s^{(n-1)}\bigl(\vec b^{(n-1)}\bigr)$ for
small $\vec b^{(n-1)}$}
 Everything  we  considered so far is valid if
    $b_0 ^{(n-1)}>\epsilon _{n-2}k^{-2l+1-2\delta }$.
 However, at each step $b_0^{(n-1)}$ is taken smaller, since the reciprocal
lattice is getting finer. For this reason we now consider $\vec
b^{(n-1)}$ being close to a vertex of $K_{n-1}$: $0<b_0
^{(n-1)}\leq \epsilon _{n-2}k^{-2l+1-2\delta }$.
As in previous steps, we  show  that for such $\vec b^{(n-1)}$ the
resolvent $\Bigl(H^{(n-1)}\left(\vec y^{(n-1)}(\varphi
)\right)-k^{2l}\Bigr)^{-1}$
has no more than two poles  in $\varPhi _{n-1}$ and even its small
neighborhood. We  surround these poles by two contours $\gamma
^{\pm \ (n-1)}_s$ and obtain an estimate for the norm of
$\Bigl(H^{(n-1)}\left(\vec y^{(n-1)}(\varphi
)\right)-k^{2l}\Bigr)^{-1}$ when $\varphi $ is outside these
contours.\footnote{If $|\vec b^{(n-1)}|=b_0^{(n-1)}$, then $\vec y
^{(n-1)}(\varphi )=\vec \varkappa _{n-1}(\varphi )+\vec
b^{(n-1)}$. For $\vec b^{(n-1)}$ being close to a vertex $\vec e$
of $K_{n-1}$ other than $(0,0)$, we take $\vec y^{(n-1)}(\varphi
)=\vec \varkappa _{n-1}(\varphi )+\vec b^{(n-1)}-\vec e$.}

By analogy with Step 3, $\tilde \varPhi _{n-1}$ is the
$\frac{r^{(n-2)}}{2}$-neighborhood of $\varPhi _{n-1}$. Obviously,
$\tilde \varPhi _{n-1} \subset \tilde \varPhi _{n-2}$. We assume
that the analog of Lemma \ref{L:3.7.1.1m} holds when we replace
the index $(2)$ by $(n-2)$. Namely, we assume that for any
$|\epsilon_0|<b_0^{(n-2)}k^{2l-1-\delta }r^{(n-1)}$ the equation
$\lambda ^{(n-2)}(\vec y ^{(n-2)}(\varphi))=k^{2l}+\epsilon_0$
 has no more than
two solutions $\varphi ^{\pm (n-2)}_{\epsilon _0}$ in $\tilde
\varPhi _{n-2}$ and the inequality $|\varphi^{\pm (n-2)}_{\epsilon
_0}-\varphi^{\pm (n-3)}_{0}|
    <r^{(n-1)}/4$ holds, here
    $\varphi^{\pm
(n-3)}_{0}$ is $\varphi^{\pm (n-3)}_{\epsilon _0}$ for $\epsilon
_0=0$. We also assume that the estimate analogous to
(\ref{3.7.1.6.1/2+}) holds when we replace $(2)$ by $(n-2)$. Let
us formulate the corresponding results for the next step of
approximation ($n-2\to n-1$). The formulations and proofs of the
following four lemmas and are analogous to those for Lemmas
\ref{L:3.7.1.1m} -- \ref{L:july9c*}. Definition \ref{D:sat} is
analogous to Definition \ref{SN1}, and  Corollary \ref{C:new1+} is
an analog of Corollary \ref{C:new1}.

\begin{lemma}\label{L:3.7.1.1++}
If $0<b_0 ^{(n-1)}\leq \epsilon _{n-2}k^{-2l+1-2\delta }$ and
$|\epsilon_0|<b_0^{(n-1)}k^{2l-1-\delta}r^{(n)}$, then the
equation
    \begin{equation}\label{3.7.1.2*+}
    \lambda ^{(n-1)}\left(\vec y ^{(n-1)}(\varphi)\right)=k^{2l}+\epsilon_0
    \end{equation}
 has no more than two
solutions, $\varphi^{\pm\ (n-1)}_{\epsilon _0}$, in $\tilde
\varPhi _{n-1}$. For any $\varphi^{\pm\ (n-1)}_{\epsilon _0}$
there is $\varphi^{\pm\ (n-2)}_{0}\in \tilde \varPhi _{n-2}$ such
that
    \begin{equation}
    |\varphi^{\pm (n-1)}_{\epsilon _0}-\varphi^{\pm (n-2)}_{0}|
    <r^{(n)}/4. \label{july5a*+}
    \end{equation}
\end{lemma}
\begin{lemma}\label{L:july5a*+} Suppose $0<b_0 ^{(n-1)}\leq \epsilon _{n-2}k^{-2l+1-2\delta }$ and
 $\varphi \in \tilde \varPhi
_{n-1}$ obeys the inequality analogous to (\ref{july5a*+}):
$|\varphi-\varphi^{\pm (n-2)}_{0}|
    <r^{(n)}$. Then,
    \begin{equation}\label{3.7.1.6.1/2+m}
    \frac{\partial}{\partial \varphi}\lambda^{(n-1)}\left(\vec
    y^{(n-1)}(\varphi)\right)=_{k\to
    \infty}\pm 2lb_0^{(n-1)}k^{2l-1}\bigl(1+o(1)\bigr).
    \end{equation}
    \end{lemma}

\begin{definition} \label{D:sat} Let $\Gamma^{\pm (n-1)}_s\bigl(\vec b^{(n-1)}\bigr)$ be the
open disks centered at $\varphi_0^{\pm (n-2)}$ with radius
$r^{(n)}$; $\gamma^{\pm (n-1)}_s\bigl(\vec b^{(n-1)}\bigr)$ be
their boundary circles and ${\cal O}_s^{(n-1)}\bigl(\vec
b^{(n-1)}\bigr)=\Gamma^{+(n-1)}_s\cup \Gamma^{-(n-1)}_s$.
\end{definition}

\begin{lemma}\label{L:3.7.1*+}
For any $\varphi$ in $\varPhi _{n-1}\setminus {\cal
O}_s^{(n-1)}(\vec b^{(n-1)})$,
    \begin{equation}\label{3.7.1.10*m}
    \left|\lambda^{(n-1)}\left(\vec
    y^{(n-1)}(\varphi)\right)-k^{2l}\right|\geq b_0^{(n-1)}k^{2l-1-\delta}r^{(n)}.
    \end{equation}
\end{lemma}
\begin{lemma}\label{L:july9c*+}
For any $\varphi \in \varPhi _{n-1}\setminus {\cal
O}_s^{(n-1)}\bigl(\vec b^{(n-1)}\bigr)$,
    \begin{equation}\label{3.7.1.10.1/2*+}
    \Bigl\|\Bigl(H^{(n-1)}\left(\vec
    y^{(n-1)}(\varphi)\right)-k^{2l}\Bigr)^{-1}\Bigr\|<\frac{16}{b_0^{(n-1)}r^{(n)}k^{2l-1-\delta}},
    \end{equation}
    \begin{equation}\label{3.7.1.13j}
    \Bigl\|\Bigl(H^{(n-1)}\left(\vec y^{(n-1)}(\varphi)\right)-k^{2l}\Bigr)^{-1}\Bigr\|_1<
    \frac{16\cdot 4^{n-2}c_0k^{2+2s_{n-1}}}{b_0^{(n-1)}r^{(n)}k^{2l-1-\delta }}.
    \end{equation}
\end{lemma}
\begin{corollary} \label{C:new1+}
If $\epsilon _{n-1}k^{-2l+1-2\delta }<b_0^{(n-1)}\leq  \epsilon
_{n-2}k^{-2l+1-2\delta }$, and $\varphi$ belongs to $\varPhi
_{n-1}\setminus {\cal O}_s^{(n-1)}\left(\vec b^{(n-1)}\right)$,
then
    \begin{equation}\label{3.7.1.10.1/2**+}
    \Bigl\|\Bigl(H^{(n-1)}\left(\vec
    y^{(n-1)}(\varphi)\right)-k^{2l}\Bigr)^{-1}\Bigr\|<\frac{1}{\epsilon _{n-1}^2},
    \end{equation}
    \begin{equation}\label{3.7.1.13**+}
    \Bigl\|\Bigl(H^{(n-1)}\left(\vec y^{(n-1)}(\varphi)\right)-k^{2l}\Bigr)^{-1}\Bigr\|_1<
    \frac{4^{n-2}c_0k^{2+2s_{n-1}}}{\epsilon _{n-1}^2}.
    \end{equation}
    \end{corollary}
    The corollary follows from the condition on $b_0^{(n-1)}$,
     estimate (\ref{k*}) and the formula $r^{(n)}=r^{(n-1)}k^{-2-4s_{n}-\delta }$.

\section{Limit-Isoenergetic Set and Eigenfunctions} \label{chapt7}

\setcounter{equation}{0}

\subsection{Limit-Isoenergetic Set and Proof of the Bethe-Sommerfeld
Conjecture} At every step $n$ we constructed  a set
$\cal{B}_n(\lambda)$, $\cal{B}_{n}(\lambda)\subset
\cal{B}_{n-1}(\lambda)\subset S_1(\lambda)$,  and a function
$\varkappa _n(\lambda,\vec{\nu})$, $\vec{\nu} \in
\cal{B}_n(\lambda)$, with the following properties. The set
$\cal{D}_{n}(\lambda )$ of vectors $\vec{\varkappa}=\varkappa
_n(\lambda ,\vec{\nu})\vec{\nu}$,
   $\vec{\nu} \in \cal{B}_{n}(\lambda )$,
    is a slightly distorted circle with holes,  see Fig.\ref{F:1}, Fig.\ref{F:2}, formula (\ref{Dn})
    and Lemmas \ref{L:2.13}, \ref{L:3.9},
\ref{L:3.9*}, \ref{L:3.9*+}.
For any $\vec \varkappa _n(\lambda,\vec{\nu})\in
\cal{D}_{n}(\lambda )$ there is a single eigenvalue of
 $H^{(n)}(\vec \varkappa _n)$
equal to $\lambda $ and  given by a perturbation series.~
\footnote{The operator $H^ {(n)}(\vec \varkappa )$ is defined for
every $\vec \varkappa \in \R^2$ as explained in Remark
\ref{R:May20}, page \pageref{R:May20}. The perturbation series is
given by a formula analogous to (\ref{3.66}), which coincides with
(\ref{3.15}) up to a shift of indices corresponding to the
parallel shift of $\vec \varkappa $ into $K_n$.} Let
    $\cal{B}_{\infty}(\lambda)=\bigcap_{n=1}^{\infty}\cal{B}_n(\lambda).$
Since $\cal{B}_{n+1} \subset \cal{B}_n$ for every $n$,
$\cal{B}_{\infty}(\lambda)$ is a unit circle with infinite number
of holes, more and more holes of smaller and smaller size
appearing at each step. \begin{lemma} \label{L:Dec9} The length of
$\cal{B}_{\infty}(\lambda)$ satisfies estimate (\ref{B}) with
$\gamma _3=\delta /2$.
\end{lemma}
\begin{proof}
 Using (\ref{theta1}),  (\ref{theta2}), (\ref{theta2*s})
and (\ref{theta2*}) and consideirng that $S_n\approx 2^ns_1$, we
easily conclude that
$L\left(\cal{B}_n\right)=\left(1+O(k^{-\delta/2})\right)$,
$k=\lambda ^{1/2l}$ uniformly in $n$. Since $\cal{B}_n$ is a
decreasing sequence of sets, (\ref{B}) holds.
\end{proof} Let us consider
    $\varkappa _{\infty}(\lambda, \vec{\nu})=\lim_{n \to \infty}\varkappa _n(\lambda,\vec{\nu}),\quad
    \vec{\nu} \in \cal{B}_{\infty}(\lambda ).$
    \begin{lemma} The limit $\varkappa _{\infty}(\lambda,
    \vec{\nu})$ exists for any $\vec{\nu} \in \cal{B}_{\infty}(\lambda
    )$ and the following estimates hold when $n\geq 1$:
    \begin{equation}\label{6.1}
    \left|\varkappa _{\infty}(\lambda, \vec{\nu})-\varkappa
    _n(\lambda,\vec{\nu})\right|<4\epsilon_{n} ^4k^{-2l+1},\ \
    \epsilon _{n}=\exp(-\frac{1}{4}k^{\eta
    s_{n}}),\ \
    s_n=2^{n-1}s_1.
    \end{equation}
    \end{lemma}
    \begin{corollary}\label{Dec18}
    For every $\vec{\nu} \in \cal{B}_{\infty}(\lambda)$ estimate (\ref{h}) holds, where\\ $\gamma
_4=(4l-3-4s_1-3\delta)/2l
    >0$.
    \end{corollary}
    The lemma easily follows from
  the estimates (\ref{3.75}), (\ref{3.75*}) and (\ref{3.75*+}). To obtain corollary we use
 (\ref{2.75}) and take into account that $\gamma
 _0=2l-2-4s_1-2\delta $.

    Estimates (\ref{3.75}), (\ref{3.75*}) and (\ref{3.75*+}) justify convergence of the series $\sum_{m=1}^{\infty}
    \frac{\partial h_n}{\partial \varphi },$ and hence,
    of the sequence $\frac{\partial \varkappa _n}{\partial \varphi }.$
    We denote the limit of this sequence by $\frac{\partial \varkappa _{\infty}}{\partial \varphi }.$
    \begin{lemma} The  estimate (\ref{Dec9a}) with $\gamma
_5=(4l-5-8s_1-4\delta)/2l >0,$ holds for any $\vec \nu \in
    \cal{B}_{\infty}(\lambda)$.
   \end{lemma}
We define $\cal{D}_{\infty}(\lambda )$ by (\ref{D}). Clearly,
$\cal{D}_{\infty}(\lambda )$ is a slightly distorted circle of
radius $k$ with infinite number of holes. We can assign a tangent
vector $\frac{\partial \varkappa }{\partial \varphi }\vec \nu
+\varkappa \vec \mu $, $\vec \mu =(-\sin \varphi ,\cos \varphi )$
to the curve $\cal{D}_{\infty}(\lambda )$, this tangent vector
being the limit of corresponding tangent vectors for  curves
$\cal{D}_{n}(\lambda )$ at points $\vec \varkappa _n(\lambda ,\vec
\nu )$ as $n\to \infty $.

 \begin{remark} \label{R:Dec9} We easily see from (\ref{6.1}), that any $\vec{\varkappa}
\in \cal{D}_{\infty}(\lambda )$ belongs to the
$\left(4\epsilon_{n} ^4k^{-2l+1}\right)$-neighborhood of
$\cal{D}_n(\lambda )$. Applying perturbation formulae for $n$-th
step, we easily obtain that  there is an
 eigenvalue  $\lambda^{(n)}(\vec \varkappa )$ of $H^{(n)}(\vec \varkappa )$
satisfying the estimate $\lambda^{(n)}(\vec \varkappa )=\lambda
+\delta _n$, $\delta _n=O\left(\epsilon _{n}^4\right)$, the
eigenvalue $\lambda^{(n)}(\vec \varkappa )$
     being given by a perturbation
series of the type (\ref{3.66}). Hence, for every $\vec{\varkappa}
\in \cal{D}_{\infty}(\lambda)$ there is a limit:
\begin{equation} \lim _{n\to \infty }\lambda^{(n)}(\vec \varkappa
)=\lambda.\label{6.2}
\end{equation}
\end{remark}

\begin{theorem}[Bethe-Sommerfeld Conjecture]
The spectrum of  operator $H$ contains a semi-axis.
\end{theorem}
\begin{proof}
    By Remark \ref{R:Dec9}, there is a point of the spectrum of
$H_n$ in the $\delta _n$-neighborhood of $\lambda $ for every
$\lambda >k_*^{2l}$, $k_*$ being introduced by (\ref{k}). Since
$\|H_n-H\|<\epsilon_{n}^4$, there is a point of the spectrum of
$H$ in the $\delta _n^*$-neighborhood of $\lambda $,  $\delta
_n^*=\delta _n+\epsilon_{n}^4$. Since it is true for every $n$ and
the spectrum of $H$ is closed, $\lambda $ is in the spectrum of
$H$.
\end{proof}

\subsection{Generalized Eigenfunctions of $H$}
A plane wave is usually denoted by $e^{i\langle \vec k, x
    \rangle}$, $\vec k \in \R^2$. Here we  use $\vec \varkappa $ instead of
    $\vec k$ to comply with our previous notations. We show that for
    every  $\vec \varkappa $ in
a set $$\cal{G} _{\infty }=\cup _{\lambda
>\lambda _*}\cal{D}_{\infty}(\lambda ),\ \ \lambda _*=k_*^{2l},$$ there is a solution $\Psi _{\infty }(\vec \varkappa , x)$
of the equation for eigenfunctions:
    \begin{equation} (-\Delta)^{2l}\Psi _{\infty}(\vec \varkappa , x)+V(x)\Psi _{\infty }(\vec \varkappa ,
    x)=\lambda _{\infty}(\vec \varkappa )\Psi _{\infty }(\vec \varkappa , x),
    \label{6.2.1}
    \end{equation}
which can be represented in the form
    \begin{equation}
    \Psi _{\infty }(\vec \varkappa , x)=e^{i\langle \vec \varkappa , x
    \rangle}\Bigl(1+u_{\infty}(\vec \varkappa , x)\Bigr),\ \ \ \
    \bigl\|u_{\infty}(\vec \varkappa , x))\bigr\| _{L_{\infty }(\R^2)}<c|\vec \varkappa |^{-\gamma_1},
   \label{6.2.1a}
    \end{equation}
where $u_{\infty}(\vec \varkappa , x)$ is a limit-periodic
function,
 $\gamma _1=2l-4-7s_1-2\delta >0$; the eigenvalue $\lambda _{\infty}(\vec \varkappa )$ satisfies the asymptotic
formula:
 \begin{equation}\lambda _{\infty}(\vec \varkappa )=|\vec \varkappa |^{2l}+O(|\vec
\varkappa |^{-\gamma _2}), \ \ \ \gamma _2=2l-2-4s_1-3\delta
>0.\label{6.2.4}
\end{equation}
We also show that the set $\cal{G} _{\infty }$ satisfies
(\ref{full}).

 In fact,   by (\ref{6.1}), any $\vec{\varkappa} \in \cal{D}_{\infty}(\lambda
)$ belongs to the $(\epsilon_n k^{-2l+1-\delta})$-neighborhood of
$\cal{D}_n(\lambda )$. Applying the perturbation formulae proved
in the previous sections, we obtain the following inequalities:
    \begin{equation}\bigl\|E^{(1)}(\vec{\varkappa})-{E}^{(0)}(\vec{\varkappa})\bigr\|_1<2k^{-\gamma_0},
    \ \ \ \ \bigl\|E^{(n+1)}(\vec{\varkappa})-\tilde{E}^{(n)}(\vec{\varkappa})\bigr\|_1<
    48\epsilon_n^3, \quad n \geq 1,\label{6.2.2}
    \end{equation}
    \begin{equation}\bigl|\lambda ^{(1)}(\vec{\varkappa})-|\vec \varkappa |^{2l}
     \bigr|
     <2k^{-\gamma _2}
    , \ \ \ \ \bigl|\lambda ^{(n+1)}(\vec \varkappa )-\lambda ^{(n)}(\vec \varkappa )\bigr|<12\epsilon _n^4,
     \quad n \geq 1,
    \label{6.2.3}
    \end{equation}
where $E^{(n+1)},\ \tilde{E}^{(n)}$ are one-dimensional spectral
projectors in $L_2(Q_{n+1})$ corresponding to potentials $W_{n+1}$
and $W_n$, respectively; $\lambda ^{(n+1)}(\vec \varkappa )$ is
the eigenvalue corresponding to $E^{(n+1)}(\vec{\varkappa})$,
${E}^{(0)}(\vec{\varkappa})$ corresponds to $V=0$ and the periods
$a_1$, $a_2$. This means that for properly chosen eigenfunctions
$\Psi _{n+1}(\vec \varkappa ,x)$:
\begin{equation}\label{psi1}
    \|\Psi _{1}-\Psi _0\|_{L_2(Q_{1})}<4k^{-\gamma_0}|Q_1|^{1/2},\ \ \ \Psi _0(x)=e^{i \langle \vec \varkappa ,x
    \rangle},
    \end{equation}
    \begin{equation}
    \|\Psi _{n+1}-\tilde \Psi _n\|_{L_2(Q_{n+1})} < 100\epsilon_n^{3}|Q_{n+1}|
    ^{1/2},\label{Dec9c}
    \end{equation}
    where $\tilde \Psi _n $
    is $\Psi _n $ extended quasi-periodically from $Q_n$ to
    $Q_{n+1}$. Eigenfunctions $\Psi _{n}$, $n\geq 1$, are chosen to
    obey
    two
    conditions. First, $\|\Psi _{n}\|_{L_2(Q_{n})}=|Q_{n}|^{1/2}$; \footnote{The condition
    $\|\Psi _{n}\|_{L_2(Q_{n})}=|Q_{n}|^{1/2}$ implies
    $\|\tilde \Psi _{n}\|_{L_2(Q_{n+1})}=|Q_{n+1}|^{1/2}$.} second
    $\text{Im}
    (\Psi _{n},\tilde \Psi _{n-1})=0$, here    $(\cdot
    ,\cdot)$ is an inner product in $L_2(Q_{n})$. These two
    conditions, obviously, provide a unique choice of each $\Psi
    _n$.
    Considering that $\Psi _{n+1}$ and $\tilde \Psi _n$ satisfy  equations
for eigenfunctions and taking into account (\ref{6.2.3}),
(\ref{Dec9c}) we obtain:
    $\|\Psi _{n+1}-\tilde \Psi _n\|_{W_2^{2l}(Q_{n+1})} <
    ck^{2l}\epsilon_n^{3}|Q_{n+1}|^{1/2}$, $n \geq
    1$
and, hence,
    $\|\Psi _{n+1}-\tilde \Psi _n\|_{L_{\infty}(Q_{n+1})}
    < ck^{2l}\epsilon_n^{3}|Q_{n+1}|^{1/2}$.  Since $\Psi _{n+1}$ and $\tilde \Psi _n$ obey the same
quasiperiodic conditions, the same inequality holds in the whole
space $\R^2$:
    \begin{equation}
    \|\Psi _{n+1}-\Psi _n\|_{L_{\infty}(\R^2)}
    < ck^{2l}\epsilon_n^{3}|Q_{n+1}|^{1/2}, \ \ n \geq 1, \label{Dec10}
    \end{equation}
where $\Psi _{n+1},\Psi _n$ are quasiperiodically extended to
$\R^2$. Obviously, we have a Cauchy sequence in
$L_{\infty}(\R^2)$. Let $\Psi _{\infty }(\vec \varkappa
,x)=\lim_{n \to \infty}\Psi _n(\vec \varkappa, x).$ This limit is
defined pointwise uniformly in $x$ and in $W_{2,loc}^{2l}(\R^2)$.

\begin{theorem} \label{T:Dec10}
For every sufficiently large $\lambda $ and $\vec{\varkappa} \in
\cal{D}_{\infty}(\lambda)$ the sequence of functions $\Psi
_n(\vec{\varkappa},x)$ converges  in $L_{\infty}(\R^2)$ and
$W_{2,loc}^{2l}(\R^2)$. The limit function $\Psi _{\infty
}(\vec{\varkappa},x)$,
    $\Psi _{\infty }(\vec{\varkappa},x)=\lim_{n\to \infty }
    \Psi _n(\vec{\varkappa},x)$,
     satisfies the equation
    \begin{equation}\label{6.7}
     (-\Delta)^{2l}\Psi _{\infty }(\vec{\varkappa}, x)+V(x)\Psi _{\infty }(\vec{\varkappa},
    x)= \lambda \Psi _{\infty }(\vec{\varkappa}, x).
    \end{equation}
 It can be represented in the form
   \begin{equation}\label{6.4}
    \Psi _{\infty }(\vec{\varkappa},x)=e^{i\langle \vec{\varkappa}, x
    \rangle}\bigl(1+u_{\infty}(\vec{\varkappa}, x)\bigr),
    \end{equation}
where $u_{\infty}(\vec{\varkappa}, x)$ is a limit-periodic
function:
    \begin{equation}\label{6.5}
    u_{\infty}(\vec{\varkappa}, x)=\sum_{n=1}^{\infty}\tilde u_n(\vec{\varkappa},
    x),
    \end{equation}
 $\tilde u_n(\vec{\varkappa}, x)$  being a periodic function with the periods
$2^{M_n-1}b_1, 2^{M_n-1}b_2$, $2^{M_n}\approx k^{2^{n-1}s_1}$,
\begin{equation}
    \|\tilde{u}_1\|_{L_{\infty}(\R^2)} <ck^{-\gamma_1}, \ \ \
    \gamma _1=2l-4-7s_1-2\delta >0, \label{6.6a}
    \end{equation}
    \begin{equation}\label{6.6}
    \|\tilde{u}_n\|_{L_{\infty}(\R^2)} <ck^{2l}\epsilon_{n-1}^{3}|Q_{n}|^{1/2}, \ \ \ n\geq
    2.\end{equation}
\end{theorem}
\begin{corollary} Function $u_{\infty}(\vec{\varkappa}, x)$ obeys the
 estimate (\ref{6.2.1a}).
\end{corollary}
\begin{remark} If $V$ is sufficiently smooth, say $V\in C^1(R)$,
then  estimate (\ref{6.6a}) and, hence, (\ref{6.2.1a}) can be
improved by replacing $\gamma _1$ by $\gamma _0$.
\end{remark}

\begin{proof}
 Let us show that $\Psi _{\infty }$ is a
limit-periodic function. Obviously,
    $ \Psi _{\infty } =\Psi _0+\sum_{n=0}^{\infty}(\Psi_{n+1}-\Psi _n)
$, the series converging in $L_{\infty}(\R^2)$ by (\ref{Dec10}).
Introducing the notation $u_{n+1}=e^{-i \langle \vec \varkappa
,x\rangle}(\Psi_{n+1}-\Psi _n)$, we arrive at (\ref{6.4}),
(\ref{6.5}). Note that
 $\tilde{u}_n$ is periodic with  the periods $2^{M_n-1}b_1,
 2^{M_n-1}b_2$. Estimate (\ref{6.6})  follows from (\ref{Dec10}). We check
 (\ref{6.6a}). Indeed,  by (\ref{psi1}), Fourier coefficients
 $(u_1)_j$, $j\in \Z^2$, satisfy the estimate
 $\left|(\tilde{u}_1)_j\right|<ck^{-\gamma_0}|Q_1|^{1/2}<ck^{-\gamma_0+s_1}$. This estimate can be
 easily improved for $j:p_j(0)>2k$: $\left|(\tilde{u}_1)_j\right|<c|j|^{-2l}$. Summarizing these
 inequalities and taking into account that the number of $j:p_j(0)\leq 2k$
 does not exceed $c_0k^{2+2s_1}$, we arrive at
 (\ref{6.6a}).
It remains to prove (\ref{6.7}). Indeed, $\Psi _n(\vec \varkappa,
x)$, $n\geq 1$, satisfy equations for eigenfunctions: $H^{n)}\Psi
_n=\lambda ^{(n)}(\vec \varkappa )\Psi _n $.
Considering that $\Psi _n(\vec \varkappa,x)$ converges to $\Psi
(\vec \varkappa, x)$ in $W_{2l,loc}^2$ and relation (\ref{6.2}),
we arrive at (\ref{6.7}).\end{proof}

\begin{theorem} Formulae (\ref{6.2.1}), (\ref{6.2.1a}) and (\ref{6.2.4}) hold for every $\vec \varkappa \in \cal{G} _{\infty }$.
The set $\cal{G} _{\infty }$ is Lebesgue measurable and satisfies
  (\ref{full})  with $\gamma _3=\delta/2 $.
\end{theorem}
\begin{proof}
By Theorem~\ref{T:Dec10}, (\ref{6.2.1}), (\ref{6.2.1a}) hold,
where $\lambda _{\infty}(\vec \varkappa )=\lambda $ for $\vec
\varkappa \in \cal{D}_{\infty}(\lambda )$. Using (\ref{h}), which
is proven in Corollary~\ref{Dec18}, with $\varkappa _{\infty
}=|\vec \varkappa |$, we easily obtain (\ref{6.2.4}). It remains
to prove (\ref{full}). Let us consider a small region $U_n(\lambda
_0)$ around an isoenergetic surface $\cal{D}_n(\lambda _0)$,
$\lambda _0>k_*^{2l}$. Namely, $U_n(\lambda _0)=\cup _{|\lambda
-\lambda _0|<r_n} \cal{D}_{n}(\lambda )$, $r_n=\epsilon
_{n-1}k^{-2\delta}$, $k=\lambda _0^{1/2l}$. Considering an
estimate of the type (\ref{3.73}) for $\lambda ^{(n)}(\vec
\varkappa )$, which holds in the
$\left(\epsilon_{n-1}k^{-2l+1-2\delta}\right)$-neighborhood of
$\cal{D}_n(\lambda _0)$, we see that $U_n(\lambda _0)$ is an open
set (a distorted ring with holes) and the width of the ring is of
order $\epsilon_{n-1}k^{-2l+1-2\delta}$. Hence, $|U_n(\lambda
_0)|=2\pi k r_n\left(1+o(k^{-\delta /2})\right)$. It easily
follows from Lemma \ref{L:3.9*+} that $U_{n+1}\subset U_n$.
Definition of $\cal{D}_{\infty }(\lambda _0)$ yield:
$\cal{D}_{\infty }(\lambda _0)=\cap _{n=1}^{\infty }U_n(\lambda
_0)$. Hence, $\cal{G}_{\infty }=\cap _{n=1}^{\infty }\cal{G}_n$,
where
 $\cal{G}_n=\cup _{\lambda
>\lambda _*}\cal{D}_n(\lambda )$. Considering that $U_{n+1}\subset U_n$
for every $\lambda _0>\lambda _*$, we obtain
$\cal{G}_{n+1}\subset \cal{G}_n$. Hence, $\left|\cal{G} _{\infty
}\cap
        \bf B_R\right|=\lim _{n\to \infty }\left|\cal{G} _{n}\cap
        \bf B_R\right|$. Summarizing volumes of the regions
        $U_{n}$, we easily conclude $\left|\cal{G} _{n}\cap
        \bf B_R\right|=|{\bf B_R}|\left(1+O(R^{-\delta /2})\right)$ uniformly in $n$. Thus, we have
        obtained (\ref{full}) with $\gamma _3=\delta/2 $.
\end{proof}

\appendix
\setcounter{equation}{0}
\section{Appendices.}
\subsection{Proof of Corollary~\ref{C:3.3}}\label{A:1}
 Let $\tau_0
\in \chi _2$. Taking into account the relation
$\lambda_j^{(1)}(\tau_0+2\pi p/N_1 a)=k^{2l}$ and the definition
of $C_2$, we see that
%
$|\lambda_j^{(1)}(\tau_0+2\pi p/N_1 a)-z| =\epsilon_1/2$.
%
Using  (\ref{3.6}) and the last equality, we easily obtain:
   $|\lambda_n^{(1)}(\tau_0+2\pi \hat{p}/N_1a)-z| \geq \epsilon _1/2$
for $ \lambda_n^{(1)}(\tau_0+2\pi \hat{p}/N_1a) \neq
\lambda_j^{(1)}(\tau_0+2\pi p/N_1a)$. Therefore, for any $z \in
C_2$,
    \begin{equation} \label{Sept23}
    \|(\tilde{H}^{(1)}(\tau_0)-z)^{-1}\| \leq 2/\epsilon_1,
    \end{equation}
    i.e., (\ref{3.10}) is proved for $\tau _0\in \chi _2$.
Now we consider $\tau$ in the complex\\ $(\epsilon_1
k^{-2l+1-\delta})-$neighborhood of $\chi_2$. By  Hilbert relation,
    \begin{align*}(\tilde{H}^{(1)}(\tau)-z)^{-1}=(\tilde{H}^{(1)}(\tau_0)-z)^{-1}
    +T_1T_2
    (\tilde{H}^{(1)}(\tau)-z)^{-1},\ \ \ \ \ \ \ \ \ \ \ &\\   T_1=
    (\tilde{H}_0(\tau_0)-z)^{-1}(\tilde{H}_0(\tau_0)+k^{2l})
,\ \ \
  T_2=(\tilde{H}_0(\tau_0)+k^{2l})^{-1}(\tilde{H}_0(\tau_0)-\tilde{H}_0(\tau)).\end{align*}
    Suppose we have checked that $\|T_1T_2\|<k^{-\delta}$. Then,
    using (\ref{Sept23}) and a standard argument as in Lemma \ref{L:2.4.1/2},
    we easily arrive at (\ref{3.10}).
 The estimate $\|T_1\|<4k^{2l}/\epsilon _1$,
easily follows from (\ref{Sept23}).   The estimate
$\|T_2\|<2l\epsilon_1 k^{-2l-\delta} $ easily follows from $|\tau
-\tau _0|<\epsilon_1 k^{-2l+1-\delta}$.  Thus,
$\|T_1T_2\|<8lk^{-\delta }$ and, hence, (\ref{3.10}) is proved.
Estimate (\ref{3.11}) follows from the Hilbert relation
$$(\tilde{H}^{(1)}(\tau)-z)^{-1}=(\tilde{H}_0(\tau)+k^{2l})^{-1}
+(\tilde{H}_0(\tau)+k^{2l})^{-1}(-W_1+k^{2l}+z)(\tilde{H}^{(1)}(\tau)-z)^{-1},$$
 an elementary estimate
$\|(\tilde{H}_0(\tau)+k^{2l})^{-1}\|_1<b_1b_2k^{-2l+2+2s_2}$ and
(\ref{3.10}).

 \subsection{Proof of estimate (\ref{3.2.30})}\label{A:2}
 Note that
        $$
        \Bigl| | \vec y_0 (\varphi)+\vec p_m(0)|_*^2-|\vec y
        (\varphi)+\vec p_m(0)|_*^2 \Bigr| =$$
        \begin{equation}\label{May18a}
        \Bigl| | \vec y_0 (\varphi)+\vec p_m(0)|_*^2-|\vec y_0
        (\varphi)+\vec h_1 (\varphi)+\vec p_m(0)|_*^2 \Bigr|=
        O\Bigl(|h_1|\cdot |\vec y_0(\varphi)+\vec p_m(0)|+|h_1|^2\Bigr).
        \end{equation}
   If $p_m(0) < 4k$, then, obviously,
    $|\vec y_0(\varphi)+\vec p_m(0)|_*^2=O(k^2)$. Considering  (\ref{2.75}), we
    arrive at
        \begin{equation*}
        \Bigl| |\vec y_0(\varphi)+\vec p_m(0)|_*^{2l}-|\vec y(\varphi)+\vec p_m(0)|_*^{2l}\Bigr|
        =O(|h_1|\cdot k^{2l-1})=O(k^{2l-2-4s_1-2\gamma_0-\delta}).
        \end{equation*}
    %
    %
    Combining this relation with (\ref{3.2.5.1}) where
    $\epsilon_0=k^{-4-6s_1-3\delta}$, we obtain
        \begin{multline*}
         \left| \frac{|\vec y_0(\varphi)+\vec p_m(0)|_*^{2l}-|\vec y(\varphi)+\vec p_m(0)
         |_*^{2l}} {|\vec y_0(\varphi)+\vec p_m(0)|_*^{2l}-k^{2l}}\right| \\
        \leq \frac{c|h_1|k^{2l-1}}{b_0k^{2l-9-12s_1-6\delta}}
        =O(b_0^{-1}|h_1|k^{8+12s_1+6\delta})=O(b_0^{-1}k^{7+8s_1-2\gamma_0+5\delta})=o(1),
        \end{multline*}
        the last estimate following from the restriction on $b_0$ in
        the statement of the lemma.
        Thus, (\ref{3.2.30}) holds for $p_m(0) < 4k$.
 Suppose $ p_m(0)\geq 4k.$ Using  (\ref{3.2.5.2}) and
(\ref{May18a}), we easily obtain:
        \begin{equation*}
         \left| \frac{|\vec y_0(\varphi)+\vec p_m(0)|_*^{2l}-|\vec y(\varphi)
         +\vec p_m(0)|_*^{2l}} {|\vec y_0(\varphi)+\vec p_m(0)|_*^{2l}-k^{2l}}\right|
        \leq 
        \frac{|h_1|}{p_m(0)}
        =O(k^{-2-4s_1-2\gamma_0-\delta}).
        \end{equation*}
         This
        means (\ref{3.2.30}) holds in this case too.

\subsection{Proof of estimate (\ref{May13c})}\label{A:3}
We use (\ref{Oct5}). First,
 we represent $\nabla \lambda^{(1)}(\vec \varkappa _1)-
     \nabla \lambda^{(2)}(\alpha,\vec \varkappa _2)$ as the sum of $A_1$ and $A_2$:
     \begin{align*}
A_1=\nabla \lambda^{(1)}(\varkappa _1\vec{\nu})-
     \nabla \lambda^{(2)}(\alpha,\varkappa _1\vec{\nu}),\ \
     \ A_2=
     \nabla \lambda^{(2)}(\alpha,\varkappa _1\vec{\nu})-\nabla \lambda^{(2)}(
     \alpha , \varkappa _2\vec{\nu}).
     \end{align*}
     Considering
(\ref{3.68}), we obtain $ |A_1 |<24\alpha \epsilon
_1^3k^{2l-1+\delta }$. Using the Mean Value Theorem and
(\ref{2.20**a}), we get: $|A_2|<ck^{2l-2}|\varkappa _2-\varkappa
_1|.$ Applying (\ref{3.71}) yields: $|A_2|<c\alpha \epsilon
_1^4k^{-1}.$ Adding the  estimates for $A_1$ and $A_2$ we arrive
at:
\begin{equation} \bigl|\nabla
\lambda^{(1)}(\varkappa _1\vec{\nu})-
     \nabla \lambda^{(2)}(\alpha,\varkappa _2\vec{\nu})
\bigr|<25\alpha \epsilon _1^3k^{2l-1+\delta }.\label{May13a}
\end{equation}
Similarly,
\begin{equation}
\bigl|\varkappa _1\nabla \lambda^{(1)}(\varkappa _1\vec{\nu})-
     \varkappa _2\nabla \lambda^{(2)}(\alpha,\varkappa _2\vec{\nu})\bigr|<
     26\alpha \epsilon _1^3k^{2l+\delta }. \label{May13e}
     \end{equation}
By Lemma \ref{L:2.13}, $\frac{\partial \varkappa _1}{\partial
\varphi}= \frac{\partial h_1 }{\partial \varphi}=O(k^{-2\gamma
_0+1})$. Substituting (\ref{May13a}), (\ref{May13e}) and the last
estimate into the formula for $F(\lambda,\vec \nu )$, we obtain
(\ref{May13c}).

\subsection{Lemma \ref{L:July5}}\label{A:4}
\begin{lemma} \label{L:July5} If $t$ is in the complex $(2k^{-1-4s_1-2\delta })$-neighborhood  of the
nonresonance set $\chi _1(\lambda ,\delta )$, then the
operator-function $\left(H^{(1)}(t)-z\right)^{-1}$ has a single
pole $z_0$ inside the contour $C_1$ and
\begin{equation}
\left \|(z-z_0)\left(H^{(1)}(t)-z\right)^{-1}\right\|<16,
\label{July3a}
\end{equation}
for every $z$ inside $C_1$.
\end{lemma}
\begin{proof}
  By
 (\ref{3.2.27}),
 \begin{equation}\Bigl|\det \frac{H^{(1)}(t)-z}{H_0(t)-z}-1 \Bigr|<
 2 \|W_1\|\left \|\left(H_0(t)-z\right)^{-1}\right\|_1 \label{r1}
 \end{equation}
 for every $z$ on the contour $C_1$.
 Using the estimate (\ref{2.8}) and considering that $H_0(t)$ is a diagonal operator,
 we easily
 obtain $\bigl\|\left(H_0(t)-z\right)^{-1}\bigr\|_1<k^{-2l+4+6s_1+\delta}$.
It immediately follows that
 the right-hand part of (\ref{r1}) is less then 1.
 Applying  Rouch\'{e}'s theorem, we conclude that the
 determinant has the same number of zeros and poles inside $C_1$.
 Considering that $\left(H_0(t)-z\right)^{-1}$ has a single pole,
 we obtain that  $\left(H^{(1)}(t)-z\right)^{-1}$ has a single pole too. We
 denote it by $z_0$. Obviously, $(z-z_0)\left(H^{(1)}(t)-z\right)^{-1}$ is holomorphic
 inside $C_1$. The definition of $C_1$, given in Corollary \ref{C:2.2},
 yields:
 $|z-z_0|<2k^{2l-2-4s_1-\delta
}$.
  Using the last inequality and estimate (\ref{2.40.2}),
  we obtain (\ref{July3a}) for any $z\in C_1$. By the Maximum
principle it can be extended inside $C_1$.
\end{proof}

\subsection{Proof of Lemma \ref{L:3.7.1.1}}\label{A:5}
By Lemma \ref{L:2.13}, Part 2, the function  $\vec \varkappa
_1(\varphi )$ is holomorphic in the
$(k^{-2-4s_1-2\delta})-$neighborhood of $\varPhi _1$ and  $
\lambda^{(1)}(\vec \varkappa _1(\varphi))=k^{2l}.$
Hence, the equation (\ref{3.7.1.2}) is equivalent to
$\lambda^{(1)}(\vec y(\varphi))=\lambda ^{(1)}(\vec
y(\varphi)-\vec b)+\epsilon_0$.
We use  perturbation formula (\ref{2.66}): $|\vec
y(\varphi)|_*^{2l}+f_1(\vec y(\varphi))
    =|\vec y(\varphi)-\vec b|_*^{2l}+f_1(\vec y(\varphi)-\vec
    b)+\epsilon_0.$ This equation can be rewritten as
    \begin{multline}\label{3.7.1.5}
    \Bigl(2\langle \vec y(\varphi),\vec b\rangle _*-|\vec b|^2\Bigr)
    \Bigl(|\vec y(\varphi)|_*^{2l-2}+\cdots+|\vec y(\varphi)-\vec
    b|_*^{2l-2}\Bigr)\\
    +f_1(\vec y(\varphi))
    -f_1(\vec y(\varphi)-\vec b)-\epsilon_0=0,
    \end{multline}
where $\langle \vec y, \vec b\rangle _*=y_1b_1+y_2b_2$. Using the
notation $\vec b=b_0(\cos \varphi_b,\sin \varphi_b)$, dividing
both sides of the equation (\ref{3.7.1.5}) by $2b_0k\Bigl(|\vec
y(\varphi)|_*^{2l-2}+\cdots+|\vec y(\varphi)-\vec
    b|_*^{2l-2}\Bigr)$, and considering that $\vec y(\varphi )=\vec \varkappa _1(\varphi)+\vec b=
    (k+h_1)\vec \nu +\vec b$, we obtain:
    \begin{equation}\label{3.7.1.6}
    \cos (\varphi-\varphi_b)-\epsilon_0g_1(\varphi)+g_2(\varphi)=0,
    \end{equation}
where $g_1(\varphi)
    =\Bigl[2b_0k\Bigl(|\vec y(\varphi)|_*^{2l-2}+\cdots+|\vec y(\varphi)-\vec
    b|_*^{2l-2}\Bigr)\Bigr]^{-1}$
    and
    $$g_2(\varphi)=\dfrac{\langle \vec h_1(\varphi),\vec b\rangle
_*}{b_0k}+\dfrac{b_0}{2k} +\Bigl(f_1(\vec y(\varphi))
    -f_1(\vec y(\varphi)-\vec b)\Bigr)g_1(\varphi),\ \ \vec h_1(\varphi)=h_1(\varphi)\vec \nu.$$
   Obviously $g_1=O\left(b_0^{-1}k^{-2l+1}\right)$.
Let us estimate $g_2(\varphi)$. Using the inequality (\ref{2.75})
for $h_1$, and considering that $b_0=o(k^{-1-4s_1-2\delta})$, we
easily obtain:
    $$
    \left|\frac{\langle \vec h_1(\varphi),\vec b\rangle
    _*}{b_0k}\right|\leq
    \frac{2h_1}{k}=O(k^{-2-4s_1-2\gamma_0-\delta}),\ \ \
    \frac{b_0}{2k}=o(k^{-2-4s_1-2\delta}).
    $$
 By (\ref{2.45}),
    $$ \left|f_1(\vec y(\varphi))-f_1(\vec y(\varphi)-\vec
    b)\right| \leq \sup |\nabla
    f_1|b_0=O(b_0k^{2l-1-2\gamma_0+\delta})=O(b_0k^{-2l+3+8s_1+5\delta}),$$
and therefore,
    $ \left|\bigl(f_1(\vec y(\varphi))-f_1(\vec y(\varphi)-\vec
    b)\bigr)g_1(\varphi)\right| =O(k^{-4l+4+8s_1+5\delta}).$
Since $s_1=\dfrac{2l-11}{32}$, we have
    $g_2(\varphi)=o(k^{-2-4s_1-2\delta}).$
Using $g_1(\varphi)=O\left(b_0^{-1}k^{-2l+1}\right)$ and
$\epsilon_0<b_0k^{2l-3-4s_1-3\delta}$, we obtain
    $\epsilon_0g_1(\varphi)=o(k^{-2-4s_1-2\delta}).$ Thus,
    \begin{equation}
    g_2(\varphi)-\epsilon_0g_1(\varphi)=o(k^{-2-4s_1-2\delta}).\label{Nov14}
    \end{equation}
    Suppose that $\varphi_b\pm
\dfrac{\pi}{2}$ is in the
$\left(\frac{1}{8}k^{-2-4s_1-2\delta}\right)$-neighborhood of
$\tilde \varPhi _1$. We draw two circles $C_{\pm}$ centered at
$\varphi_b\pm \dfrac{\pi}{2}$ with the radius
$\frac{1}{8}k^{-2-4s_1-2\delta}$. They are both inside the complex
$\frac{3}{4}k^{-2-4s_1-2\delta}$-neighborhood of $\varPhi _1$, the
perturbation series converging and the estimate (\ref{Nov14})
holds. For any $\varphi$ on $C_{\pm}$, $|\varphi-(\varphi_b \pm
\pi/2)|=\frac{1}{8}k^{-2-4s_1-2\delta}$ and, therefore,
    $|\cos(\varphi-\varphi_b)|>\frac{1}{16}k^{-2-4s_1-2\delta}>|g_2(\varphi)-\epsilon_0g_1(\varphi)|$
    for any $\varphi \in C_{\pm}$.
By Rouch\'{e}'s Theorem, there is only one solution of the
equation (\ref{3.7.1.6}) inside each $C_{\pm}$. If $\varphi_b+\pi
/2$ is not in the $(\frac{1}{8}k^{-2-4s_1-2\delta})-$neighborhood
of $\tilde \varPhi _1$, then
$|\cos(\varphi-\varphi_b)|>\frac{1}{16}k^{-2-4s_1-2\delta}$ in
$\varPhi _1$ and, hence,  equation (\ref{3.7.1.6}) has no
solution.
 Thus, there are at most two solutions in $\tilde \varPhi _1$ and
    $|\varphi^{\pm }_{\epsilon _0}-(\varphi_b \pm \pi /2)|<\frac{1}{8}k^{-2-4s_1-2\delta}.$


    \subsection{Proof of Lemma \ref{L:july5a}}\label{A:6}
Using the perturbation formula (\ref{2.66}), we obtain:
    \begin{align}\label{3.7.1.7}
    \lefteqn{
    \frac{\partial}{\partial \varphi}\lambda^{(1)}\bigl(\vec y(\varphi)\bigr)
    =\frac{\partial}{\partial \varphi}\left[\lambda^{(1)}\bigl(\vec
    y(\varphi)\bigr)-k^{2l}\right]=
    \frac{\partial}{\partial \varphi}\left[\lambda^{(1)}\bigl(\vec
    y(\varphi)\bigr)-\lambda^{(1)}\bigl(\vec y(\varphi)-\vec b\bigr)\right]=
    }& \notag \\
    &\hspace{1cm}
    \left \langle \nabla_{\vec y}\lambda^{(1)}\bigl(\vec
    y(\varphi)\bigr)-\nabla_{\vec y}\lambda^{(1)}\bigl(\vec y(\varphi)-\vec
    b\bigr),\frac{\partial}{\partial \varphi}\vec y(\varphi)\right \rangle
    _*=
    \notag \\
    &\hspace{1cm}
    \left\langle \nabla \bigl|\vec y(\varphi)\bigr|_*^{2l}-\nabla \bigl|\vec y(\varphi)-\vec
    b\bigr|_*^{2l},(k+h_1)\vec{\mu}+h_1^{\prime}\vec{\nu}\right \rangle _* +\notag\\
    &\hspace{2cm}
    \left\langle \nabla f_1\bigl(\vec y(\varphi)\bigr)-\nabla f_1\bigl(\vec y(\varphi)-\vec
    b\bigr),(k+h_1)\vec{\mu}+h_1^{\prime}\vec{\nu}\right \rangle _*,
    \end{align}
where $\vec{\nu}=(\cos \varphi, \sin \varphi)$ and
$\vec{\mu}=\vec{\nu}^{\prime}=(-\sin \varphi,\cos \varphi)$. Note
that
    \begin{multline}\label{3.7.1.8}
    \nabla \bigl|\vec y(\varphi)\bigr|_*^{2l}-\nabla \bigl|\vec y(\varphi)-\vec
    b\bigr|_*^{2l}=2l\bigl|\vec y(\varphi)\bigr|_*^{2l-2}\vec y(\varphi)-2l\bigl|\vec y(\varphi)-\vec
    b\bigr|_*^{2l-2}\bigl(\vec y(\varphi)-\vec b\bigr)\\
    =2l\bigl|\vec y(\varphi)\bigr|_*^{2l-2}\vec b
    +2l\Bigl[\bigl|\vec y(\varphi)\bigr|_*^{2l-2}-\bigl|\vec y(\varphi)-\vec b\bigr|_*^{2l-2}\Bigr]
    \bigl(\vec y(\varphi)-\vec b\bigr).
    \end{multline}
Substituting (\ref{3.7.1.8}) into (\ref{3.7.1.7}), we get $
    \frac{\partial}{\partial \varphi}\lambda_j^{(1)}\bigl(\vec
    y(\varphi)\bigr)=T_1+T_2+T_3 ,$
    \begin{align*}
    T_1 &
    =2l \bigl| \vec y(\varphi) \bigr| _*^{2l-2} \left\langle \vec
    b,(k+h_1)\vec{\mu}+h_1^{\prime} \vec{\nu} \right\rangle _* ,\\
    T_2 & =2l\left[ \bigl| \vec y(\varphi) \bigr| _*^{2l-2}-\bigl| \vec
    y(\varphi)-\vec b\bigr| _*^{2l-2} \right]
    \left\langle \vec y(\varphi)-\vec
    b,(k+h_1)\vec{\mu}+h_1^{\prime}\vec{\nu} \right\rangle _* ,\\
    T_3 & =\left \langle \nabla f_1\bigl( \vec y(\varphi) \bigr) - \nabla
    f_1\bigl( \vec y(\varphi)-\vec
    b \bigr),(k+h_1)\vec{\mu}+h_1^{\prime}\vec{\nu}\right\rangle _* .
    \end{align*}
Considering that $\varphi$ is close to $\varphi_b \pm \pi /2$, we
readily obtain:
    $\langle \vec b,\vec{\nu}\rangle _*=o(b_0),$
 $\langle \vec
b,\vec{\mu}\rangle _*=\pm b_0(1+o(1))$.  Using also estimates
(\ref{2.75}) for $h_1$, we get $T_1=\pm 2lb_0k^{2l-1}(1+o(1))$.
Substituting   $\vec y(\varphi)-\vec b=(k+h_1(\varphi)) \vec{\nu}$
into the formula for $T_2$ and taking into account that $\langle
\vec{\mu},\vec{\nu}\rangle _*=0$, we arrive at the estimate
$T_2=o\left(b_0k^{2l-1}\right)$. By (\ref{2.67a}) with $|m|=2$,
    $\Bigl|\nabla f_1\bigl(\vec y(\varphi)\bigr)-\nabla f_1\bigl(\vec y(\varphi)-\vec
    b\bigr)\Bigr|=O(b_0k^{-2l+4+12s_1+7\delta}).$
Hence, $T_3=o\left(b_0k^{2l-1}\right)$. Adding the estimates for
$T_1,T_2,T_3$, we get (\ref{3.7.1.6.1/2}).


%

\end{document}